\documentclass[12pt,preprint]{aastex}

\usepackage{epsfig}
\usepackage{lscape}
\usepackage{hyperref}

        \shorttitle{Intermediate Mass Binary Light Curves}
        \shortauthors{Williams et al.}

\newcommand{\noprint}[1]{}

\begin{document}

%%\received{}
%%\accepted{}

\title{ASAS Light Curves of Intermediate Mass Eclipsing Binary Stars and the Parameters of HI Mon}

\author{S. J. Williams\footnotemark[1], D. R. Gies}
\affil{Center for High Angular Resolution Astronomy and \\
 Department of Physics and Astronomy,\\
 Georgia State University, P. O. Box 4106, Atlanta, GA 30302-4106; \\
 swilliams@chara.gsu.edu, gies@chara.gsu.edu}

\author{J. W. Helsel}
\affil{Department of Physics, Furman University,\\
 3300 Poinsett Hwy. Greenville, SC 29613;\\
 john.helsel@furman.edu}

\author{R. A. Matson, S. Caballero-Nieves\footnotemark[1]}
\affil{Center for High Angular Resolution Astronomy and \\
 Department of Physics and Astronomy,\\
 Georgia State University, P. O. Box 4106, Atlanta, GA 30302-4106; \\
 rmatson@chara.gsu.edu, scaballero@chara.gsu.edu}

\footnotetext[1]{Visiting astronomer, Cerro Tololo 
Inter-American Observatory, National Optical Astronomy Observatory, 
which is operated by the Association of Universities for Research 
in Astronomy, under contract with the National Science Foundation.}

\begin{abstract}
We present a catalog of 56 candidate intermediate mass 
eclipsing binary systems extracted from the 3rd data 
release of the All Sky 
Automated Survey. We gather pertinent observational
data and derive orbital properties, including ephemerides, 
for these systems as a prelude to anticipated spectroscopic 
observations. We find that 37 of the 56, or $\sim$66\% of the
systems are not identified in the Simbad Astronomical Database
as known binaries. As a specific example, we show spectroscopic 
data obtained for the system HI Mon (B0 V + B0.5 V) observed 
at key orbital phases based on the computed ephemeris, and we
present a combined spectroscopic and photometric solution
for the system and give stellar parameters for each component. 

\end{abstract}
\keywords{stars: binaries -- eclipsing, 
  stars: individual (HI Mon)}

%%%%%%%%%%%%%%%%%%%%%%%%%%%%%%%%%%%%%%%%%%%%%%%%%%%%%%%%%%%%%

\setcounter{footnote}{1}

\section{Introduction}

Eclipsing binaries are a major source of fundamental astrophysical
parameters of stars such as mass and radius. These parameters
can help us better understand and constrain the overall picture of
stellar evolution and its consequences, including but
not limited to exoplanet host star parameters and studies of 
Galactic and extragalactic star formation and abundance evolution. 
In a recent review, \citet{tor10} list 190 stars with mass and
radius known to better than $\pm3$\%, only 26 (or $\sim14$\%) 
of which have a mass greater than 3 $M_{\rm \odot}$. This paucity
of data for intermediate to high mass stars in binary systems
is alarming. In particular, binary star evolution of systems
of this type are one formation channel for type Ia supernovae
\citep{men10},
one of the most studied phenomenon in determining cosmological
parameters. In recent years
large scale photometric surveys such as the All Sky Automated 
Survey \citep[ASAS;][]{poj02} have preceded other surveys 
looking for transiting exoplanets. The byproduct of these
searches will be a very large number of well sampled eclipsing
binary light curves. The impetus behind this
work is to find candidate intermediate and high mass binary 
systems with the
future goal of obtaining fundamental parameters by observing these
systems spectroscopically and resolving double lines. The 
applicability and validity of this technique has been 
demonstrated for lower mass
systems in the recent effort by \citet{deb11}.

We are starting a 
systematic observing program of newly identified or neglected
intermediate mass eclipsing binaries.
As an example of the kind of analyses we wish to pursue, 
we show here a case study of the binary system HI Mon (HD 51076). 
This system's variability
was first noted in \citet{wac68}. The system was classified as an
``OB'' type system in \citet{vog76} and has been listed as an
early B-type system since \citet{egg78}, although no spectra
for the system have been published. Spectroscopic
investigations of such binaries with hot components are a 
critical part of any parametric analysis since the
photometric colors are similar for all hot stars and are 
influenced by interstellar reddening. 

The extraction and selection of candidate massive eclipsing 
binary light curves is described in Section 2. This is followed
in Section 3 by a discussion of the analyses of these
systems and estimation of uncertainties associated with
computed values. Section 4 discusses the methods and results
of the combined light curve and spectroscopic analysis
performed on HI Mon.

%%%%%%%%%%%%%%%%%%%%%%%%%%%%%%%%%%%%%%%%%%%%%%%%%%%%%%%%%%%%%

\section{Sample Selection}

The sample selection criteria are based upon an ongoing 
spectroscopic observing program with the DeVeny Spectrograph 
(formerly the White Spectrograph on the 2.1 m telescope 
at Kitt Peak National Observatory) on the 1.8 m Perkins 
telescope at the Anderson Mesa Station of Lowell Observatory.
We searched the All Sky Automated Survey
Catalog of Variable Stars \citep{poj02,poj03,poj04,
poj05a,poj05b} which is a subset of the 3rd data release
\citep{poj02}.\footnote{http://www.astrouw.edu.pl/asas/?page=main}
The ASAS Catalog of Variable Stars lists preliminary
orbital periods for the identified eclipsing binaries.
Furthermore, Pilecki, Pojma\'{n}ski, \& Szczygie\l~have
prepared an on-line 
catalog\footnote{http:/www.astrouw.edu.pl/asas/?page=eclipsing}
that provides estimates of the system parameters based upon the
$V$ and $I$-band light curves.
We visually inspected several thousand individual light
curves and extracted photometry for those systems that met 
our specific criteria. 
First, systems were selected that were typically north of $-25\arcdeg$
with a few exceptions near this cutoff limit.
The most southern object is HD 122026 at $-27\arcdeg$
while the most northern object is DR Vul with a declination
of $+26\arcdeg$. The data releases so far from the ASAS have
reached a northern limit of $+28\arcdeg$, 
a hard limit for the time being.
In order to meet the exposure time limits for our spectral 
resolving power of $R = \lambda / \Delta\lambda \gtrsim 5000$,
we adopted a limiting $V$ magnitude of 11.
We then selected only light curves that had a primary eclipse
depth greater than 0.1 mag to include only systems with 
well defined photometric variations. In addition, to ensure that 
the secondary spectral lines will be readily detected, we chose 
light curves where the ratio of eclipse depths was not extreme. 
We avoided contact systems and selected those with 
``EA'' and ``EB'' light curves in the definition of \citet{giu83}
because we want to study systems where the stars have yet to 
interact with each other and have evolved thus far
as individual stars.

With this first cut of light curves, we then found the
corresponding $J$,$H$,$K_s$ magnitudes from the 2MASS catalog 
\citep{skr06} for each object. Using the maximum brightness
from the ASAS $V-$band light curve and the 2MASS $K$ magnitude, 
we computed $V-K$ colors for each candidate object. For many 
objects there is an unknown amount of interstellar reddening,
which will cause the $V-K$ color to increase and indicate a 
star with a cooler spectral type. Therefore,
to look for intermediate mass binary candidates 
($\gtrsim 3 M_{\rm \odot}$),
we used Table A5 in \citet{ken95} to eliminate later spectral
types based on the apparent $V-K < 0.7$, corresponding to
a spectral type earlier than F0 V. To add an additional 
constraint, we also computed the reddening-free $Q$ parameter
in the infrared from \citet{com02} with the 2MASS values. 
Stars with $Q < 0.1$ usually correspond to main sequence
stars with mass greater than 1.4 $M_{\odot}$. 

Observational parameters for candidate systems meeting all 
criteria are listed in Table 1. Column 1 lists the ASAS
identification tag in the ASAS catalog of variable stars.
In column 2 we list an alternate object name, typically
found in the Simbad database. Following this, columns 3 and
4 contain the right ascension and declination of the objects
in the 2000.0 epoch as found from the Simbad Astronomical 
Database. Columns 5 and 6 list the $V-K$ and $Q$ values
as discussed above. Next is column 7, showing the
spectral classification found in the literature for each
object while column 8 lists the reference for each entry.
Column 7 demonstrates that most of the systems are 
intermediate mass binaries with an A- or B-type primary
star. 
Column 9 lists whether the system was known to be a
binary before the ASAS Catalog of Variable Stars and
gives the reference to the work where the 
system was first discovered to be a binary. We wish
to note that only 21 of our 56 candidates listed in 
Table 1 are contained in the on-line atlas of Pilecki, 
Pojma\'{n}ski, \& Szczygie\l. 

%%%%%%%%%%%%%%%%%%%%%%%%%%%%%%%%%%%%%%%%%%%%%%%%%%%%%%%%%%%%%

\section{Light Curve Analysis}

For each system, we fit the $V-$band light curve using the 
Eclipsing
Light Curve (ELC) code of \citet{oro00}. The program solves 
for orbital and physical parameters relevant to the 
light curve. These parameters include the epoch 
of mid-eclipse of the primary star $T_{\rm o}$,
the orbital period $P$, the orbital inclination $i$, the 
Roche lobe filling factor of each star, the temperature 
of each star, and the eccentricity and longitude of 
periastron of the primary $\omega$ 
in cases with elliptical orbits. In ELC, the primary is
defined as the more massive star.
It should be noted that the sizes and temperatures of the
stars are relative, and without spectroscopic 
analysis, we cannot determine their true values. 

In finding a best fit light curve for each system, 
our solutions with ELC create $\gtrsim40,000$ simulated 
light curves and record the
$\chi^{2}$ statistic associated with each light curve. We 
project $\chi^{2}$ over the range of each parameter of 
interest and focus on the region around 
$\chi^{2}_{\rm min}$. The resulting $\chi^{2}$ curves are 
not symmetric about the minimum, so our final estimate
for the uncertainty of any particular parameter was the 
larger of the two adjoining $\chi^{2}_{\rm min} + 1$ values. 
The best fit
values are listed in Table 2, with uncertainties listed in 
parentheses in units of the last digit quoted.
Column 1 in Table 2 lists the ASAS ID and column 2 gives the
alternate ID. The orbital period for each system is given 
in column 3. The period uncertainties are smaller in the
shorter period systems
due to the large number of orbits covered in the ASAS.
Similarly, column 4 is the epoch of mid-eclipse of the 
primary. Note that this is different than the epoch of
periastron for binaries with eccentric orbits. The ASAS
Catalog of Variable Stars also lists $P$ and 
$T_{\rm o}$. Our orbital periods do not differ greatly,
but our reference epochs were chosen to be more recent
than those listed in \citet{poj02}, reflecting times
closer to when spectroscopic observations may be taken.
Next is the orbital inclination in column 5, 
followed by the eccentricity of the orbit in column 6,
and the longitude of periastron in column 7, where applicable.
The value of the eccentricity found for BD+11 3569 of 
$0.006 \pm 0.002$ is only 3$\sigma$ different than zero. We 
therefore estimate that given the typical 
uncertainty in the ASAS data, any system listed with
$e=0.0$ probably has an eccentricity no greater than 0.01. 
It should also be noted
that the longitude of periastron given here follows the
spectroscopic orbit definition, i.e., measured from the plane 
of the sky through the center of mass to the primary. 

To check the accuracy of our results, we can compare 
our values listed in Table 2 with those previously published.
For example, a detailed light curve study of YY Sgr was
made by \citet{lac93}. His fit gives a longitude of 
periastron of 214.52 $\pm$ 0.46 deg, matching with
our value of 214.1 $\pm$ 0.5 deg within uncertainties. The
eccentricity in Lacy's work of 0.1573 $\pm$ 0.0008 and
the inclination of 88.89 $\pm$ 0.12 deg are close
to our derived values of 0.117 $\pm$ 0.001 and 88.1
$\pm$ 0.1 deg, respectively, but still differ by more 
than 3$\sigma$. 
This is most likely due to the differences in the 
data analyzed. The data set in \citet{lac93}
consists of 717 measures most of which were made during 
eclipses where the above parameters are best constrained. 
Our data set
includes 601 measures more or less evenly distributed 
around the orbit, and thus eclipses are not as well
measured and our resulting parameters are not as 
accurate nor as well 
constrained. Future observations, specifically targeted
to get data during eclipses, will help resolve these 
differences.

One of our goals in selecting the candidates in Tables 1 and
2 is to collect a set of data for early-type eclipsing 
binaries. 
Using previously published spectral classifications we
construct a histogram of the number of objects as a
function of spectral type in Figure 1. This plot may be
somewhat biased, as several spectral types are from
\citet{can24}. Subtle differences in the He I 4471 \AA~
line are the determinant for discerning late-B from early-A
spectral types. Early photographic plate spectra may not have
been sufficiently sensitive to the changes in this 
temperature region, and we suspect several systems are
consequently lumped into the ``A0'' category. Regardless, this
plot shows that for the objects with published spectra,
we have met our goal of finding systems earlier than F0.
In fact, we have more stars (30) hotter than A0 than
stars (17) cooler than A0. Note that HD 60389, with a
spectral type estimate of G6 V does not appear in this
figure. Note also that not all systems discussed here
have spectral types, and some previously known systems
have spectral type estimates for both stars plotted in
the histogram. Light curve solutions for the full sample
are given in the Appendix.

%%%%%%%%%%%%%%%%%%%%%%%%%%%%%%%%%%%%%%%%%%%%%%%%%%%%%%%%%%%%

\section{HI Mon Spectra}

Spectra for HI Mon were obtained at the Cerro Tololo
Inter-American Observatory (CTIO) 4-m Blanco Telescope between
2009 Dec 23 and 2009 Dec 27.
The RC-spectrograph was used with the G380 (1200 grooves 
mm$^{-1}$) grating in second order and blue collimator 
giving wavelength coverage from 3932 to 4750 \AA~with a
three pixel resolution of $R = \lambda / \delta\lambda =
5400$. Exposure times ranged from 150 to 180 s in order to
obtain a signal-to-noise ratio (S/N) $>100$~pixel$^{-1}$. 
Helium-Neon-Argon comparison lamp spectra were obtained before
and after each target observation, and bias and flat-field
spectra were obtained nightly. The spectra were extracted
and calibrated using standard routines in IRAF\footnote{IRAF 
is distributed by the National Optical Astronomy Observatory, 
which is operated by the Association of Universities for 
Research in Astronomy (AURA) under cooperative agreement 
with the National Science Foundation.}. Each 
continuum-rectified spectrum was then transformed to a
common heliocentric wavelength grid in log $\lambda$ 
increments.

To compute important astrophysical parameters for individual
stars in double-lined spectroscopic binaries that are also
eclipsing, spectra need to be obtained near quadrature
points (times of maximum velocity separation). This will 
allow for the estimation of velocity semiamplitudes for each
star in the system, and leads directly to an estimate of
the mass ratio. To observe the binary system in such a
configuration, one needs to observe the system at very 
specific times in its orbit. Therefore, an accurate
ephemeris from the more easily obtained photometric light 
curve data is desirable. Using this ephemeris for HI Mon
as computed above, we observed the system near quadrature 
phases in order to test both the accuracy of our ephemerides
and to test whether this approach can yield accurate ($<3\%$
uncertain) astrophysical parameters for stars. 
Two such observations
near quadrature points (orbital phase 0.31 and 0.75) 
are shown in Figure 2 along with identified spectral 
features.

\subsection{Radial Velocities}

Radial velocities were measured using four spectral lines,
\ion{He}{1} $\lambda$4026, H$\delta$ $\lambda$4101,
\ion{He}{1} $\lambda$4143, and H$\gamma$ $\lambda$4340.
A template-fitting scheme \citep{gie02} was used that measures
velocities by using model spectra weighted by a flux
ratio to match both the shifts and line depths in the 
observed spectra. We found no evidence for emission or 
intrinsic line asymmetries in any spectral feature.

The BSTAR2006 grid of stellar models from \citet{lan07} was
used to derive template spectra. These models are based on
the line-blanketed, non local thermodynamic equilibrium
(NLTE), plane-parallel, hydrostatic atmosphere code TLUSTY
and the radiative transfer code SYNSPEC 
\citep{hub88,hub95,hub98}. The process of finding templates
began by making initial estimates for temperatures, 
gravities, projected
rotational velocities, and relative flux contributions from
each star. Parameters for model templates were then checked
against spectral features in the tomographically reconstructed
spectra of each star (Section 4.2). The parameters were
changed and new templates made after initial fits to the
light and radial velocity curves (Section 4.3) indicated
that slightly different values were more appropriate. The
velocity analysis was performed again until the best fit
was obtained. Based on the relative line depths of
spectral components, we were able to obtain an observational
monochromatic flux ratio in the blue of 
$F_2/F_1 = 0.70 \pm 0.05$. 

The four values for velocity from each spectrum for each
component were averaged, and the standard deviation of the
mean was calculated. Each of these values are listed in 
Table 3, along with the observation date, orbital phase,
and observed minus calculated values. 

\subsection{Tomographic Reconstruction}

We used the Doppler tomography algorithm of \citet{bag94}
to separate the primary and secondary spectra of HI~Mon.
This iterative method uses the eight observed composite
spectra from outside eclipse phases, 
their observed velocity shifts, and an assumed monochromatic
flux ratio to derive individual component spectra. The 
best flux ratio was the one that best matched the line
depths in the reconstructions with those in the model spectra.
Figure 3
shows the reconstructed spectra for the primary and secondary,
along with the best fit model spectra for each. The region
containing the diffuse interstellar absorption band around 
$\lambda$4428 \AA, and other interstellar features were 
removed when performing the
reconstruction, as otherwise this introduced noise into the 
final reconstructed spectra. Of particular importance are
the fits of the wings of the H$\gamma$ $\lambda$4340 line,
as these are produced by linear Stark broadening and are
good estimators for the gravities of the stars. The relative
depths of the \ion{He}{1} $\lambda$4471 and \ion{Mg}{2}
$\lambda$4481 lines give an indication of the temperature in
early B-type stars. Note that fast rotation
can lead to changing depths of these features, but these stars
are not seen to be rotating quickly. 

The final reconstructed spectra were fitted with 
TLUSTY/SYNSPEC model synthesis spectra (see Section 4.1). 
Estimates listed in Table 4 were made by comparing the
reconstructed and model spectral line profiles against a
grid of test values \citep[see][]{wil08,wil09}. Spectral
type estimates were obtained by comparing the effective 
temperature against Table 2 of \citet{boh81} and Table 3
of \citet{har88}. The gravities of each star indicate they
are main sequence objects, so final estimates for the 
spectral classifications are B0 V and B0.5 V. 

\subsection{Combined Radial Velocity and Light-Curve Solution}

ELC was used to fit both the light curve
and the radial velocity curves simultaneously. The parameters
allowed to vary included the orbital period, epoch of
primary eclipse, $T_{\rm 0}$, 
inclination, mass ratio, primary velocity semiamplitude,
effective temperatures of each star, and Roche lobe filling
factor for each star. The Roche lobe filling factor is
defined by \citet{oro00} as the ratio of the radius of the
star toward the inner Lagrangian point ($L_1$) to the 
distance to $L_1$ from the center of the star, 
$f \equiv x_{\rm point}/x_{\rm L1}$. Nonzero eccentricities
for HI~Mon were explored during initial fitting, but
rejected based on the higher $\chi^2$ values for those
fits. HI~Mon is therefore a ``false member'' of candidate
eccentric systems listed in \citet{heg88}. The resulting
best fit velocity curves are shown in Figure 4
and the best fit light curve is shown in Figure 5. 
During eclipses (at phase 0.0 and 0.5 in both Figures 4 
and 5), bumps in the radial velocity curve can be seen
due to the rotational Doppler shifts of the unobscured
portions of the eclipsed star. This
changes the shape of the observed spectral features,
and gives rise to the velocity change, a perturbation
known as the Rossiter-McLaughlin effect.

To estimate uncertainties based on our best fit, we used the
nearly 40,000 recorded $\chi^2$ values for each fit to the
light and radial velocity curves performed by ELC. We
projected the well explored $\chi^2$ surface onto each
parameter of interest. The lowest $\chi^2$ value is
found for each parameter and the 1-$\sigma$ uncertainty
estimated from the region where 
$\chi^2\le\chi^{2}_{\rm min}+1$. These values are listed
in Table 5 for the circular orbital solution. Astrophysical
parameters of interest are listed in Table 6. 

Comparing our results with those from the on-line catalog
of Pilecki, Pojma\'{n}ski, \& Szczygie\l~reveals a few 
differences. Our inclination of 80.0 $\pm$ 0.2 agrees
well with their inclination of 80.9. Also matching
reasonably well are our temperature ratios, 0.963 from
our best fit, and 0.971 in their catalog. Their estimations
for polar radius of each star relative to its Roche lobe polar
radius are 0.793 and 0.830 for the primary and secondary,
respectively. These values are systematically larger
than the values we derive from our best ELC fit, most
likely due to the additional constraints from the
spectra. 
The most discrepant set of values are for the temperatures
of the individual stars. They list 8110 K for the primary 
and 7870 K for the secondary. They arrive at the value
for the primary based on the $V-I$ color index. There are
several problems with estimating the temperature of
intermediate mass stars with only a color index. First,
interstellar extinction has not been taken into account.
Distances for mid to early B-type stars of the magnitude ranges
covered in the ASAS ($V\sim$7-12) are going to be in excess
of one kpc. Since these intermediate mass stars are typically
found in or near the galactic plane, where extinction is 
more pronounced, any estimation of the temperature based
on photometry will be effected by an unknown amount. Also,
because the temperatures of intermediate mass binaries are high,
their photometric colors sample the Rayleigh-Jeans tail of the 
energy distribution and the color differences are small. 
For example, the difference in $V-I$ is only 0.05 mag between
a 20 kK star and a 25 kK star \citep{bes98}.
Both of these factors illustrate the need for spectroscopy
to understand the nature of these intermediate mass
binaries.

\subsection{Discussion}

With a modest number of spectra (11 total), we have 
determined the masses and radii of each component in HI~Mon
to better than 3$\%$ accuracy, as is seen in Table 6. Both
stars are within their Roche lobes but experience tidal
distortion, as evidenced in the light curve in Figure 5.
Tidal evolution in the HI~Mon system is also seen in the
fact that the projected rotational velocities from the
tomographic reconstructions (Table 4) and from the 
synchronous rates from ELC
(Table 6) are consistent within uncertainties. This 
indicates that the system has had sufficient time to 
undergo tidal evolution in order to attain synchronous 
rotation.

To make a more quantitative estimate of the age of the
system, we used the effective temperatures and 
effective radii of
each component from Table 6 and plotted the two stars of 
HI~Mon on a theoretical Hertzsprung-Russell (H-R) diagram
to compare their locations with evolutionary tracks. Since
the stars are tidally distorted, the effective radius of 
each star is the radius of a sphere with the same volume.
Figure 6
shows evolutionary tracks from \citet{sch92} for stars
of 12 and 15 $M_{\odot}$ and isochrones from \citet{lej01}
for solar metallicity for ages of 1.5, 2.5, and 3.5 Myr. 
The ages of the stars are most consistent with an age of
2.5 Myr.
The evolutionary tracks from \citet{sch92} are for 
nonrotating stellar models. The ratios of spin angular
velocity to critical angular velocity are 
$\Omega/\Omega_{crit} =$ 0.42 for the primary and 0.43
for the secondary. The evolutionary tracks for these 
moderate rotation rates are only slightly steeper
and more extended in time than those illustrated 
\citep{eks08}. Thus, our derived age is a lower limit. 

We can also estimate the distance to the system by fitting
a spectral energy distribution (SED) to photometric
measurements from the literature (since our spectra were
not flux calibrated). We used measurements from \citet{vog76}
for $U$, $B$, and $V$ values, noting that the $V$-band 
measurement of maximum light from the ASAS light curve
is consistent with these measures, meaning they were likely
not taken during eclipse. In addition, we used 2MASS 
\citep{skr06} measurements to constrain the near-IR part
of the SED. Str\"{o}mgren photometric system measurements
reported by \citet{egg78} were not consistent with the other
photometry and were not used in the calculation of the SED. 
The SED fit is shown in Figure 7
with the $U$, $B$, $V$, $J$, $H$, and $K_s$ magnitudes
transformed to fluxes using the calibrations of \citet{col96}
and \citet{coh03}. Model spectra for each star from 
\citet{lan07} with parameters for our best fit from ELC (Table 6)
were scaled in the blue by the flux ratio (also Table 6)
of 0.81 $\pm$ 0.09 and added to form the total SED of the
system. This fit of the SED results in a 
limb-darkened angular diameter for the primary of 
$\theta_{\rm LD} = 21.5 \pm 0.7$ $\mu$as and a reddening
of $E(B-V)=0.35 \pm 0.02$ mag and a ratio of 
total-to-selective extinction of $R=3.2 \pm 0.1$. 
This reddening value matches reasonably well with the
values in the literature of 0.30 \citep{egg78} and
0.38 \citep{vog76}. Combining
this angular diameter with the physical radius for the 
primary from Table 6, 
we estimate the distance to HI~Mon to be 2.26 $\pm$ 0.04 kpc.
By contrast, this value is not consistent with previous
photometric distance estimates for HI~Mon of 
3.89 kpc \citep{egg78} and 3.98 kpc \citep{vog76}. 

HI~Mon has galactic coordinates of $\ell = 217\fdg03$ and
$b = -0\fdg87$ \citep{ree05}. The closest galactic cluster
is NGC 2311 that is located at $\ell = 217\fdg76$ and 
$b = -0\fdg69$ \citep{pia10}. \citet{pia10} find a distance 
to the cluster of 2.2 $\pm$ 0.4 kpc and a reddening of 
$E(B-V) = 0.25 \pm 0.05$. However, they find an age for
the cluster of $\sim$100 Myr, making HI~Mon far too young 
to be associated with NGC 2311. The distance we find for 
HI~Mon and its location suggest it is part of ``Group A'' 
as defined by \citet{vog76}. This group is an association
of luminous stars in the constellation of Monoceros at a
distance of about 2.2 kpc that is part of the local arm
of the galaxy. 

%%%%%%%%%%%%%%%%%%%%%%%%%%%%%%%%%%%%%%%%%%%%%%%%%%%%%%%%%%%%%

\acknowledgments

This material is based on work supported by the National
Science Foundation under grants AST$-$0606861 and 
AST$-$1009080.
This publication makes use of data products from the Two 
Micron All Sky Survey, which is a joint project of the 
University of Massachusetts and the Infrared Processing 
and Analysis Center/California Institute of Technology, 
funded by the National Aeronautics and Space Administration 
and the National Science Foundation. This research has made
use of the SIMBAD database, operated at CDS, Strasbourg,
France.

{\it Facility:} \facility{Blanco}

%%%%%%%%%%%%%%%%%%%%%%%%%%%%%%%%%%%%%%%%%%%%%%%%%%%%%%%%%%%%%

% Appendix

\appendix
\section{Light Curves of Intermediate Mass Binaries in the ASAS}

Figures of the light curves and fits for each of these 
systems are included in the
on-line material as Figure Set A. These plots contain 
the ASAS data with uncertainties, phased to
our derived orbits listed in Table 2. These best fit 
orbits are shown by green lines passing through the 
data between orbital phases $-0.2$ to 1.2 to show
the primary eclipse in its entirety. There are several
systems that have interesting features in
their light curves. For example, Figure %A\_19% 
26 shows the light curve for NSV 3433. It shows one of the
larger temperature differences between components in
this sample. It also shows an illumination effect on 
the secondary by the flux of the hotter primary. 
The out-of-eclipse
flux is higher just before and after the secondary eclipse,
indicating that the side of the secondary facing the
primary is slightly hotter and giving off more light at
this orbital phase. Figure %A\_29% 
36 illustrates the light curve of HD 63818. This
system has a moderately high eccentricity (0.132) with 
$\omega=343\arcdeg$, meaning the periastron of 
the orbit is close to the plane of the sky.  This
causes the secondary eclipse to be shifted from the
circular case where it would be at 0.5 in orbital phase.
A particularly illustrative example of this is
TYC 1025-1524-1 seen in Figure %A\_50%
57. The widths of the eclipses are different owing
to an eccentricity of 0.367. Periastron occurs near
phase 0.02 ($\omega=280\arcdeg$), and the deeper,
narrow eclipse at phase 0.0 results from the rapid and
close passage of the secondary star in front of the
primary star. 
Several systems have very well constrained 
light curves such as HD 48866 in Figure %A\_9% 
16 and TT Pyx seen in Figure %A\_33% 
40 that will make for highly accurate determinations
of mass and radius when combined with double-lined
spectroscopic orbits, and therefore, they represent the 
jewels of the collection.

%%%%%%%%%%%%%%%%%%%%%%%%%%%%%%%%%%%%%%%%%%%%%%%%%%%%%%%%%%%%%

% Bibliography

%%%%%%%%%%%%%%%%%%%%%%%%%%%%%%%%%%%%%%%%%%%%%%%%%%%%%%%%%%%%%

\clearpage

\begin{deluxetable}{lccrrrccc}
\tabletypesize{\scriptsize}
\tablewidth{0pt}
\rotate
\tablecaption{Observational System Parameters\label{obsparams}}
\tablehead{
\colhead{}                        &
\colhead{}                        &
\colhead{RA}                      &
\colhead{Declination}             &
\colhead{$V-K$}                   &
\colhead{$Q$}                     &
\colhead{Spectral}                &
\colhead{Classification}          &
\colhead{Discovery}               \\
\colhead{ASAS ID}                 &
\colhead{Other ID}                &
\colhead{(J2000.0)}               &
\colhead{(J2000.0)}               &
\colhead{(mag)}                   &
\colhead{(mag)}                   &
\colhead{Classification}          &
\colhead{Reference}               &
\colhead{Reference}               \\
\colhead{(1)}                     &
\colhead{(2)}                     &
\colhead{(3)}                     &
\colhead{(4)}                     &
\colhead{(5)}                     &
\colhead{(6)}                     &
\colhead{(7)}                     &
\colhead{(8)}                     &
\colhead{(9)}                     }
\startdata
053838+0901.2   & HD 37396                  & 05 38 38.1 & +09 01 10.7   &   0.33   &  $-$0.10 & A0          & 1       & 14    \\
054816+2046.1   & HD 247740                 & 05 48 16.5 & +20 46 10.3   &   0.18   &  $-$0.14 & B8          & 1       & ASAS  \\
060857+1128.9   & HD 252416                 & 06 08 57.2 & +11 25 56.2   &   0.19   &   0.00   & B8          & 1       & ASAS  \\
060927$-$1501.7 & TYC 5933$-$142$-$1        & 06 09 26.6 & $-$15 01 42.7 &   0.31   &  $-$0.10 & \nodata     & \nodata & ASAS  \\
062556$-$1254.5 & HD 45263                  & 06 25 56.1 & $-$12 54 29.1 &   0.26   &  $-$0.07 & B9 IV       & 2       & ASAS  \\
063347$-$1410.5 & HD 46621                  & 06 33 46.8 & $-$14 10 30.9 &   0.28   &   0.14   & A1 IV/V     & 2       & ASAS  \\
064010$-$1140.3 & HD 47845                  & 06 40 09.4 & $-$11 40 17.8 &   0.47   &   0.05   & A2 IV       & 3       & ASAS  \\
064118$-$0551.1 & 2MASS06411762$-$0551065   & 06 41 17.6 & $-$05 51 06.9 &   0.43   &  $-$0.04 & \nodata     & \nodata & ASAS  \\
064538+0219.4   & HD 48866                  & 06 45 37.8 & +02 19 21.1   &  $-$0.04 &  $-$0.10 & A (V)+A (V) & 3       & 15    \\
064609$-$1923.8 & HD 49125                  & 06 46 09.4 & $-$19 23 50.1 &  $-$0.03 &  $-$0.08 & B9 V        & 2       & ASAS  \\
064715+0225.6   & HD 289072                 & 06 47 14.6 & +02 25 34.3   &   0.37   &  $-$0.03 & B5          & 4       & ASAS  \\
064745+0122.3   & V448 Mon                  & 06 47 45.0 & +01 22 18.0   &   0.14   &   0.09   & B5          & 4       & 16    \\
065534$-$1013.2 & HD 51082                  & 06 55 33.9 & $-$10 13 12.6 &   0.02   &  $-$0.15 & A0 V        & 3       & 17    \\
065549$-$0402.6 & HI Mon                    & 06 55 49.1 & $-$04 02 35.8 &   0.22   &  $-$0.04 & B3/5 II     & 3       & 18    \\
070105$-$0358.2 & HD 52433                  & 07 01 05.1 & $-$03 58 15.5 &  $-$0.17 &   0.02   & B3 III      & 3       & 19    \\
070238+1347.0   & HD 52637                  & 07 02 38.2 & +13 46 58.6   &   0.23   &   0.03   & A0          & 1       & 20    \\
070636$-$0437.4 & AO Mon                    & 07 06 36.3 & $-$04 37 24.5 &  $-$0.43 &  $-$0.04 & B3+B5       & 5       & 21    \\
070943+2341.7   & BD +23 1621               & 07 09 43.3 & +23 41 42.8   &   0.70   &   0.01   & \nodata     & \nodata & ASAS  \\
070946$-$2005.5 & NSV 3433                  & 07 09 46.2 & $-$20 05 35.2 &  $-$0.09 &   0.01   & \nodata     & \nodata & 17    \\
071010$-$0035.1 & HD 54780                  & 07 10 10.4 & $-$00 35 08.7 &   0.22   &  $-$0.17 & B5/7 (III)  & 3       & 15    \\
071203$-$0139.1 & HD 55236                  & 07 12 03.5 & $-$01 39 04.2 &   0.54   &   0.04   & A2 III/IV   & 3       & ASAS  \\
071702$-$1034.9 & HD 56544\tablenotemark{a} & 07 17 02.4 & $-$10 34 56.6 &  $-$0.30 &   0.02   & A0          & 1       & ASAS  \\
072201$-$2552.6 & CX CMa                    & 07 22 01.0 & $-$25 52 35.9 &  $-$0.30 &  $-$0.10 & B5 V        & 6       & 21    \\
073053+0513.7   & HD 59607                  & 07 30 53.5 & +05 13 37.3   &  $-$0.27 &  $-$0.11 & B8          & 1       & ASAS  \\
073348$-$0940.9 & HD 60389                  & 07 33 48.4 & $-$09 40 52.6 &   0.45   &  $-$0.02 & G6 V        & 3       & 22    \\
074355$-$2517.9 & HD 62607                  & 07 43 55.2 & $-$25 17 53.7 &   0.08   &  $-$0.08 & B9.5 V      & 2       & ASAS  \\
074717$-$0519.8 & HD 63141                  & 07 47 17.3 & $-$05 19 51.1 &   0.29   &  $-$0.01 & A0 V        & 3       & 17    \\
074928$-$0721.6 & BD $-$06 2317             & 07 49 27.5 & $-$07 21 36.0 &   0.23   &  $-$0.05 & A0          & 7       & ASAS  \\
075052+0048.0   & HD 63818                  & 07 50 52.4 & +00 48 04.1   &  $-$0.06 &  $-$0.07 & A0          & 1       & ASAS  \\
080617$-$0426.8 & V871 Mon                  & 08 06 17.3 & $-$04 26 46.8 &   0.64   &   0.02   & A3/5mA7-F0  & 3       & 23    \\
081749$-$2659.7 & HD 69797                  & 08 17 48.7 & $-$26 59 37.5 &   0.66   &  $-$0.22 & A1 V        & 8       & ASAS  \\
083245+0247.3   & BD +03 2001               & 08 32 45.3 & +02 47 16.5   &   0.79   &   0.07   & A2          & 9       & ASAS  \\
084831$-$2609.8 & TT Pyx                    & 08 48 30.9 & $-$26 09 47.8 &   0.16   &  $-$0.08 & B9.5 IV     & 2       & 24    \\
101120$-$1956.3 & HD 88409                  & 10 11 19.5 & $-$19 56 20.4 &   0.49   &   0.04   & A5 II/III   & 2       & ASAS  \\
135949$-$2745.5 & HD 122026                 & 13 59 49.3 & $-$27 45 26.4 &   0.52   &   0.03   & A1/2 III/IV & 8       & ASAS  \\
160851$-$2351.0 & TYC 6780$-$1523$-$1       & 16 08 51.0 & $-$23 51 01.4 &   1.56   &   0.09   & \nodata     & \nodata & ASAS  \\
165354$-$1301.9 & HD 152451                 & 16 53 54.2 & $-$13 01 57.5 &   1.85   &   0.14   & A9 V        & 2       & ASAS  \\
170158+2348.4   & HD 154010                 & 17 01 57.5 & +23 48 22.5   &   0.56   &  $-$0.01 & A2          & 1       & ASAS  \\
173421$-$1836.3 & HD 159246                 & 17 34 21.1 & $-$18 36 21.7 &   0.57   &  $-$0.08 & B9 III      & 2       & ASAS  \\
174104+0747.1   & V506 Oph                  & 17 41 04.3 & +07 47 04.3   &   1.12   &   0.10   & A9          & 10      & 25    \\
175659$-$2012.2 & HD 312444                 & 17 56 58.8 & $-$20 12 05.7 &   1.46   &  $-$0.04 & A3          & 3       & ASAS  \\
175859$-$2323.1 & HD 313508                 & 17 58 59.3 & $-$23 23 07.9 &   0.68   &  $-$0.76 & B8          & 3       & ASAS  \\
180903$-$1824.5 & HD 165890                 & 18 09 02.9 & $-$18 24 29.5 &   0.66   &  $-$0.01 & B7/8 Ib     & 2       & ASAS  \\
181025+0047.7   & HD 166383                 & 18 10 25.7 & +00 47 47.0   &   0.88   &   0.15   & B3/5 II/III & 3       & ASAS  \\
181328$-$2214.3 & HD 166851                 & 18 13 28.0 & $-$22 14 07.2 &   0.67   &  $-$0.03 & B9 III/IV   & 2       & ASAS  \\
181909$-$1410.0 & HD 168207                 & 18 19 09.0 & $-$14 10 00.6 &   1.05   &   0.05   & B1/2(N)     & 2       & ASAS  \\
183129$-$1918.8 & BD $-$19 5039             & 18 31 28.9 & $-$19 18 47.3 &   0.90   &   0.02   & B3 V        & 11      & ASAS  \\
183219$-$1117.4 & BD $-$11 4667             & 18 32 19.0 & $-$11 17 23.6 &   2.21   &   0.02   & B1:V:pe     & 12      & 26    \\
184223+1158.9   & BD +11 3569               & 18 42 23.2 & +11 58 57.4   &  $-$0.24 &   0.01   & A2          & 9       & ASAS  \\
184327+0841.5   & TYC 1025$-$1524$-$1       & 18 43 26.9 & +08 41 32.1   &   0.53   &   0.00   & \nodata     & \nodata & ASAS  \\
184436$-$1923.4 & YY Sgr                    & 18 44 35.9 & $-$19 23 22.7 &   0.41   &   0.12   & B5+B6       & 13      & 27    \\
185051$-$1354.6 & HD 174397                 & 18 50 51.3 & $-$13 54 33.1 &   1.40   &   0.01   & A9          & 2       & ASAS  \\
194334$-$0904.0 & V1461 Aql                 & 19 43 34.1 & $-$09 04 02.1 &   1.17   &   0.19   & A0 III      & 3       & 20    \\
195342+0205.4   & HD 188153                 & 19 53 42.0 & +02 05 21.2   &   0.32   &  $-$0.01 & A0          & 1       & ASAS  \\
195613+1630.9   & HD 354110                 & 19 56 13.1 & +16 30 59.1   &   0.19   &  $-$0.01 & B8          & 3       & ASAS  \\
205642+1153.0   & HD 199428                 & 20 56 42.0 & +11 53 02.3   &   0.54   &  $-$0.02 & A2          & 1       & ASAS  \\
\enddata

\tablenotetext{a}{Misidentified in link between the Simbad Database and 2MASS catalog.}
\begin{center}
References: 1$-$\citet{can24}, 2$-$\citet{hou88}, 3$-$\citet{hou99}, 4$-$\citet{nes95},
5$-$\citet{str45}, 6$-$\citet{mil86}, 7$-$\citet{och80}, 8$-$\citet{hou82}, 9$-$\citet{hec75},
10$-$\citet{bra80}, 11$-$\citet{wal60}, 12$-$\citet{mor55}, 13$-$\citet{lac93}, 
14$-$\citet{vog04}, 15$-$\citet{por05}, 16$-$\citet{wac64}, 17$-$\citet{str65}, 18$-$\citet{wac68},
19$-$\citet{ote04}, 20$-$\citet{kar99}, 21$-$\citet{hof31},  
22$-$\citet{str66}, 23$-$\citet{wil03}, 24$-$\citet{hof33}, 25$-$\citet{hof35}, 26$-$\citet{bid82},
27$-$\citet{zin30}
\end{center}
\end{deluxetable}
\clearpage

%%%%%%%%%%%%%%%%%%%%%%%%%%%%%%%%%%%%%%%%%%%%%%%%%%%%%%%%%%%%%%%%%%%%

\begin{deluxetable}{lcccrrr}
\tabletypesize{\scriptsize}
\tablewidth{0pt}

\tablecaption{Derived System Parameters\label{derivparams}}
\tablehead{
\colhead{}                        &
\colhead{}                        &
\colhead{$P$}                     &
\colhead{$T_{\rm 0}$}             &
\colhead{$i$}                     &
\colhead{}                        &
\colhead{$\omega$}                \\
\colhead{ASAS ID}                 &
\colhead{Other ID}                &
\colhead{(d)}                     &
\colhead{(HJD$-$2,400,000)}       &
\colhead{($\arcdeg$)}             &
\colhead{$e$}                     &
\colhead{($\arcdeg$)}             \\
\colhead{(1)}                     &
\colhead{(2)}                     &
\colhead{(3)}                     &
\colhead{(4)}                     &
\colhead{(5)}                     &
\colhead{(6)}                     &
\colhead{(7)}                     }
\startdata
053838+0901.2   & HD 37396                &   1.22667(1)   & 54900.712(1)   & 77.5(8)   & 0.0      & \nodata  \\
054816+2046.1   & HD 247740               &   2.43242(2)   & 54882.262(1)   & 82.5(2)   & 0.0      & \nodata  \\
060857+1128.9   & HD 252416               &   1.9064700(1) & 54937.313(1)   & 80.2(3)   & 0.0      & \nodata  \\
060927$-$1501.7 & TYC 5933$-$142$-$1      &   0.8770820(3) & 54900.6813(5)  & 75.8(2)   & 0.0      & \nodata  \\
062556$-$1254.5 & HD 45263                &   5.059500(4)  & 54952.100(1)   & 89.6(3)   & 0.0      & \nodata  \\
063347$-$1410.5 & HD 46621                &   1.5958525(7) & 54953.6023(5)  & 84.3(6)   & 0.0      & \nodata  \\
064010$-$1140.3 & HD 47845                &   5.285085(4)  & 53847.994(2)   & 88.2(1)   & 0.0      & \nodata  \\
064118$-$0551.1 & 2MASS06411762$-$0551065 &   2.757425(6)  & 54931.327(2)   & 79.2(3)   & 0.0      & \nodata  \\
064538+0219.4   & HD 48866                &   1.568196(2)  & 54905.362(2)   & 74.2(3)   & 0.0      & \nodata  \\
064609$-$1923.8 & HD 49125                &   2.9038975(3) & 54908.0579(1)  & 88.6(1)   & 0.050(2) & 53.1(2)  \\
064715+0225.6   & HD 289072               &   3.47257(1)   & 54931.700(3)   & 76.1(3)   & 0.035(20)& 124(9)   \\
064745+0122.3   & V448 Mon                &   1.118478(2)  & 54933.555(1)   & 74.6(7)   & 0.0      & \nodata  \\
065534$-$1013.2 & HD 51082                &   2.186885(3)  & 54586.2521(8)  & 81.2(2)   & 0.0      & \nodata  \\
065549$-$0402.6 & HI Mon                  &   1.5744300(8) & 54935.531(1)   & 80.1(2)   & 0.0      & \nodata  \\
070105$-$0358.2 & HD 52433                &   5.95576(2)   & 54603.120(3)   & 77.8(1)   & 0.067(8) & 54(3)    \\
070238+1347.0   & HD 52637                &   1.918694(4)  & 54809.538(3)   & 75.1(3)   & 0.0      & \nodata  \\
070636$-$0437.4 & AO Mon                  &   1.8847476(1) & 54965.2835(5)  & 85.5(1)   & 0.027(2) & 78(3)    \\
070943+2341.7   & BD +23 1621             &   1.67610(1)   & 54944.095(2)   & 72.5(2)   & 0.0      & \nodata  \\
070946$-$2005.5 & NSV 3433                &   1.7568950(1) & 55002.5074(6)  & 85.2(7)   & 0.0      & \nodata  \\
071010$-$0035.1 & HD 54780                &   2.228760(8)  & 54963.204(2)   & 71.8(3)   & 0.0      & \nodata  \\
071203$-$0139.1 & HD 55236                &   1.2834150(5) & 54944.918(1)   & 78.0(3)   & 0.0      & \nodata  \\
071702$-$1034.9 & HD 56544                &   1.276465(2)  & 55001.2242(7)  & 76.6(5)   & 0.0      & \nodata  \\
072201$-$2552.6 & CX CMa                  &   0.95462500(2)& 55007.3938(7)  & 81.8(2)   & 0.0      & \nodata  \\
073053+0513.7   & HD 59607                &   4.13935(3)   & 54962.503(4)   & 84(2)     & 0.0      & \nodata  \\
073348$-$0940.9 & HD 60389                &   2.549608(5)  & 54573.822(2)   & 76.4(3)   & 0.0      & \nodata  \\
074355$-$2517.9 & HD 62607                &   3.863847(9)  & 54606.582(2)   & 82.5(4)   & 0.0      & \nodata  \\
074717$-$0519.8 & HD 63141                &   4.593955(8)  & 54958.149(4)   & 85.2(1)   & 0.0      & \nodata  \\
074928$-$0721.6 & BD $-$06 2317           &   2.152188(1)  & 54966.082(1)   & 76.2(3)   & 0.0      & \nodata  \\
075052+0048.0   & HD 63818                &   2.0543180(7) & 54966.584(2)   & 80.2(2)   & 0.132(1) & 343(3)   \\
080617$-$0426.8 & V871 Mon                &   4.335950(3)  & 54962.888(1)   & 82.6(2)   & 0.023(9) & 272(5)   \\
081749$-$2659.7 & HD 69797                &   1.743810(5)  & 54931.5934(6)  & 86.9(5)   & 0.0      & \nodata  \\
083245+0247.3   & BD +03 2001             &   1.657373(2)  & 54937.461(1)   & 78.7(2)   & 0.0      & \nodata  \\
084831$-$2609.8 & TT Pyx                  &   3.031555(1)  & 54987.7396(7)  & 89.5(5)   & 0.0      & \nodata  \\
101120$-$1956.3 & HD 88409                &   1.563820(3)  & 54902.837(1)   & 77.8(7)   & 0.0      & \nodata  \\
135949$-$2745.5 & HD 122026               &   6.133165(7)  & 55002.2808(9)  & 88.2(6)   & 0.0      & \nodata  \\
160851$-$2351.0 & TYC 6780$-$1523$-$1     &   4.66007(1)   & 54801.465(3)   & 81.9(4)   & 0.0      & \nodata  \\
165354$-$1301.9 & HD 152451               &   2.207586(1)  & 54804.5134(4)  & 83.1(2)   & 0.215(3) & 157(2)   \\
170158+2348.4   & HD 154010               &   4.76991(1)   & 54700.588(2)   & 86.2(2)   & 0.0      & \nodata  \\
173421$-$1836.3 & HD 159246               &   5.216895(4)  & 54807.627(1)   & 89.8(3)   & 0.0      & \nodata  \\
174104+0747.1   & V506 Oph                &   1.060427(1)  & 54806.6804(4)  & 88.5(6)   & 0.0      & \nodata  \\
175659$-$2012.2 & HD 312444               &   3.089365(5)  & 54803.486(1)   & 82.8(4)   & 0.0      & \nodata  \\
175859$-$2323.1 & HD 313508               &   1.540495(5)  & 55006.029(1)   & 79.0(6)   & 0.0      & \nodata  \\
180903$-$1824.5 & HD 165890               &   3.70752(1)   & 54807.316(5)   & 80.9(3)   & 0.0      & \nodata  \\
181025+0047.7   & HD 166383               &   3.18867(1)   & 54808.146(7)   & 71.7(3)   & 0.0      & \nodata  \\
181328$-$2214.3 & HD 166851               &   4.47977(3)   & 54803.532(14)  & 88(2)     & 0.0      & \nodata  \\
181909$-$1410.0 & HD 168207               &   1.786820(1)  & 54803.012(2)   & 70.1(2)   & 0.0      & \nodata  \\
183129$-$1918.8 & BD $-$19 5039           &   2.455005(6)  & 54949.211(2)   & 74.7(3)   & 0.0      & \nodata  \\
183219$-$1117.4 & BD $-$11 4667           &   4.95239(2)   & 55038.863(5)   & 83.1(9)   & 0.0      & \nodata  \\
184223+1158.9   & BD +11 3569             &   1.497118(3)  & 55008.3686(9)  & 81.1(4)   & 0.006(2) & 144(31)  \\
184327+0841.5   & TYC 1025$-$1524$-$1     &   2.1392200(8) & 55049.120(1)   & 84.6(1)   & 0.367(5) & 280.0(2) \\
184436$-$1923.4 & YY Sgr                  &   2.6284575(6) & 55026.2658(5)  & 88.1(1)   & 0.117(1) & 214.1(5) \\
185051$-$1354.6 & HD 174397               &   3.145935(8)  & 55047.304(5)   & 78.0(7)   & 0.0      & \nodata  \\
194334$-$0904.0 & V1461 Aql               &   1.76306(1)   & 55040.5159(8)  & 82.4(5)   & 0.0      & \nodata  \\
195342+0205.4   & HD 188153               &   1.608120(2)  & 55040.431(2)   & 73.0(2)   & 0.040(2) & 27(3)    \\
195613+1630.9   & HD 354110               &   3.444589(5)  & 55036.802(3)   & 85.5(1)   & 0.037(29)& 102(2)   \\
205642+1153.0   & HD 199428               &   4.124500(5)  & 55047.514(1)   & 86.2(4)   & 0.219(1) & 328.2(2) \\
\enddata
\end{deluxetable}
\clearpage

%%%%%%%%%%%%%%%%%%%%%%%%% Velocities for HI Mon %%%%%%%%%%%%%%%%%%%%%%%%%%%%

\begin{deluxetable}{cccccccc}
\tabletypesize{\scriptsize}
\tablewidth{0pt}
\tablecaption{HI~Mon Radial Velocity Measurements\label{rvs}}
\tablehead{
\colhead{Date}          &
\colhead{Orbital}       &
\colhead{$V_1$}         &
\colhead{$\sigma_{1}$}  &
\colhead{$(O-C)_1$}     &
\colhead{$V_2$}         &
\colhead{$\sigma_{2}$}  &
\colhead{$(O-C)_2$}     \\
\colhead{(HJD$-$2,400,000)}        &
\colhead{Phase}  &
\colhead{(km s$^{-1}$)} &
\colhead{(km s$^{-1}$)} &
\colhead{(km s$^{-1}$)} &
\colhead{(km s$^{-1}$)} &
\colhead{(km s$^{-1}$)} &
\colhead{(km s$^{-1}$)} }
\startdata
 55190.776 & 0.119 & \phn--90.2    & \phn4.5 &  15.7   & \phn257.3 & \phn2.4 & \phn\phn3.0 \\
 55190.837 & 0.157 & --131.9       & \phn4.9 &  11.9   & \phn306.7 & \phn7.2 & \phn\phn8.0 \\
 55191.769 & 0.750 & \phn312.2     & \phn8.2 & \phn0.8 & --224.0   & \phn7.1 & \phn\phn1.9 \\
 55191.818 & 0.780 & \phn306.7     & \phn8.2 & --0.8   & --219.4   & \phn8.2 & \phn\phn1.2 \\
 55191.846 & 0.798 & \phn298.4     & \phn7.9 & --2.8   & --211.1   & \phn8.5 & \phn\phn1.8 \\
 55192.648 & 0.308 & --166.6       & \phn2.6 & --0.3   & \phn329.9 & \phn3.8 & \phn\phn1.7 \\
 55192.680 & 0.328 & --157.6       & \phn2.2 & --4.1   & \phn311.5 & \phn4.8 & \phn--2.1   \\
 55192.797 & 0.402 & \phn--82.2    & \phn3.6 & --3.9   & \phn221.1 & \phn4.1 & \phn--5.8   \\
 55192.838 & 0.429 & \phn--52.1    & \phn5.4 & --9.2   & \phn185.7 & \phn6.5 & --17.0      \\
 55193.777 & 0.024 & \phn\phn--0.9 & 13.5    &  17.3   & \phn107.4 & 12.4    & \phn\phn3.6 \\
 55193.828 & 0.057 & \phn--40.4    & \phn7.8 &  12.0   & \phn156.4 & \phn8.2 & \phn--4.6   \\
\enddata

\end{deluxetable}

%%%%%%%%%%%%%%%%%%%%%%%%% Tomographic Values %%%%%%%%%%%%%%%%%%%%%%%%%%%%%

\begin{deluxetable}{lcc}
\tablewidth{0pc}
\tablecaption{Tomographic Spectral Reconstruction Parameters\label{tomo}}
\tablehead{
  \colhead{Parameter} &
  \colhead{Primary}   &
  \colhead{Secondary}}
\startdata
Spectral Type\tablenotemark{a}\dotfill & B0 V              & B0.5 V\\
$T_{\rm eff}$ (kK)\dotfill             & 30.0 $\pm$ 0.5    & 29.0 $\pm$ 0.5\\
log $g$ (cgs)\dotfill                  & \phn4.1 $\pm$ 0.2 & \phn4.1 $\pm$ 0.2\\
$V~\sin~i$ (km s$^{-1}$)\dotfill       & 150 $\pm$ 25      & \phn150 $\pm$ 25\\
$F_{\rm 2} / F_{\rm 1}$ (blue)         & \multicolumn{2}{c}{0.70 $\pm$ 0.05}
\enddata
\tablenotetext{a}{These spectral types are estimated from derived values of 
$T_{\rm eff}$ and log $g$.}
\end{deluxetable}

%%%%%%%%%%%%%%%%%%%%%%% Circular Orbital Solution %%%%%%%%%%%%%%%%%%%%%%%%

\begin{deluxetable}{lc}
\tablewidth{0pc}
\tablecaption{Circular Orbital Solution for HI~Mon\label{orb}}
\tablehead{
  \colhead{Element} &
  \colhead{Value}   }
\startdata
$P$~(days)\dotfill                     & \phn\phn\phn\phn1.5744300 $\pm$ 0.0000008\\
$T_{\rm 0}$ (HJD--2,400,000)\dotfill   & 54935.5331 $\pm$ 0.0005\\
$K_{\rm 1}$ (km s$^{-1}$)\dotfill      & \phn\phn248.8 $\pm$ 1.9 \\
$K_{\rm 2}$ (km s$^{-1}$)\dotfill      & \phn\phn288.2 $\pm$ 2.4 \\
$\gamma_{\rm 1}$ (km s$^{-1}$)\dotfill & \phn\phn\phn64.5 $\pm$ 1.3\\
$\gamma_{\rm 2}$ (km s$^{-1}$)\dotfill & \phn\phn\phn59.8 $\pm$ 1.2\\
rms~(primary)~(km s$^{-1}$)\dotfill    & \phs\phs9.6 \\
rms~(secondary)~(km s$^{-1}$)\dotfill  & \phs\phs6.5 \\
rms~(photometry)~(mag)\dotfill         & \phs\phs0.02 \\
\enddata
\end{deluxetable}

%%%%%%%%%%%%%%%%%%%%%% ELC Output Paramters %%%%%%%%%%%%%%%%%%%%%%%%%%%%%

\begin{deluxetable}{lcc}
\tablewidth{0pc}
\tablecaption{ELC Model Parameters for HI~Mon\label{ELCparms}}
\tablehead{
  \colhead{Parameter}  &
  \colhead{Primary}    &
  \colhead{Secondary}  }
\startdata
Inclination (deg)\dotfill                              & \multicolumn{2}{c}{80.0 $\pm$ 0.2}\\
$M~(M_{\odot})$\dotfill                                & \phn14.2 $\pm$ 0.3      & \phn12.2 $\pm$ 0.2\\
$R_{\rm eff}~(R_{\odot})$\dotfill                      & \phn\phn5.13 $\pm$ 0.11 & \phn\phn4.99 $\pm$ 0.07\\
$R_{\rm pole}$\tablenotemark{a}~($R_{\odot}$)\dotfill  & \phn\phn4.99 $\pm$ 0.11 & \phn\phn4.83 $\pm$ 0.07\\
$R_{\rm point}$\tablenotemark{b}~($R_{\odot}$)\dotfill & \phn\phn5.40 $\pm$ 0.11 & \phn\phn5.30 $\pm$ 0.07\\
$V_{\rm sync}~\sin~i$ (km s$^{-1}$)\dotfill            & 162.5 $\pm$ 7.2         & 157.9 $\pm$ 2.7\\
log $g$ (cgs)\dotfill                                  & \phn\phn4.17 $\pm$ 0.04 & \phn\phn4.13 $\pm$ 0.04\\
$T_{\rm eff}$ (kK)\dotfill                             & \phn29.5 $\pm$ 0.6      & \phn28.4 $\pm$ 0.4\\
Filling factor\dotfill                                 & \phn\phn0.62 $\pm$ 0.01 & \phn\phn0.65 $\pm$ 0.02\\
$a_{\rm tot}~(R_{\odot})$\dotfill                      & \multicolumn{2}{c}{16.96 $\pm$ 0.11} \\
$F_{\rm 2} / F_{\rm 1}$ (blue)\dotfill                 & \multicolumn{2}{c}{\phn0.81 $\pm$ 0.09}
\enddata
\tablenotetext{a}{Polar radius.}
\tablenotetext{b}{Radius toward the inner Lagrangian point.}
\end{deluxetable}

%
% Figures!!
%

\input{epsf}
\begin{figure}
\begin{center}
{\includegraphics[angle=90,height=12cm]{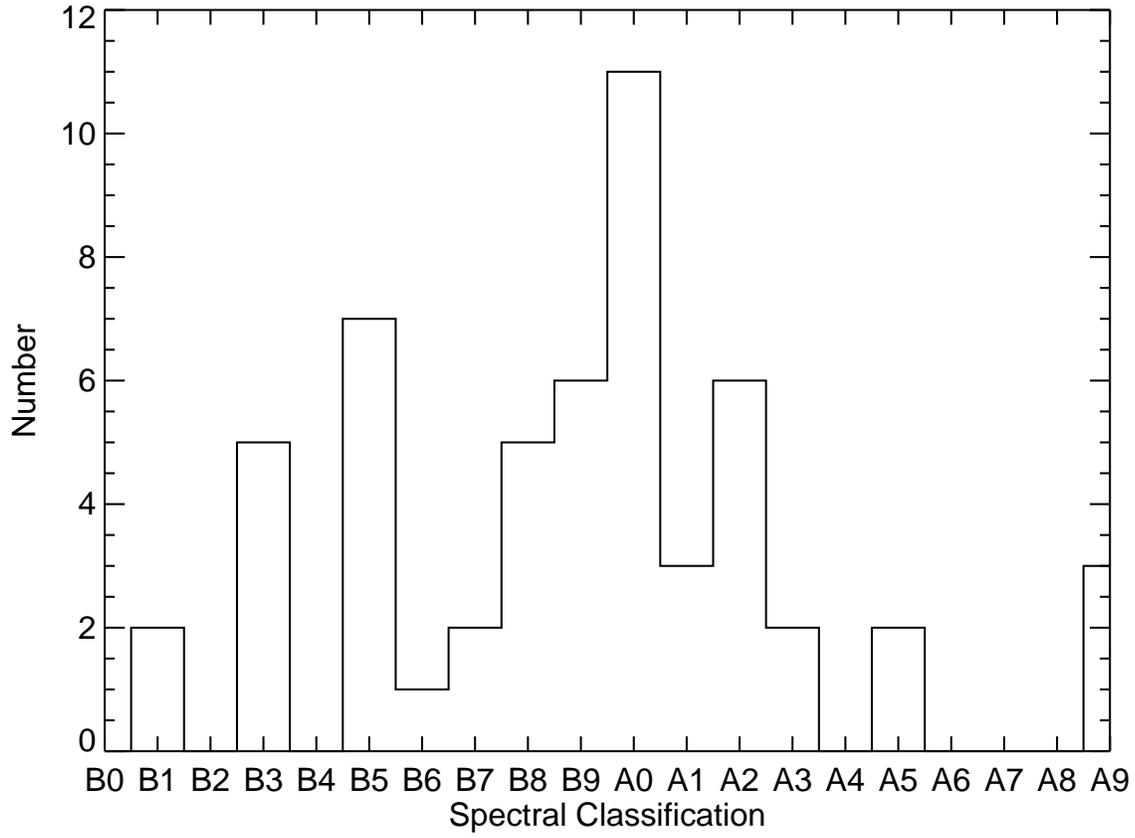}}
\end{center}
\caption{A histogram of published spectral types for our target objects.}
\label{fig1}
\end{figure}
\clearpage

\begin{figure}
\begin{center}
{\includegraphics[angle=90,height=12cm]{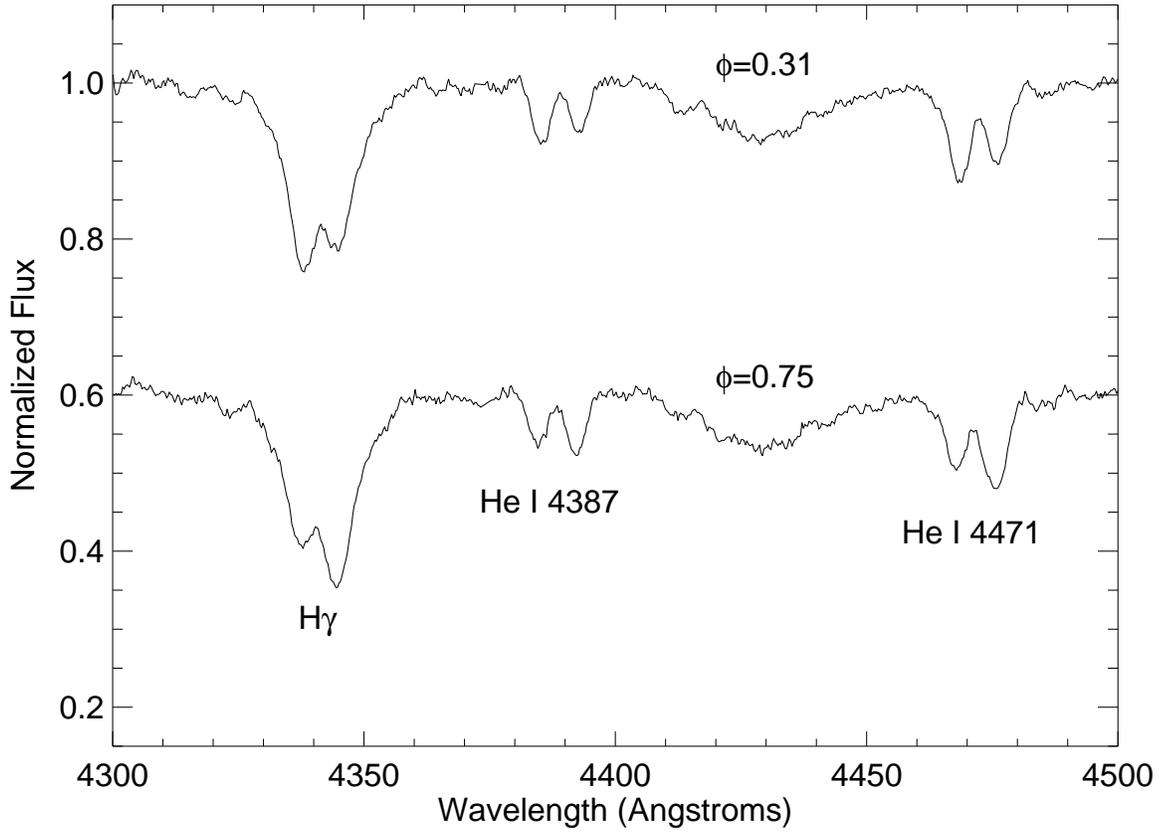}}
\end{center}
\caption{Two spectra of HI~Mon near opposing quadrature phases (offset for clarity).
  The primary in each spectrum is represented by the deeper lines.}
\end{figure}
\clearpage

\begin{figure}
\begin{center}
{\includegraphics[angle=90,height=12cm]{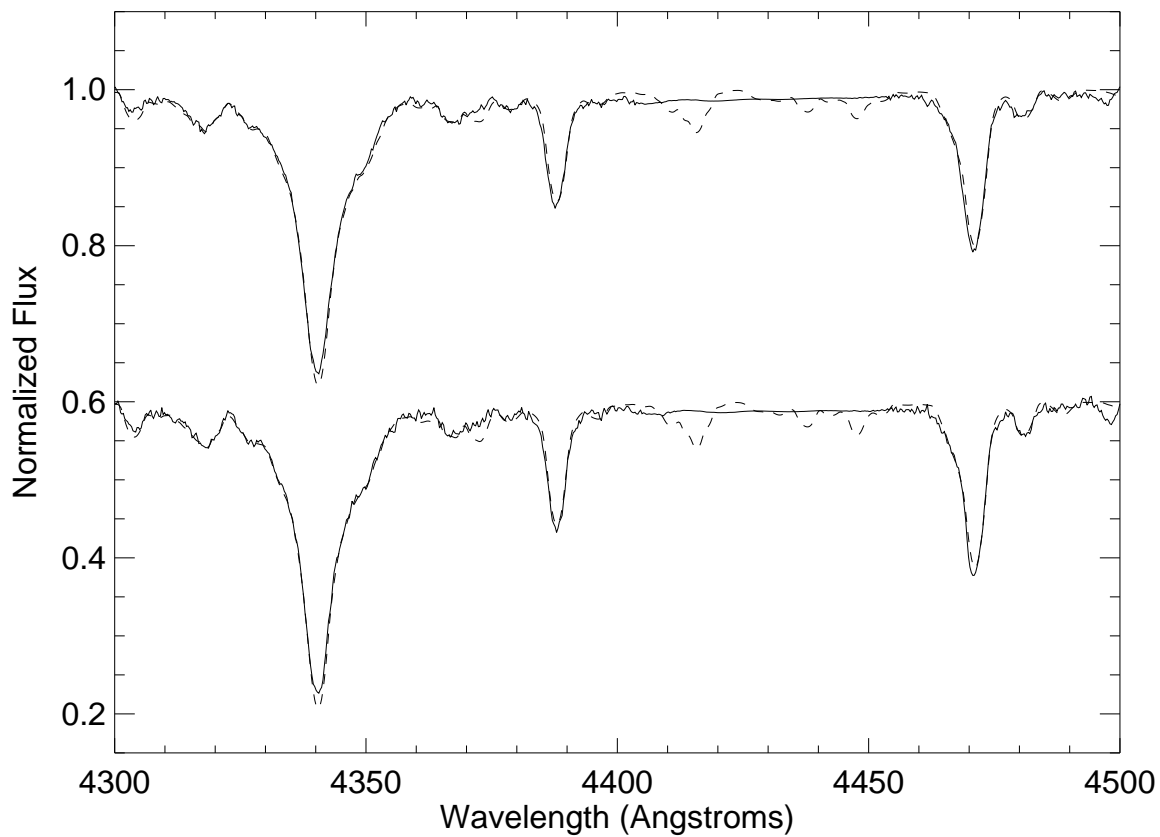}}
\end{center}
\caption{Tomographic reconstructions of the components of HI~Mon based on the 9 
  spectra obtained outside of eclipse phases. The top solid line represents the
  primary and the bottom solid line is the reconstructed secondary spectrum
  offset by 0.4 for clarity. Overplotted for both are the model spectra for
  each shown by dashed lines. The stellar parameters for the model spectra
  are given in Table 4. Note also that the region containing the 
  diffuse interstellar band near 4428\AA~has been removed for the tomographic
  reconstruction.}
\end{figure}
\clearpage

\begin{figure}
\begin{center}
{\includegraphics[angle=90,height=12cm]{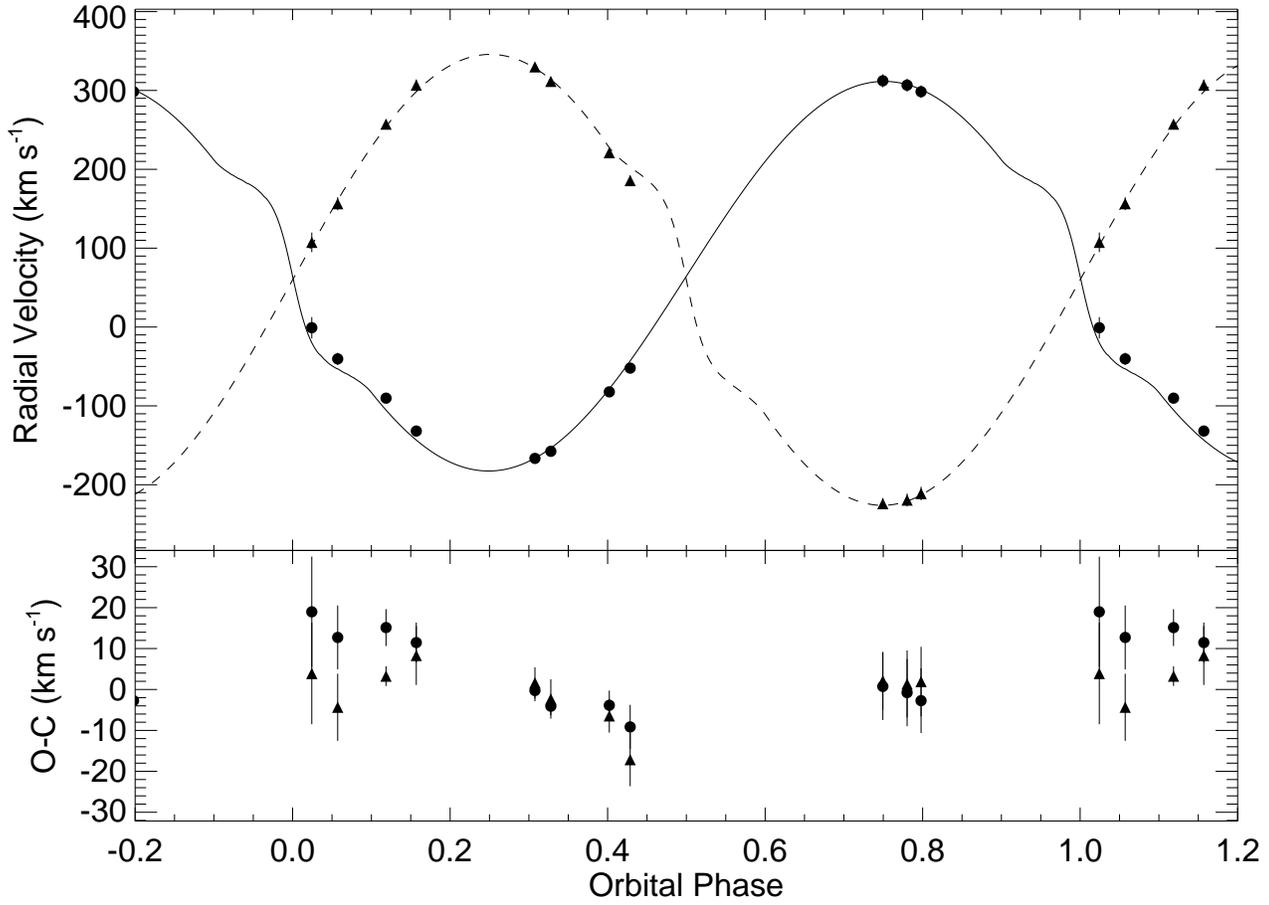}}
\end{center}
\caption{Radial velocity curves for HI~Mon. Primary radial velocities are 
  represented by filled dots and secondary radial velocities by filled 
  triangles with associated uncertainties shown as line segments for 
  both. The solid line is the best-fit solution for the primary and the
  dashed line is the best-fit solution for the secondary. The lower
  panel shows the observed minus calculated values for each measurement
  with uncertainties.}
\end{figure}
\clearpage

\begin{figure}
\begin{center}
{\includegraphics[angle=90,height=12cm]{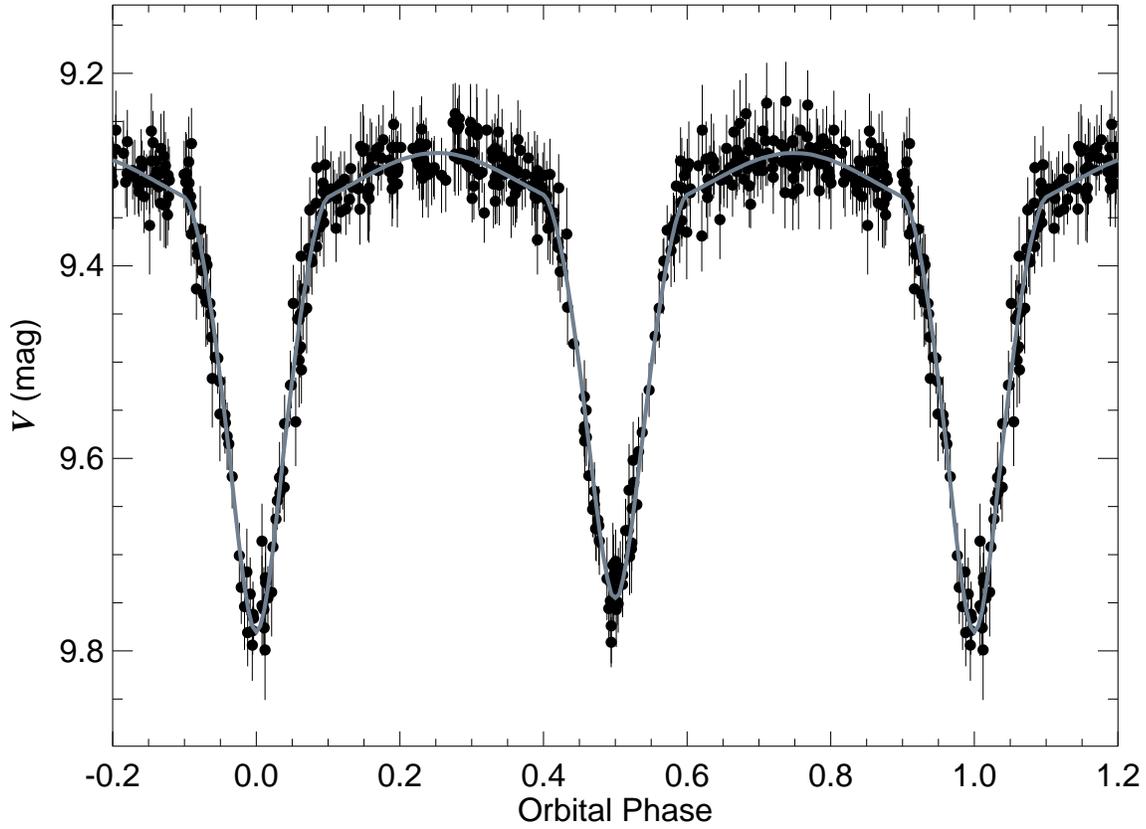}}
\end{center}
\caption{$V$-band light curve for HI~Mon taken from the ASAS database 
  \citep{poj02} and presented here in phase according to our best-fit
  solution. The model is the thick gray line and data are the filled
  dots with uncertainties represented by line segments. Phase zero for
  this plot corresponds to mid-eclipse of the primary star.}
\end{figure}
\clearpage

\begin{figure}
\begin{center}
{\includegraphics[angle=90,height=12cm]{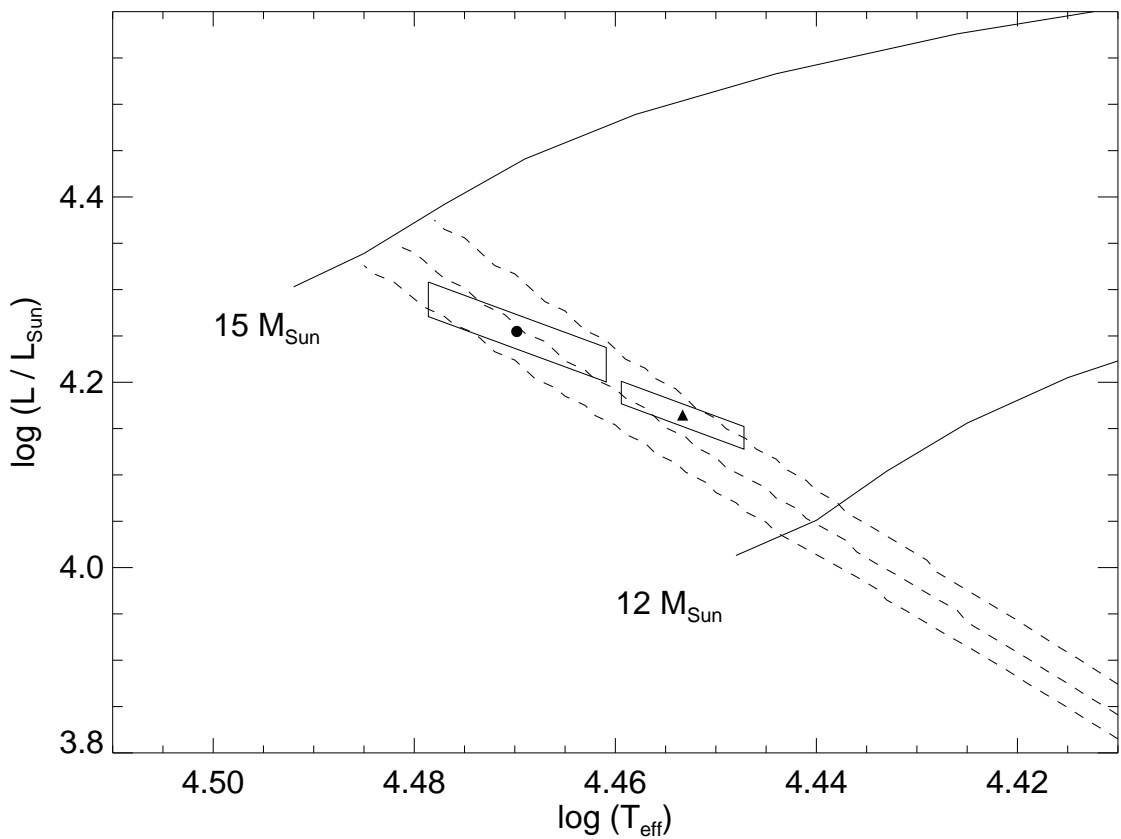}}
\end{center}
\caption{Theoretical H-R diagram showing the location of the primary star
  (filled circle) and secondary star (filled triangle) of HI~Mon including
  uncertainty regions for each. Also plotted are the evolutionary tracks 
  for stars of 12 $M_{\odot}$ and 15 $M_{\odot}$ from \citet{sch92} and 
  isochrones from \citet{lej01} for solar metallicity with ages of 1.5,
  2.5, and 3.5 Myr going from lower left to upper right. The positions 
  of the two components of HI~Mon are consistent with an age of 
  $\sim$2.5 Myr.}
\end{figure}
\clearpage

\begin{figure}
\begin{center}
{\includegraphics[angle=90,height=12cm]{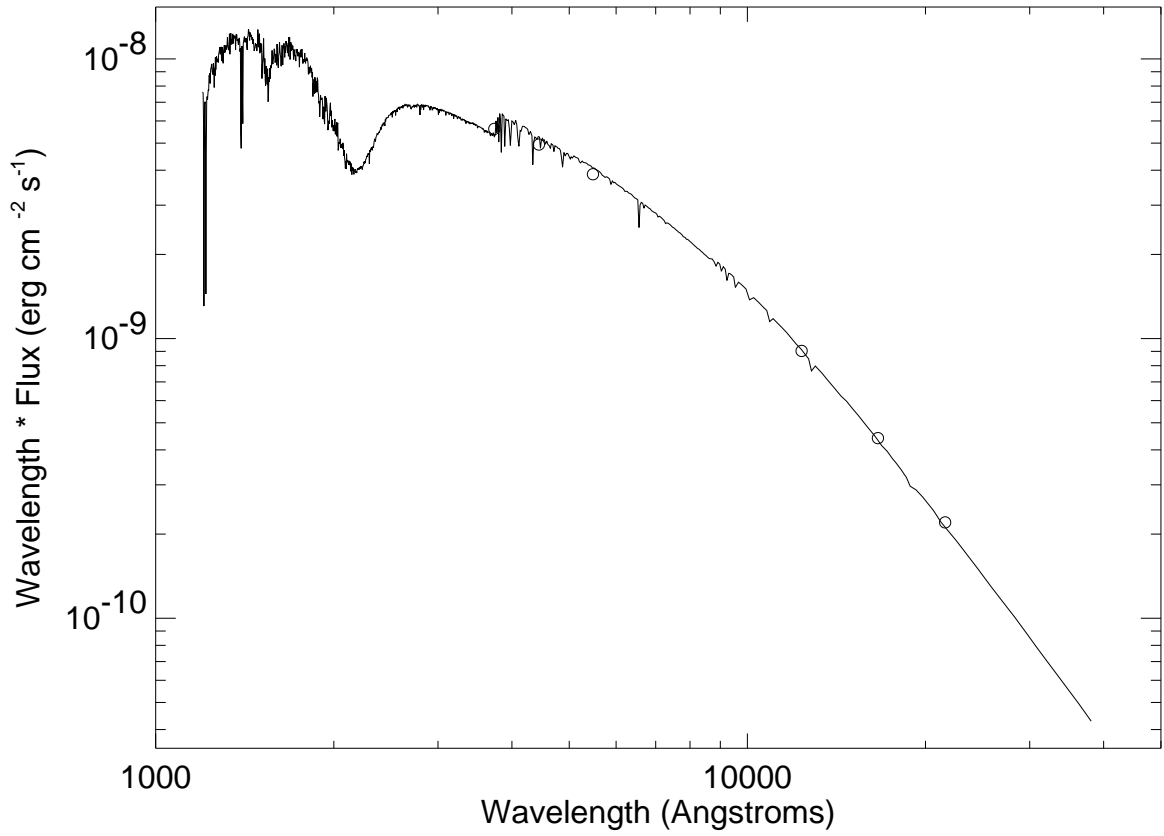}}
\end{center}
\caption{Spectral energy distribution and fit for the combined light of 
  the HI~Mon components (solid line) to Johnson $U$, $B$, $V$, $J$, $H$, 
  $K_S$ photometry (open circles).}
\end{figure}
\clearpage

%%%%%%%%%%%%%%%%%%%%%%%%%%%%%%%%%% Figure Set %%%%%%%%%%%%%%%%%%%%%%%%%%%%%%%

%\figsetstart
%\figsetnum{2}
%\figsettitle{ASAS Light Curves}
%
%\figsetgrpstart
%\figsetgrpnum{2.1}
%\figsetgrptitle{ASAS Light Curve for HD 37396}
%\figsetplot{f2_1.eps}
%\figsetgrpnote{The ASAS $V-$band light curve for HD 37396. Filled circles with lines
%  represent data with associated uncertainties. The best fit orbital solution
%  listed in Table 2 is shown as a solid line passing through the data.}
%\figsetgrpend
%
%\figsetgrpstart
%\figsetgrpnum{2.2}
%\figsetgrptitle{ASAS Light Curve for HD 244740}
%\figsetplot{f2_2.eps}
%\figsetgrpnote{The ASAS $V-$band light curve for HD 244740. Filled circles with lines
%  represent data with associated uncertainties. The best fit orbital solution
%  listed in Table 2 is shown as a solid line passing through the data.}
%\figsetgrpend
%
%\figsetend

%%%%%%%%%%%%%%%%%%%%%%%%%%%%%%%%%% End Figure Set %%%%%%%%%%%%%%%%%%%%%%%%%%%

\begin{figure}
\begin{center}
{\includegraphics[angle=90,height=12cm]{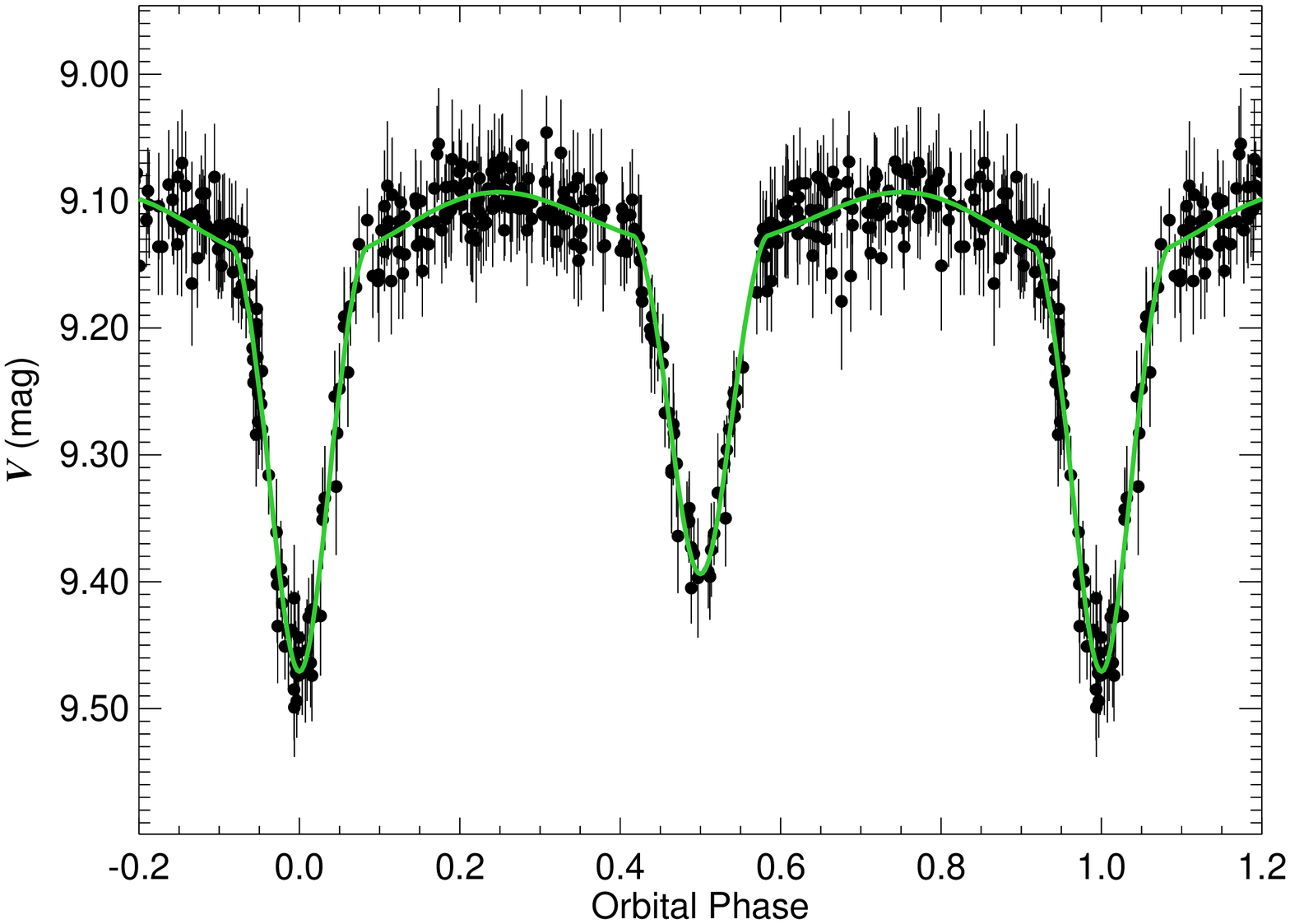}}
\end{center}
\caption{The ASAS $V-$band light curve for ASAS 053838+0901.2
  (HD 37396). Filled circles with lines
  represent data with associated uncertainties. The best fit orbital solution
  listed in Table 2 is shown as a solid line passing through the data.}
\end{figure}
\clearpage

\begin{figure}
\begin{center}
{\includegraphics[angle=90,height=12cm]{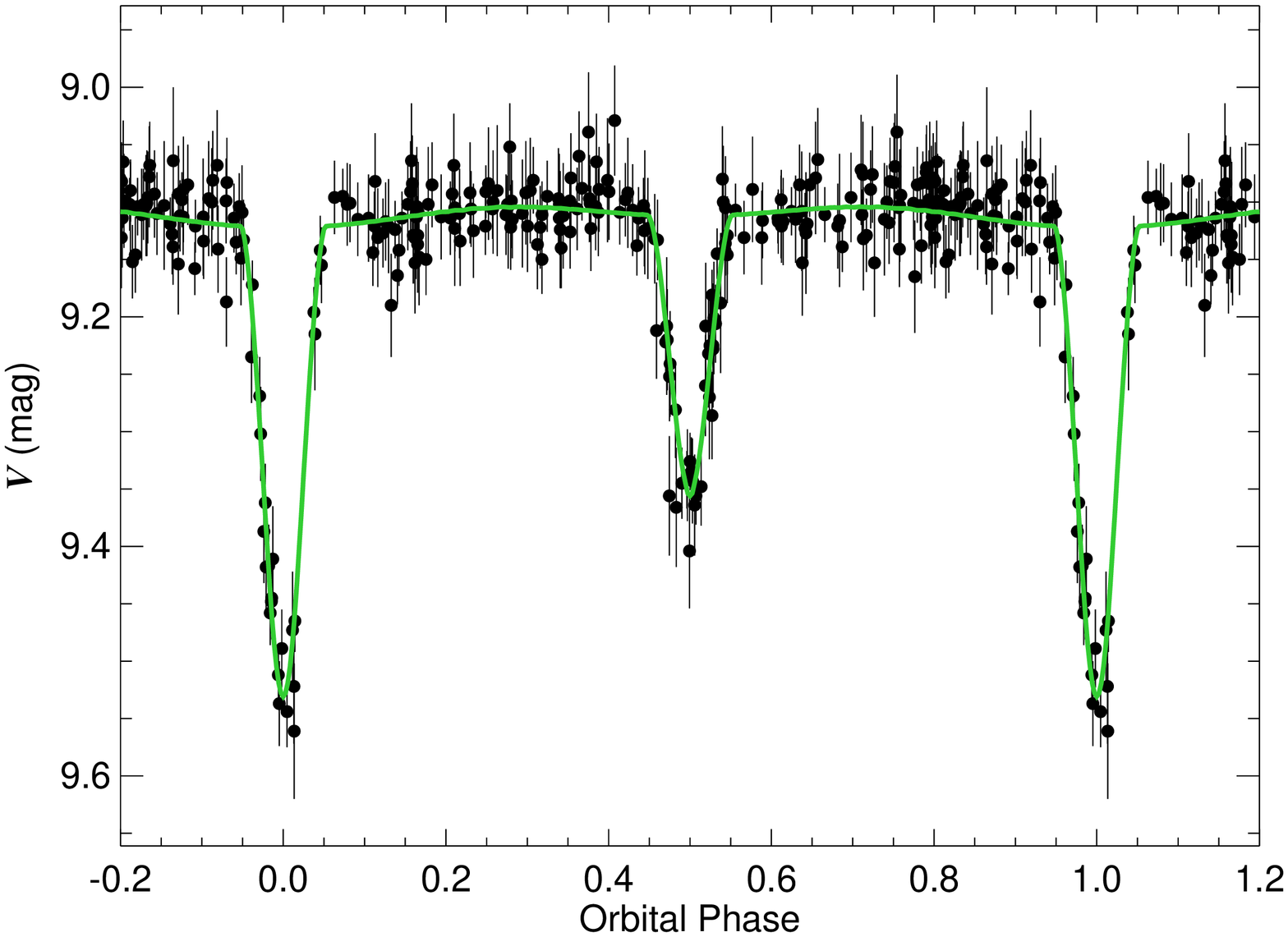}}
\end{center}
\caption{The ASAS $V-$band light curve for ASAS 054816+2046.1
  (HD 247740). Filled circles with lines
  represent data with associated uncertainties. The best fit orbital solution
  listed in Table 2 is shown as a solid line passing through the data.}
\end{figure}
\clearpage

\begin{figure}
\begin{center}
{\includegraphics[angle=90,height=12cm]{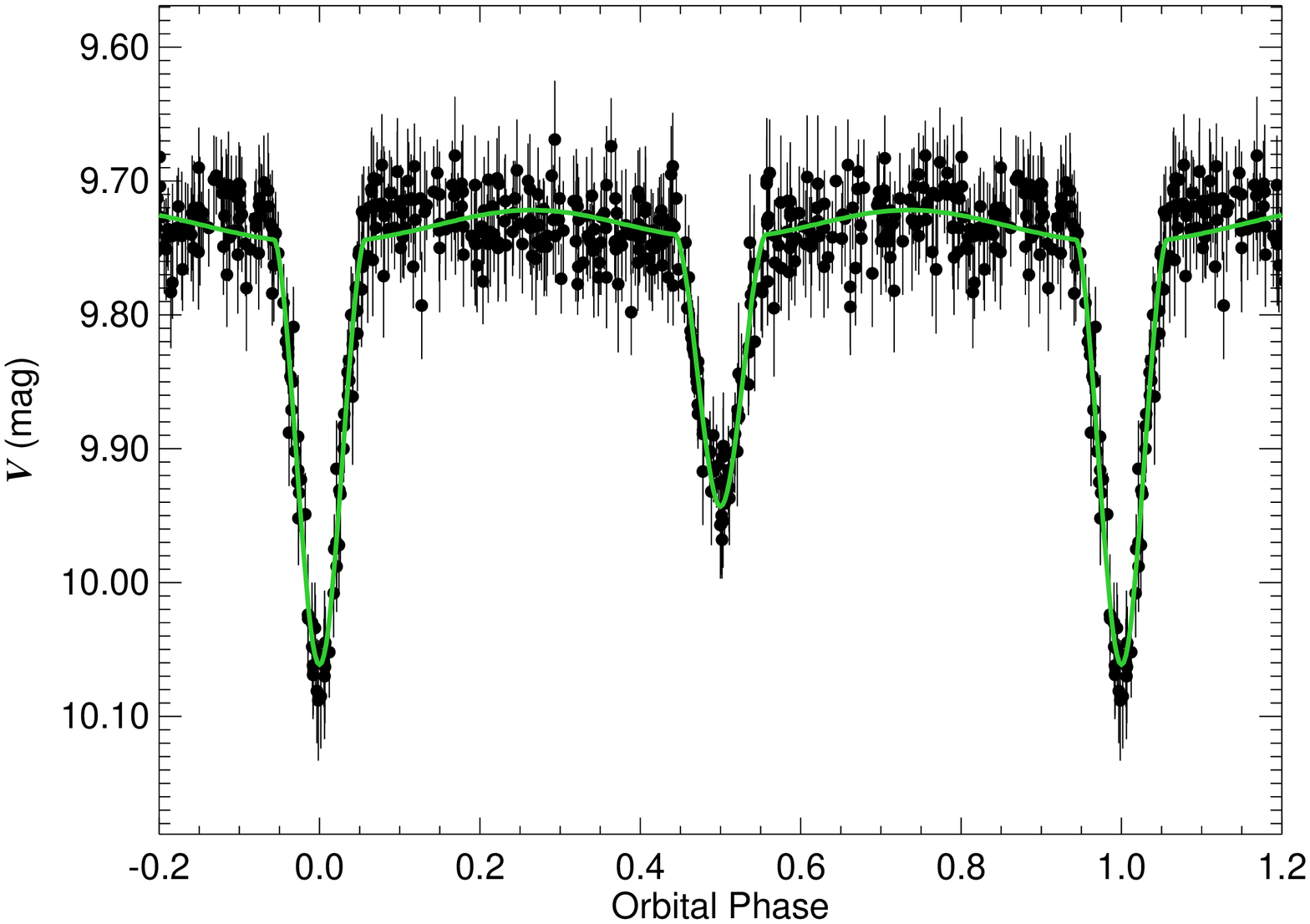}}
\end{center}
\caption{The ASAS $V-$band light curve for ASAS 060857+1128.9
  (HD 252416). Filled circles with lines
  represent data with associated uncertainties. The best fit orbital solution
  listed in Table 2 is shown as a solid line passing through the data.}
\end{figure}
\clearpage

\begin{figure}
\begin{center}
{\includegraphics[angle=90,height=12cm]{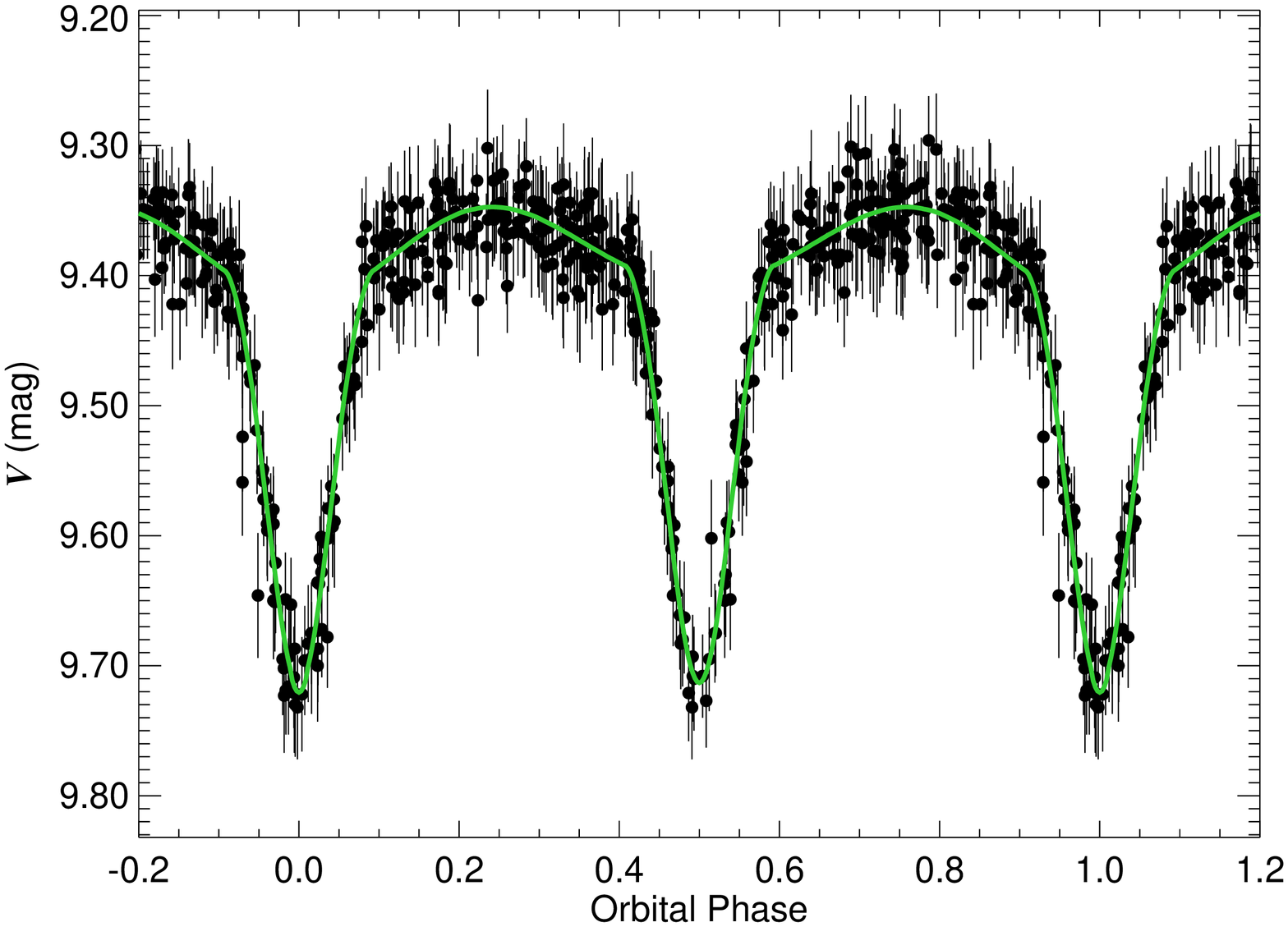}}
\end{center}
\caption{The ASAS $V-$band light curve for ASAS 060927$-$1501.7
  (TYC 5933$-$142$-$1). Filled circles with lines
  represent data with associated uncertainties. The best fit orbital solution
  listed in Table 2 is shown as a solid line passing through the data.}
\end{figure}
\clearpage

\begin{figure}
\begin{center}
{\includegraphics[angle=90,height=12cm]{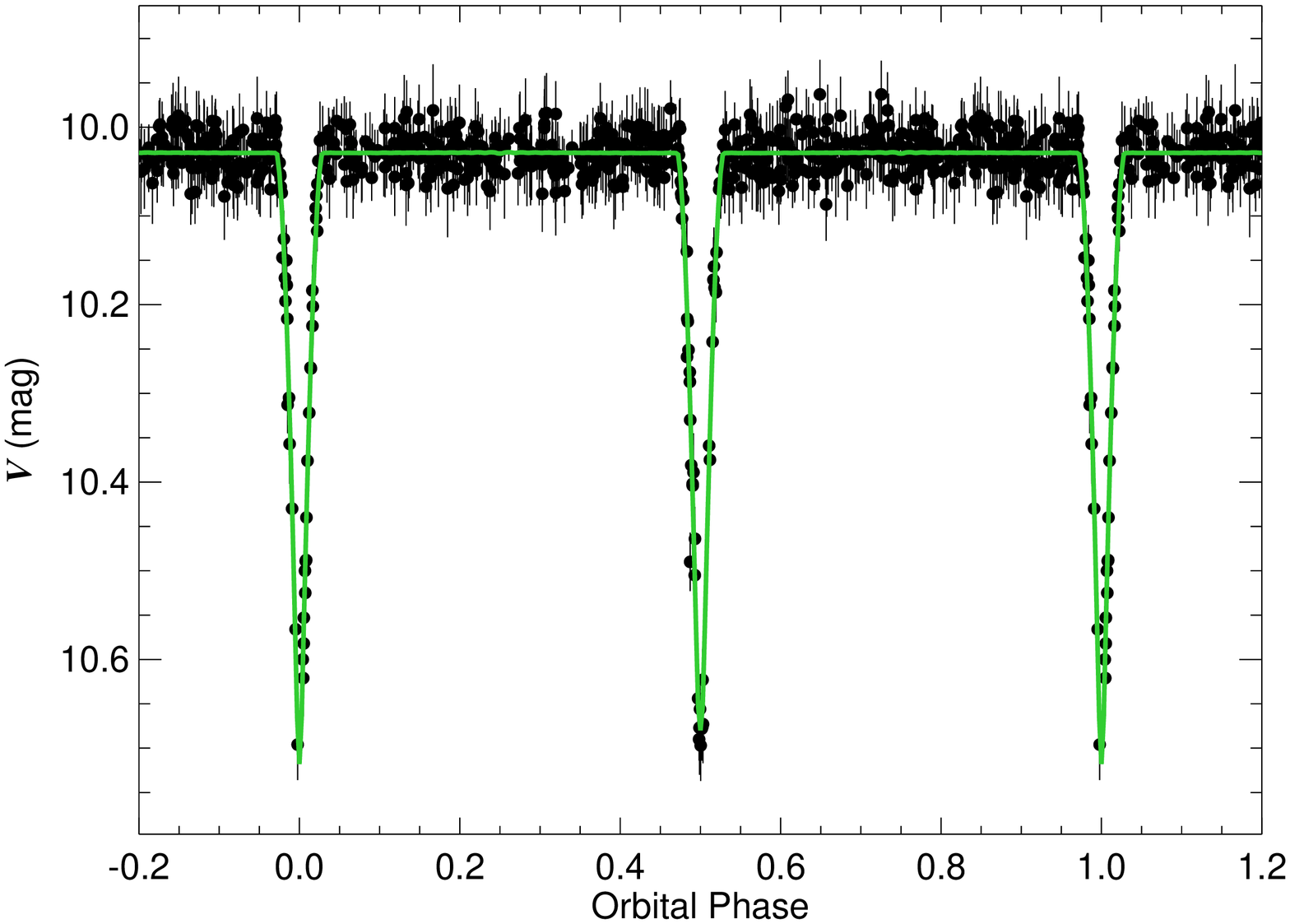}}
\end{center}
\caption{The ASAS $V$ band light curve for ASAS 062556$-$1254.5
  (HD 45263). Filled circles with lines
  represent data with associated uncertainties. The best fit orbital solution
  listed in Table 2 is shown as a solid line passing through the data.}
\end{figure}
\clearpage

\begin{figure}
\begin{center}
{\includegraphics[angle=90,height=12cm]{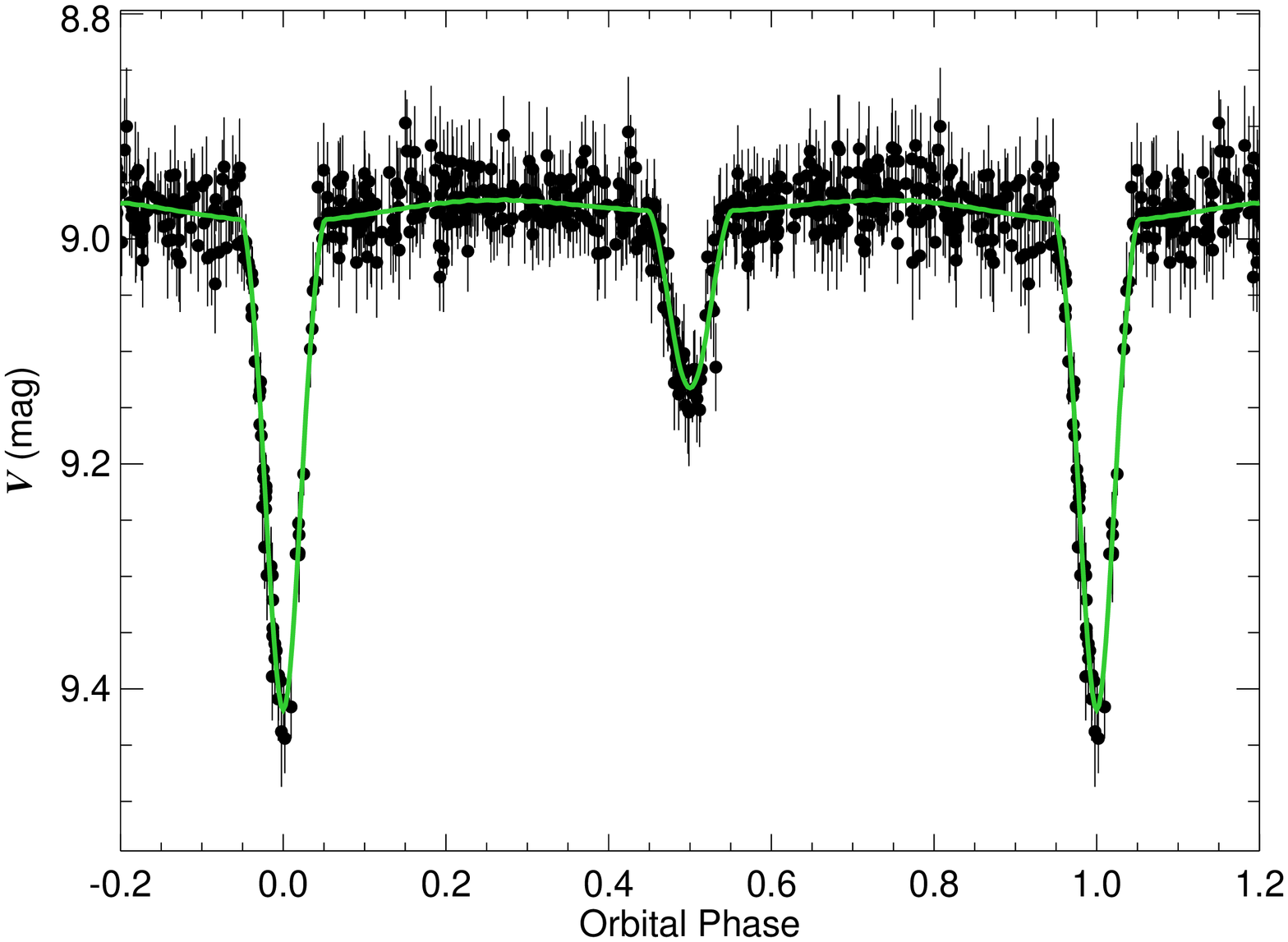}}
\end{center}
\caption{The ASAS $V-$band light curve for ASAS 063347$-$1410.5
  (HD 46621). Filled circles with lines
  represent data with associated uncertainties. The best fit orbital solution
  listed in Table 2 is shown as a solid line passing through the data.}
\end{figure}
\clearpage

\begin{figure}
\begin{center}
{\includegraphics[angle=90,height=12cm]{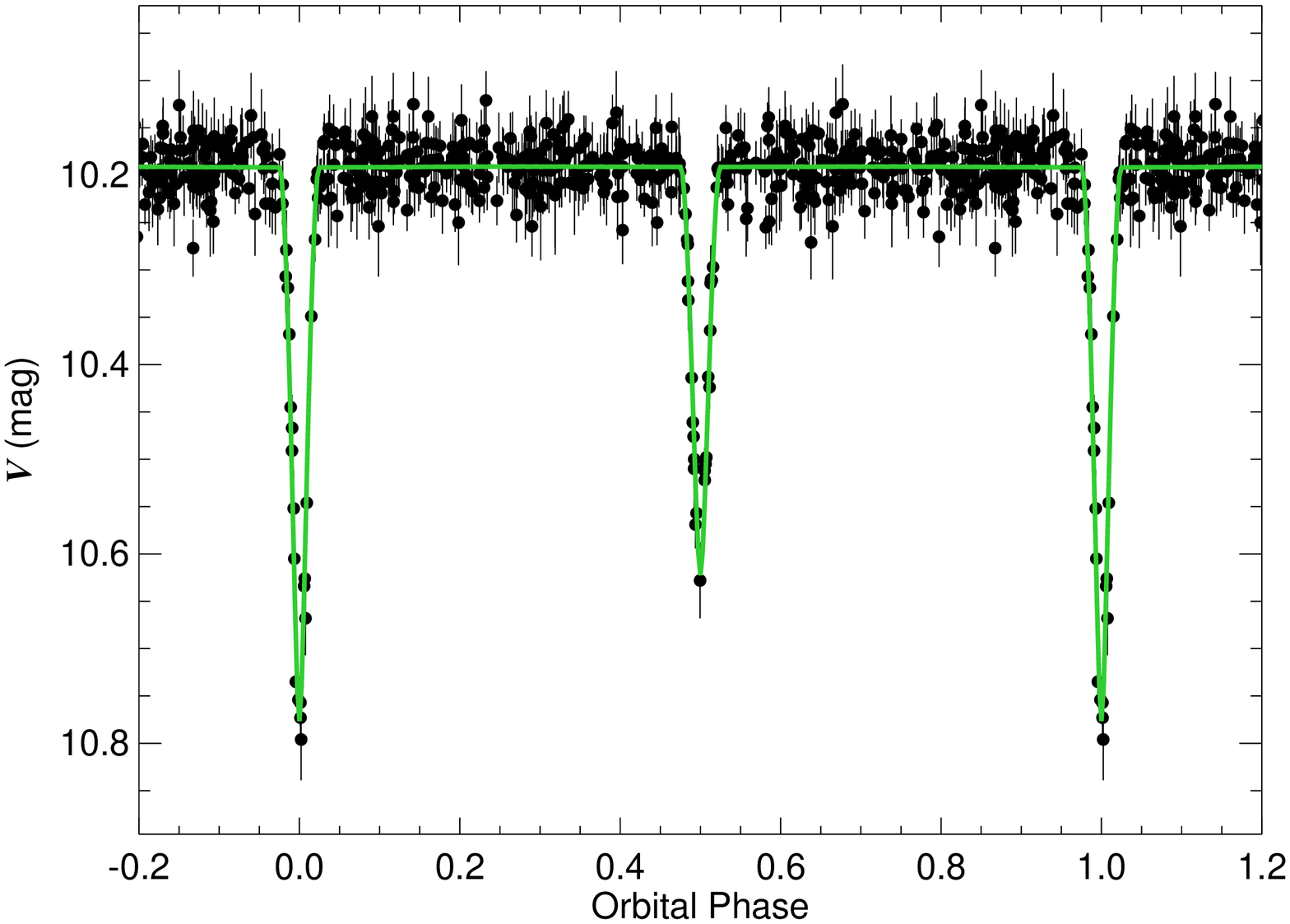}}
\end{center}
\caption{The ASAS $V-$band light curve for ASAS064010$-$1140.3
  (HD 47845). Filled circles with lines
  represent data with associated uncertainties. The best fit orbital solution
  listed in Table 2 is shown as a solid line passing through the data.}
\end{figure}
\clearpage

\begin{figure}
\begin{center}
{\includegraphics[angle=90,height=12cm]{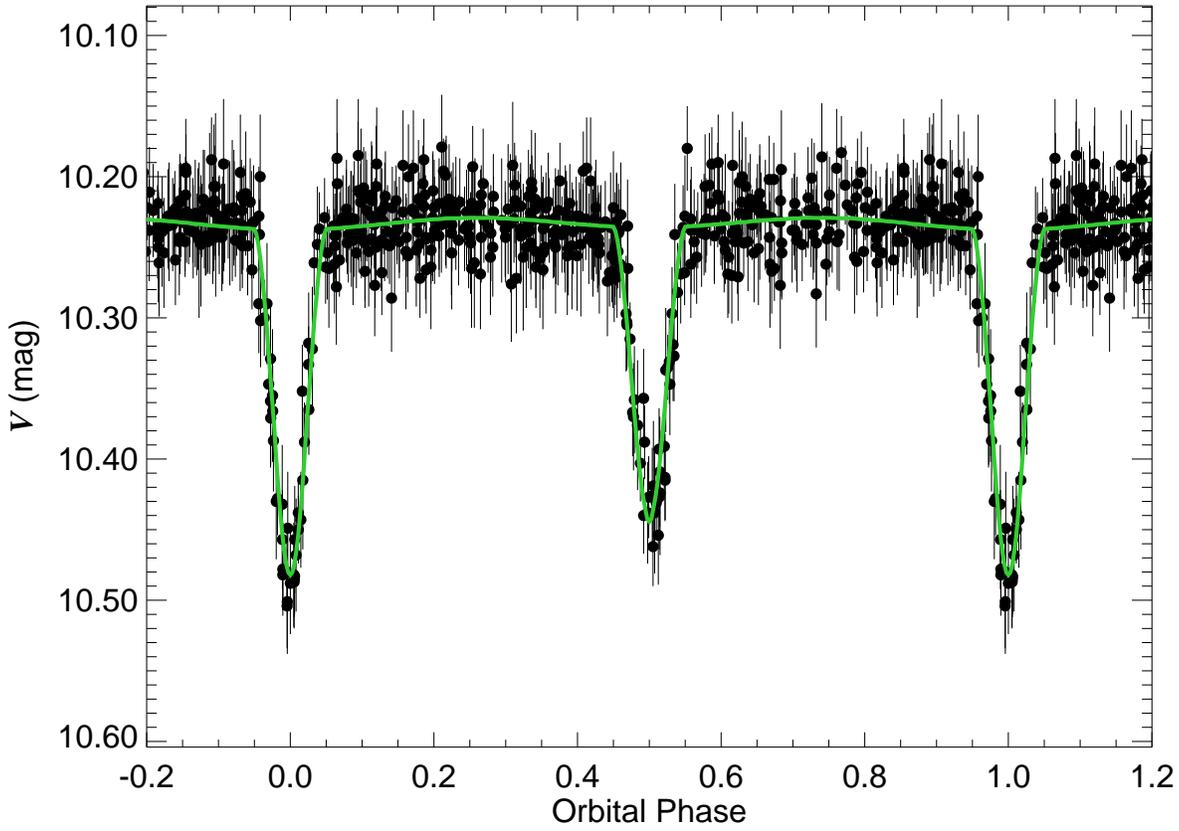}}
\end{center}
\caption{The ASAS $V-$band light curve for ASAS 064118$-$0551
  (2MASS 06411762$-$0551065). Filled circles with lines
  represent data with associated uncertainties. The best fit orbital solution
  listed in Table 2 is shown as a solid line passing through the data.}
\end{figure}
\clearpage

\begin{figure}
\begin{center}
{\includegraphics[angle=90,height=12cm]{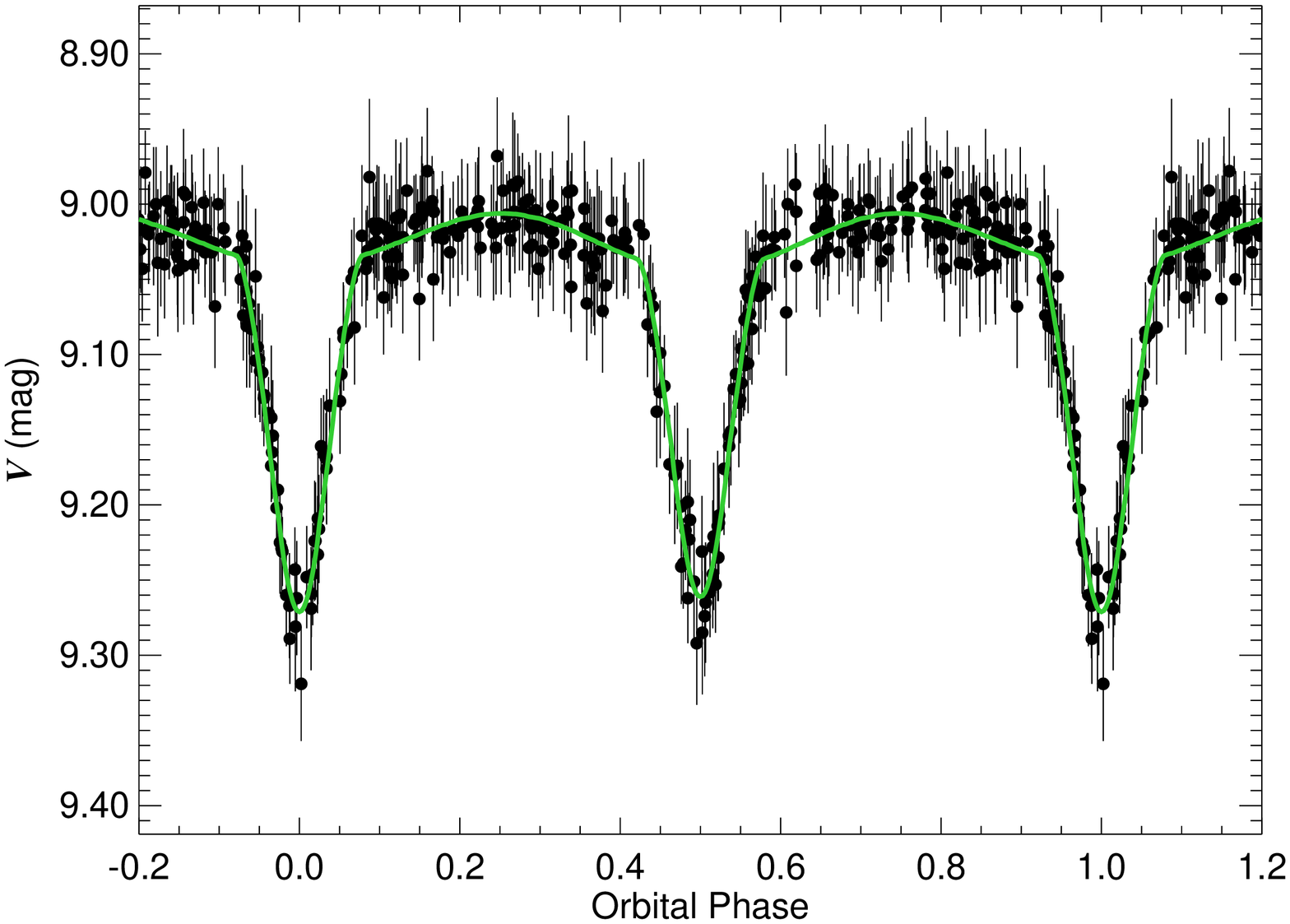}}
\end{center}
\caption{The ASAS $V-$band light curve for ASAS 064539+0219.4
  (HD 48866). Filled circles with lines
  represent data with associated uncertainties. The best fit orbital solution
  listed in Table 2 is shown as a solid line passing through the data.}
\end{figure}
\clearpage

\begin{figure}
\begin{center}
{\includegraphics[angle=90,height=12cm]{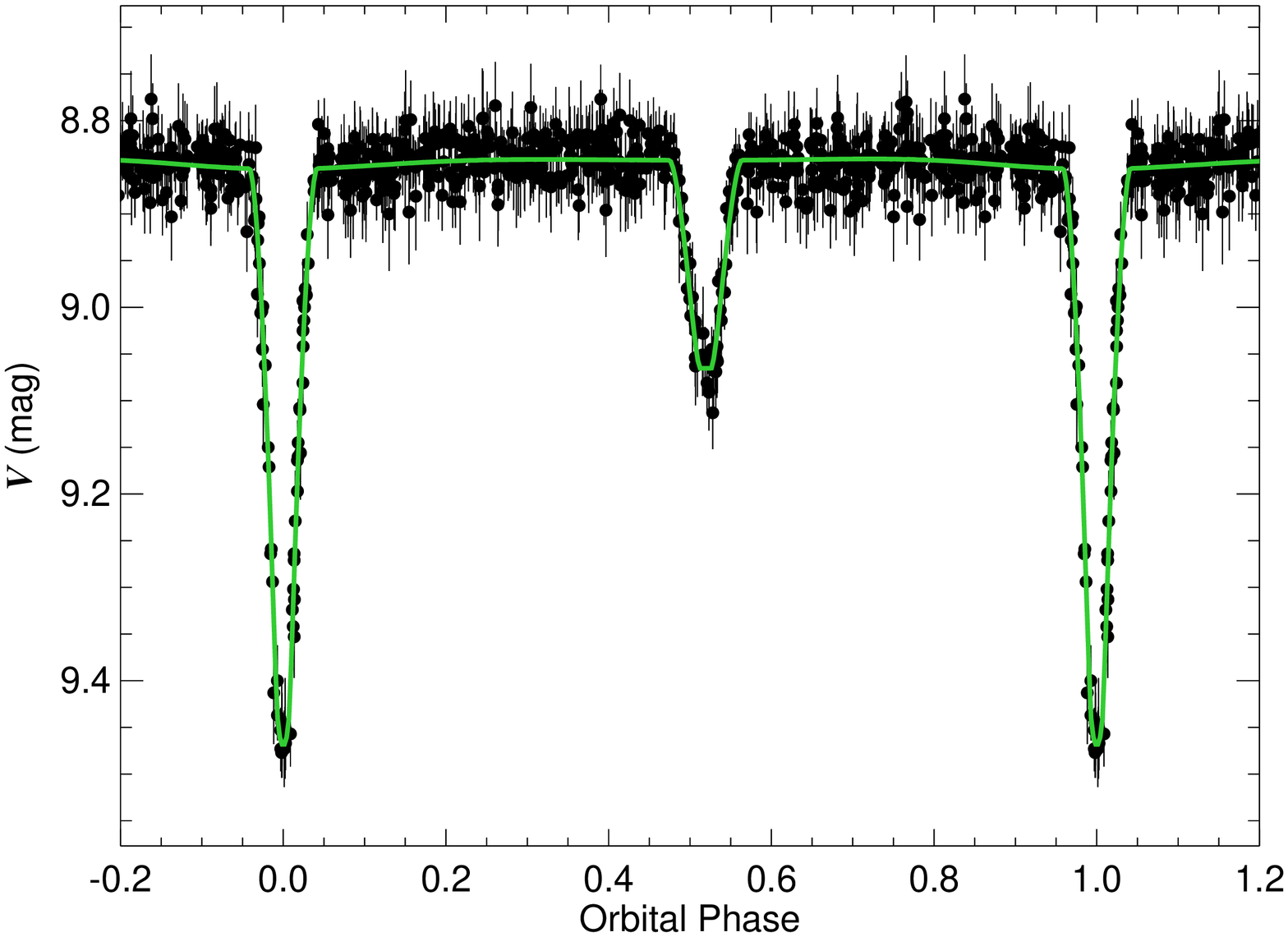}}
\end{center}
\caption{The ASAS $V-$band light curve for ASAS 064609$-$1923.8
  (HD 49125). Filled circles with lines
  represent data with associated uncertainties. The best fit orbital solution
  listed in Table 2 is shown as a solid line passing through the data.}
\end{figure}
\clearpage

\begin{figure}
\begin{center}
{\includegraphics[angle=90,height=12cm]{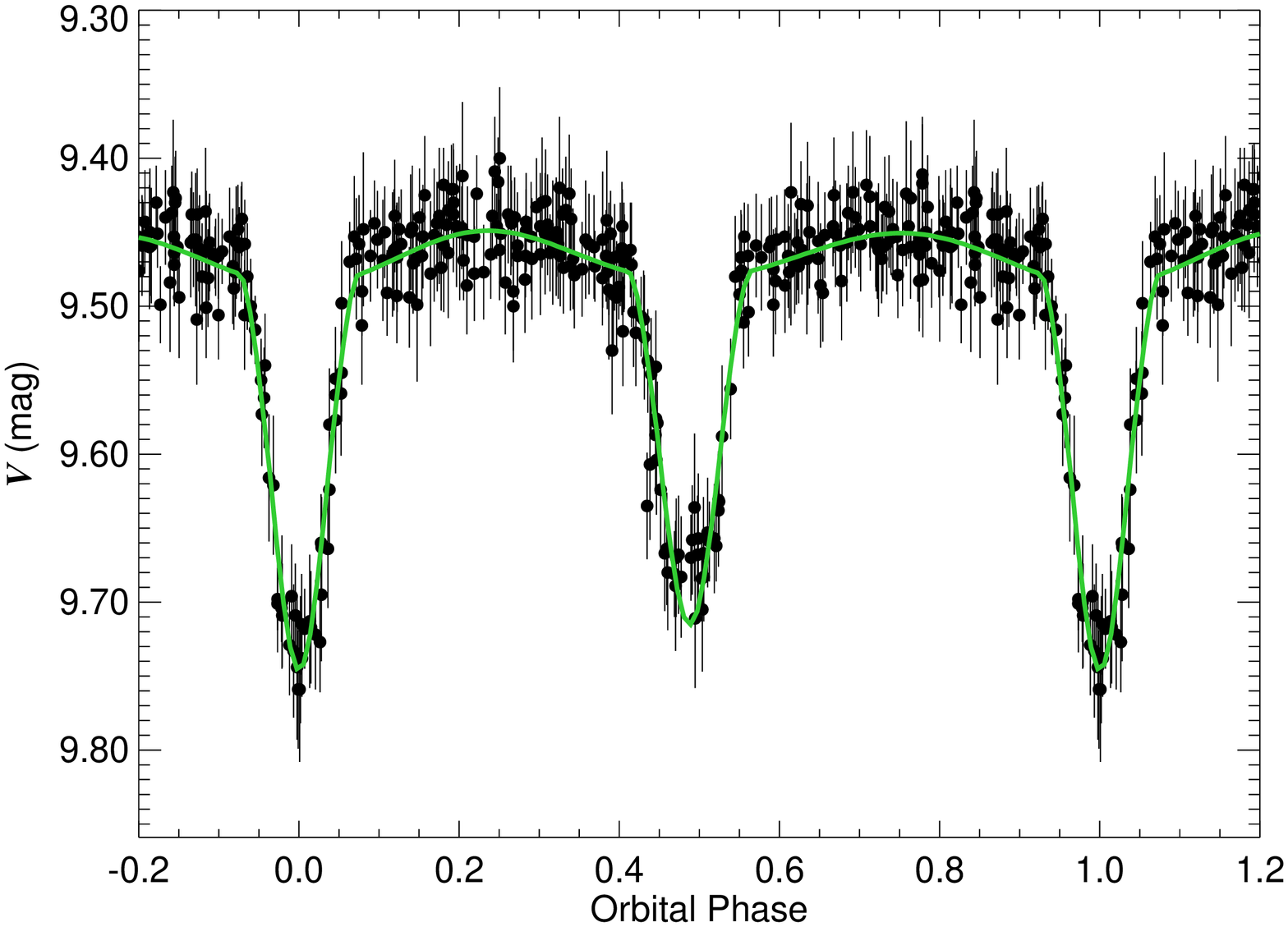}}
\end{center}
\caption{The ASAS $V-$band light curve for ASAS 064715+0225.6
  (HD 289072). Filled circles with lines
  represent data with associated uncertainties. The best fit orbital solution
  listed in Table 2 is shown as a solid line passing through the data.}
\end{figure}
\clearpage

\begin{figure}
\begin{center}
{\includegraphics[angle=90,height=12cm]{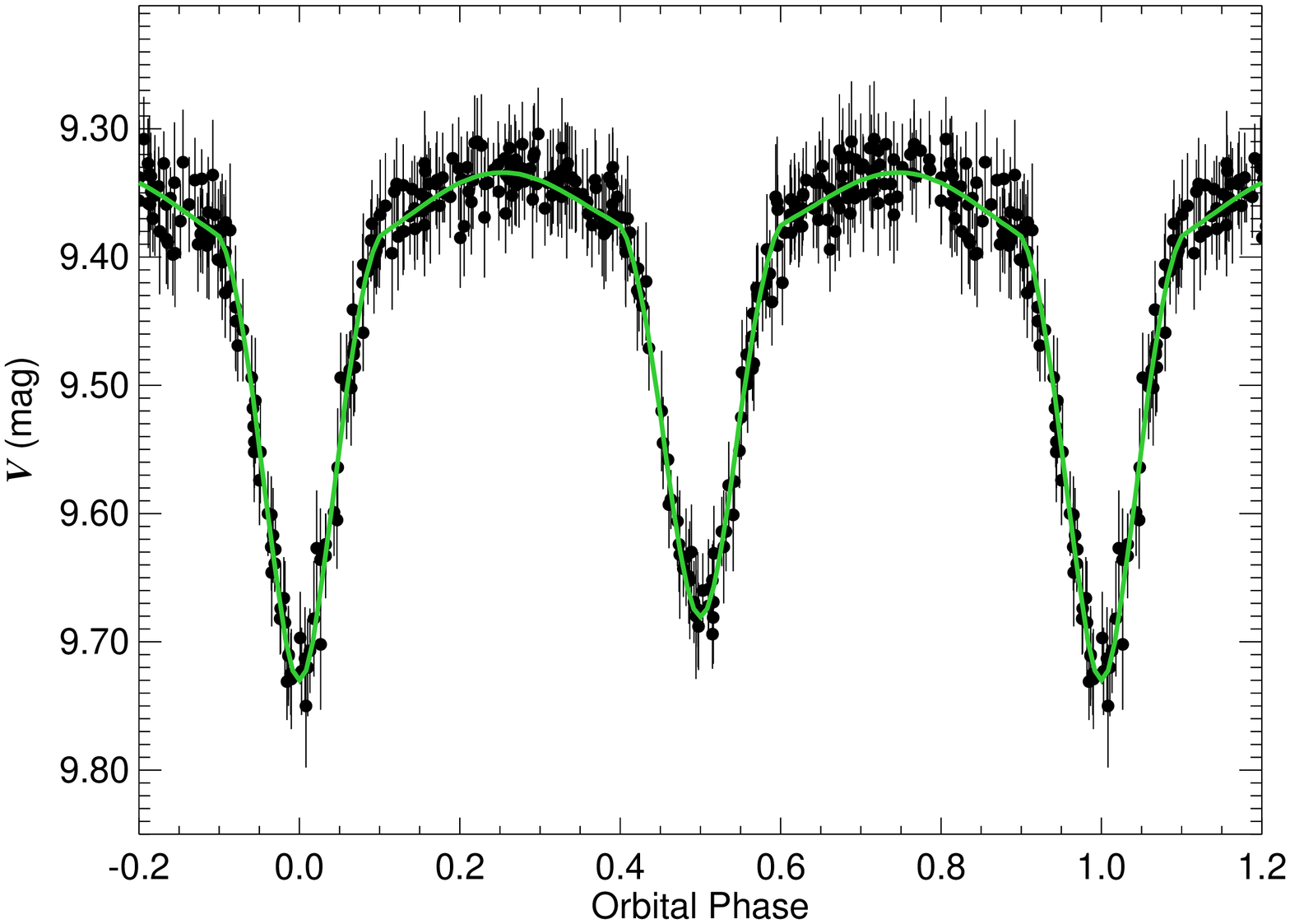}}
\end{center}
\caption{The ASAS $V-$band light curve for ASAS 064745+0122.3
  (V448 Mon). Filled circles with lines
  represent data with associated uncertainties. The best fit orbital solution
  listed in Table 2 is shown as a solid line passing through the data.}
\end{figure}
\clearpage

\begin{figure}
\begin{center}
{\includegraphics[angle=90,height=12cm]{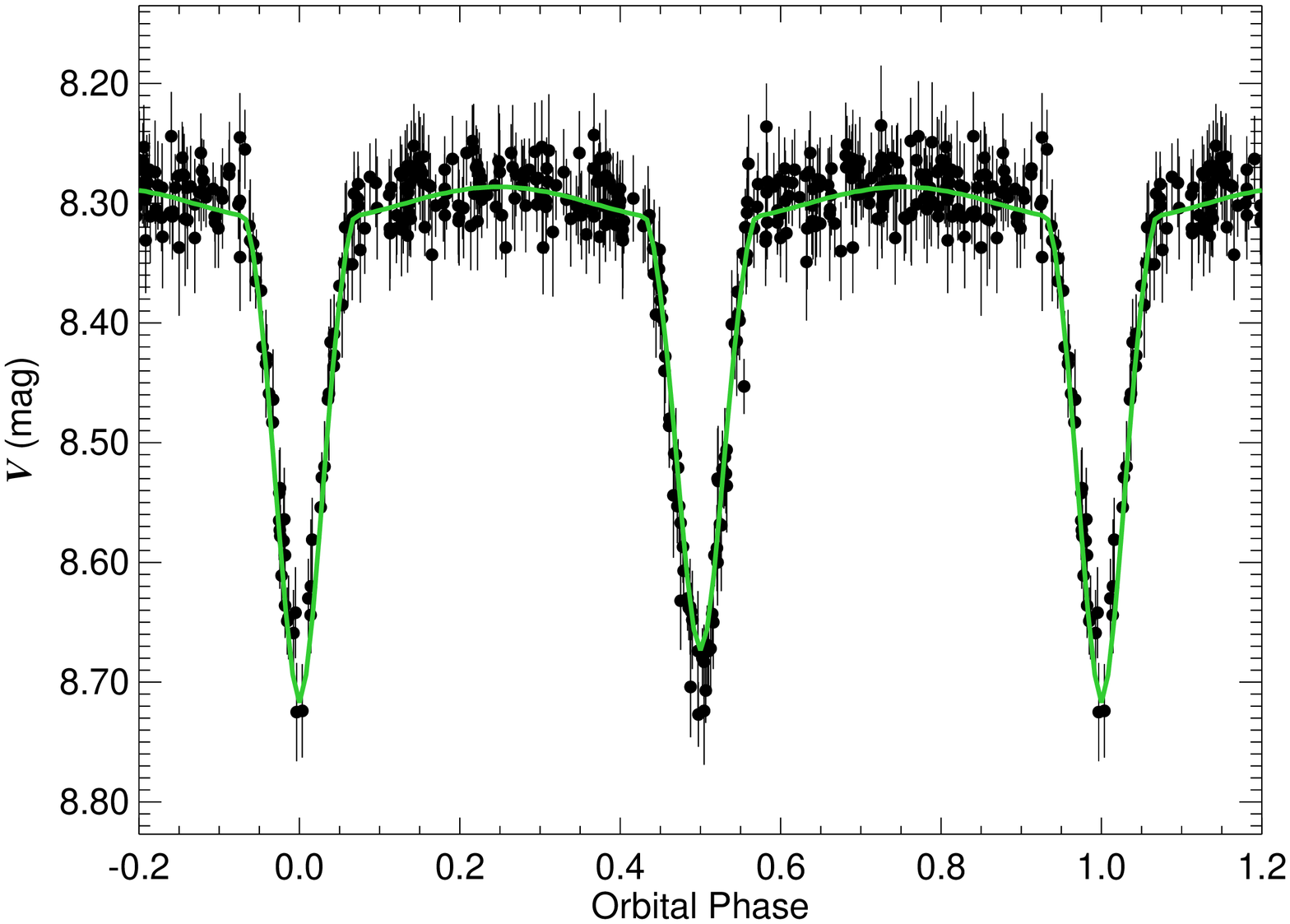}}
\end{center}
\caption{The ASAS $V-$band light curve for ASAS 065534$-$1013.2
  (HD 51082). Filled circles with lines
  represent data with associated uncertainties. The best fit orbital solution
  listed in Table 2 is shown as a solid line passing through the data.}
\end{figure}
\clearpage

\begin{figure}
\begin{center}
{\includegraphics[angle=90,height=12cm]{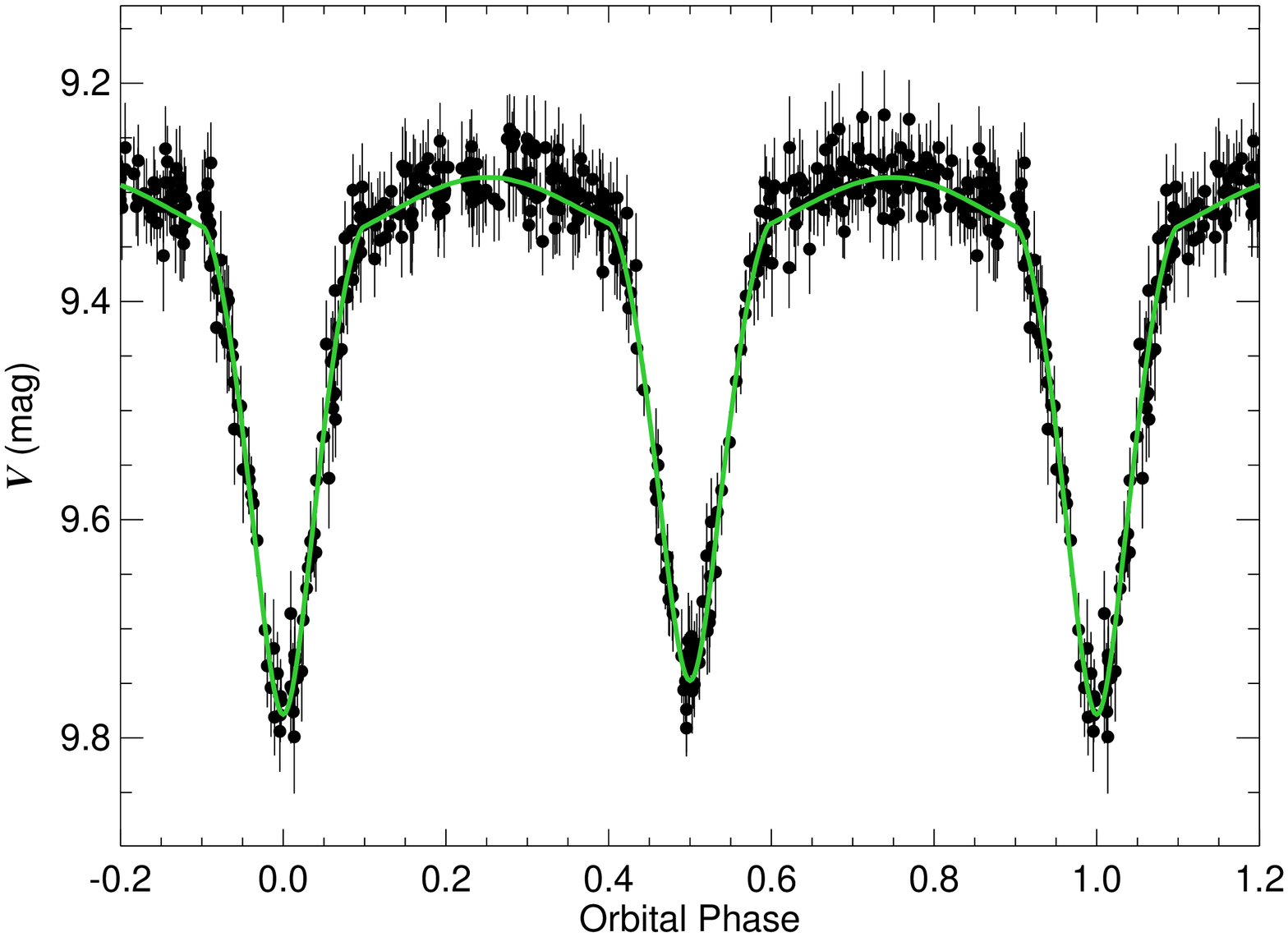}}
\end{center}
\caption{The ASAS $V-$band light curve for ASAS 065549$-$0402.6
  (HI Mon). Filled circles with lines
  represent data with associated uncertainties. The best fit orbital solution
  listed in Table 2 is shown as a solid line passing through the data.}
\end{figure}
\clearpage

\begin{figure}
\begin{center}
{\includegraphics[angle=90,height=12cm]{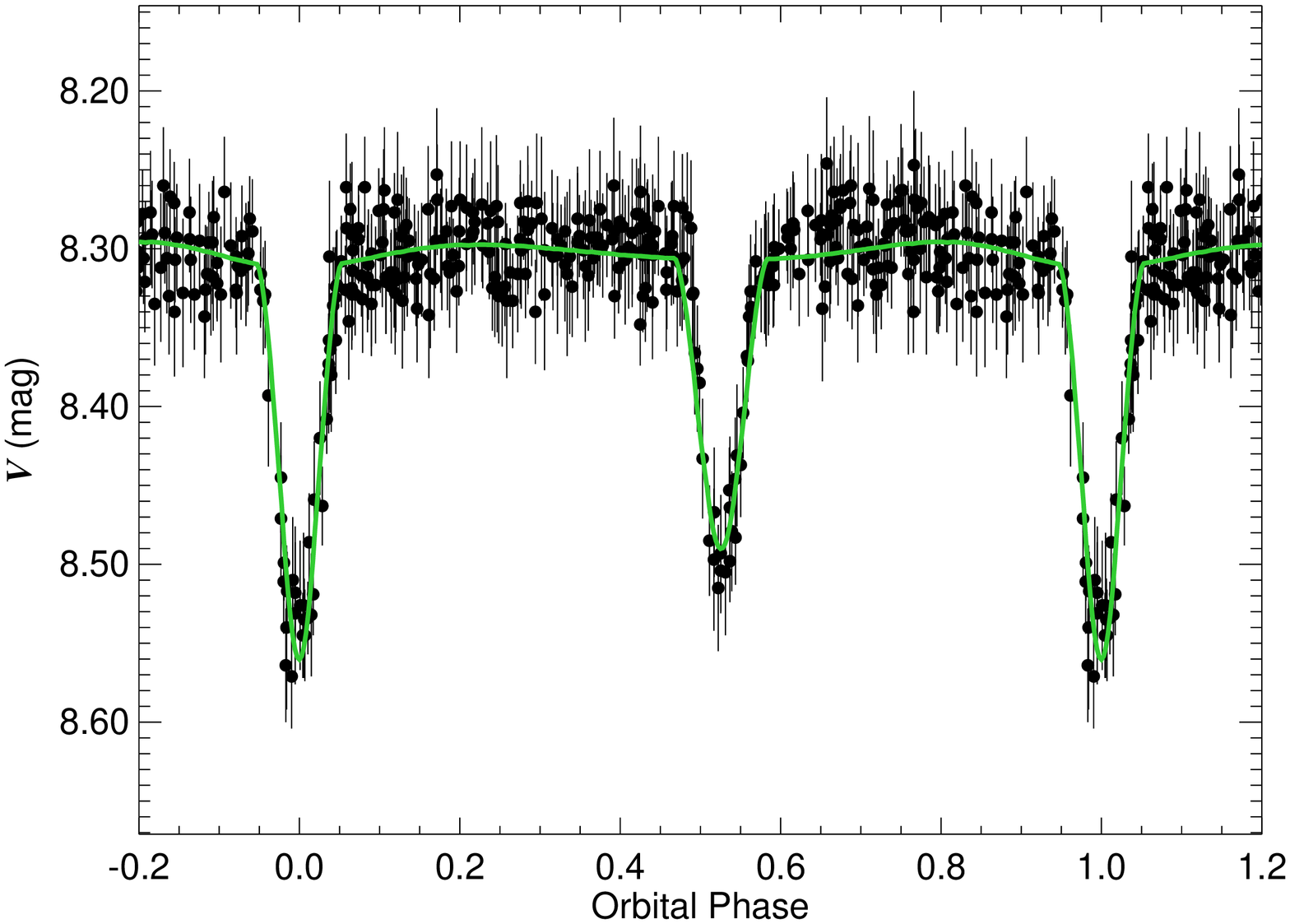}}
\end{center}
\caption{The ASAS $V-$band light curve for ASAS 070105$-$0358.2
  (HD 52433). Filled circles with lines
  represent data with associated uncertainties. The best fit orbital solution
  listed in Table 2 is shown as a solid line passing through the data.}
\end{figure}
\clearpage

\begin{figure}
\begin{center}
{\includegraphics[angle=90,height=12cm]{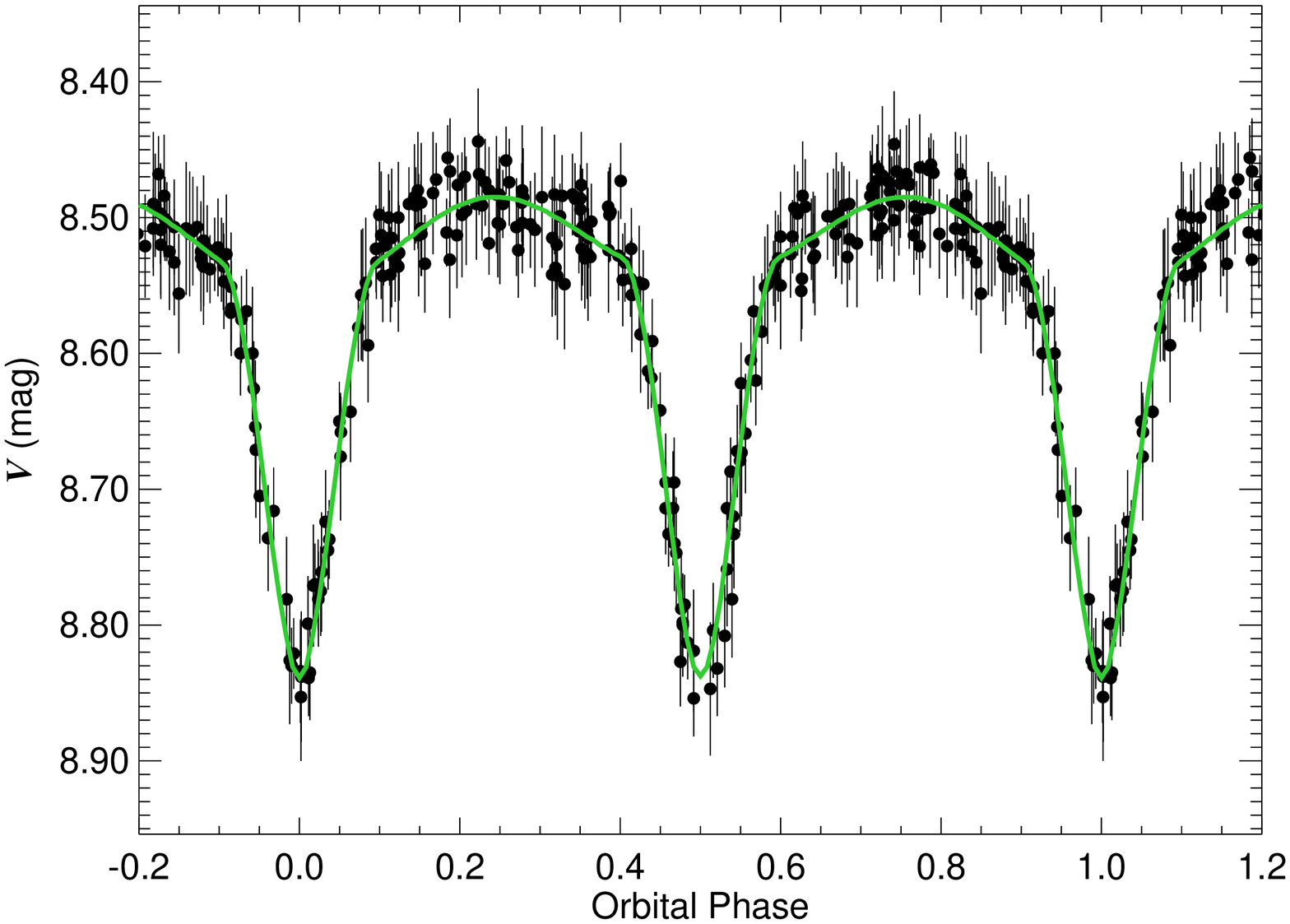}}
\end{center}
\caption{The ASAS $V-$band light curve for ASAS 070238+1347.0
  (HD 52637). Filled circles with lines
  represent data with associated uncertainties. The best fit orbital solution
  listed in Table 2 is shown as a solid line passing through the data.}
\end{figure}
\clearpage

\begin{figure}
\begin{center}
{\includegraphics[angle=90,height=12cm]{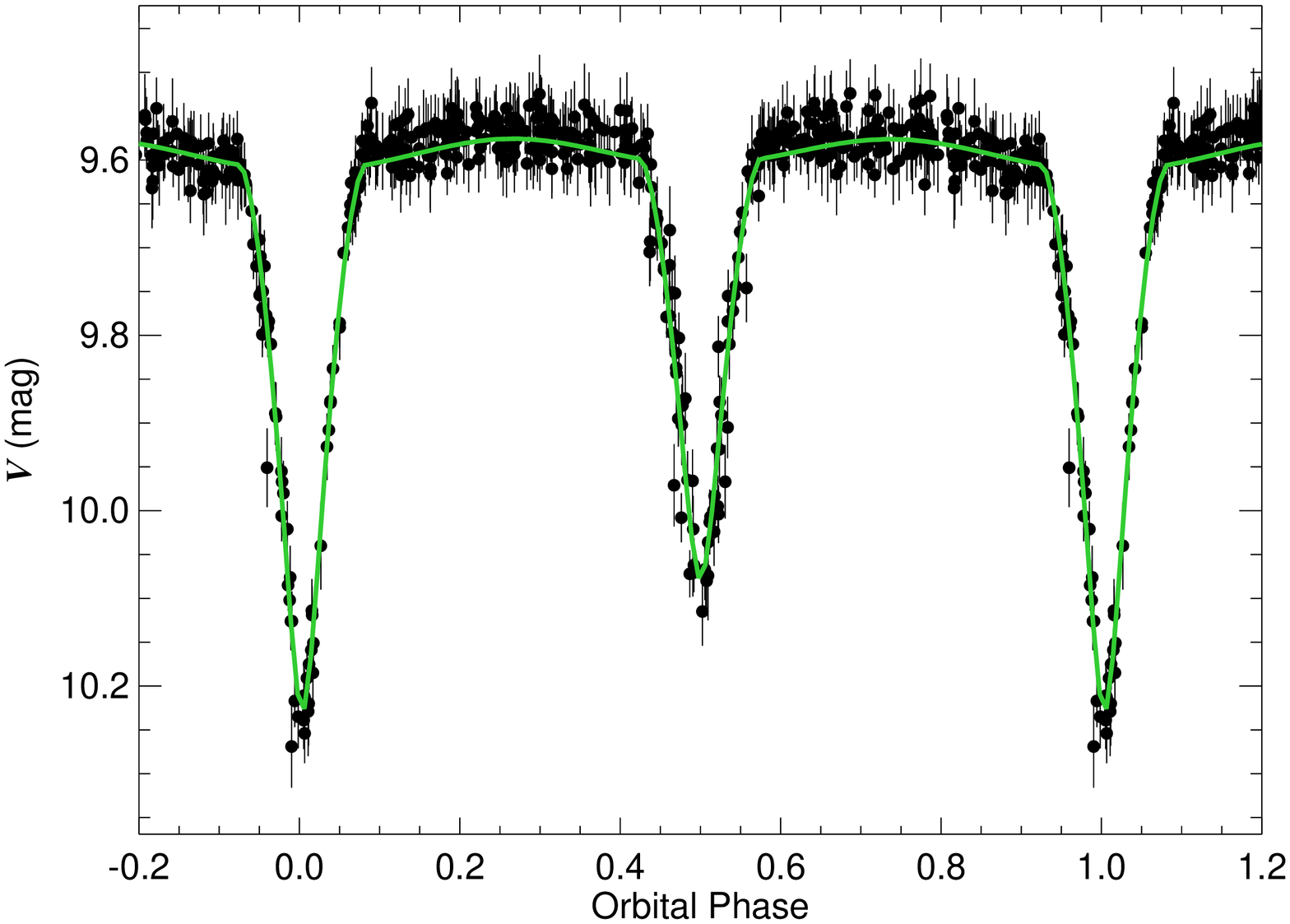}}
\end{center}
\caption{The ASAS $V-$band light curve for ASAS 070636$-$0437.4
  (AO Mon). Filled circles with lines
  represent data with associated uncertainties. The best fit orbital solution
  listed in Table 2 is shown as a solid line passing through the data.}
\end{figure}
\clearpage

\begin{figure}
\begin{center}
{\includegraphics[angle=90,height=12cm]{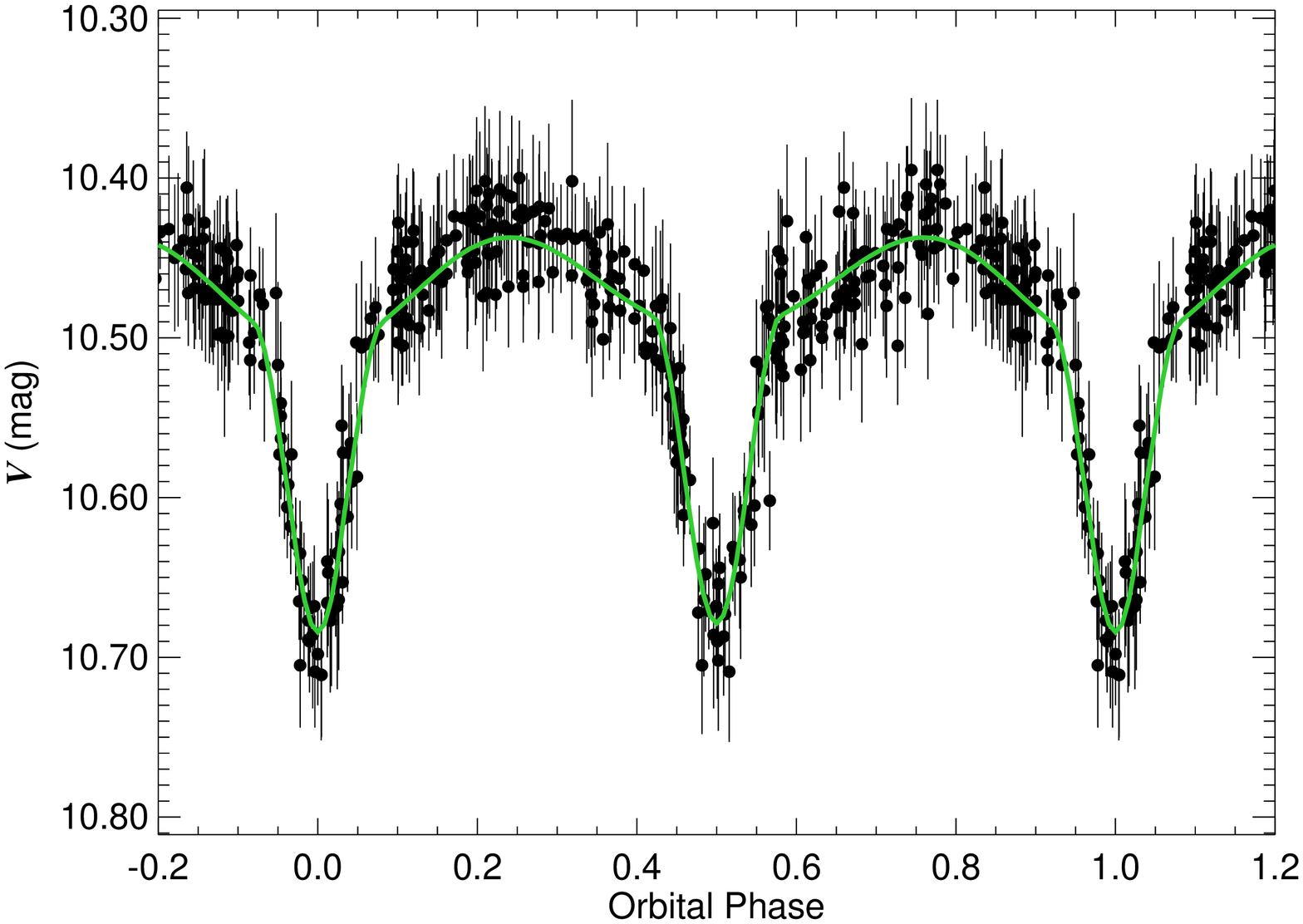}}
\end{center}
\caption{The ASAS $V-$band light curve for ASAS 070943+2341.7
  (BD +23 1621). Filled circles with lines
  represent data with associated uncertainties. The best fit orbital solution
  listed in Table 2 is shown as a solid line passing through the data.}
\end{figure}
\clearpage

\begin{figure}
\begin{center}
{\includegraphics[angle=90,height=12cm]{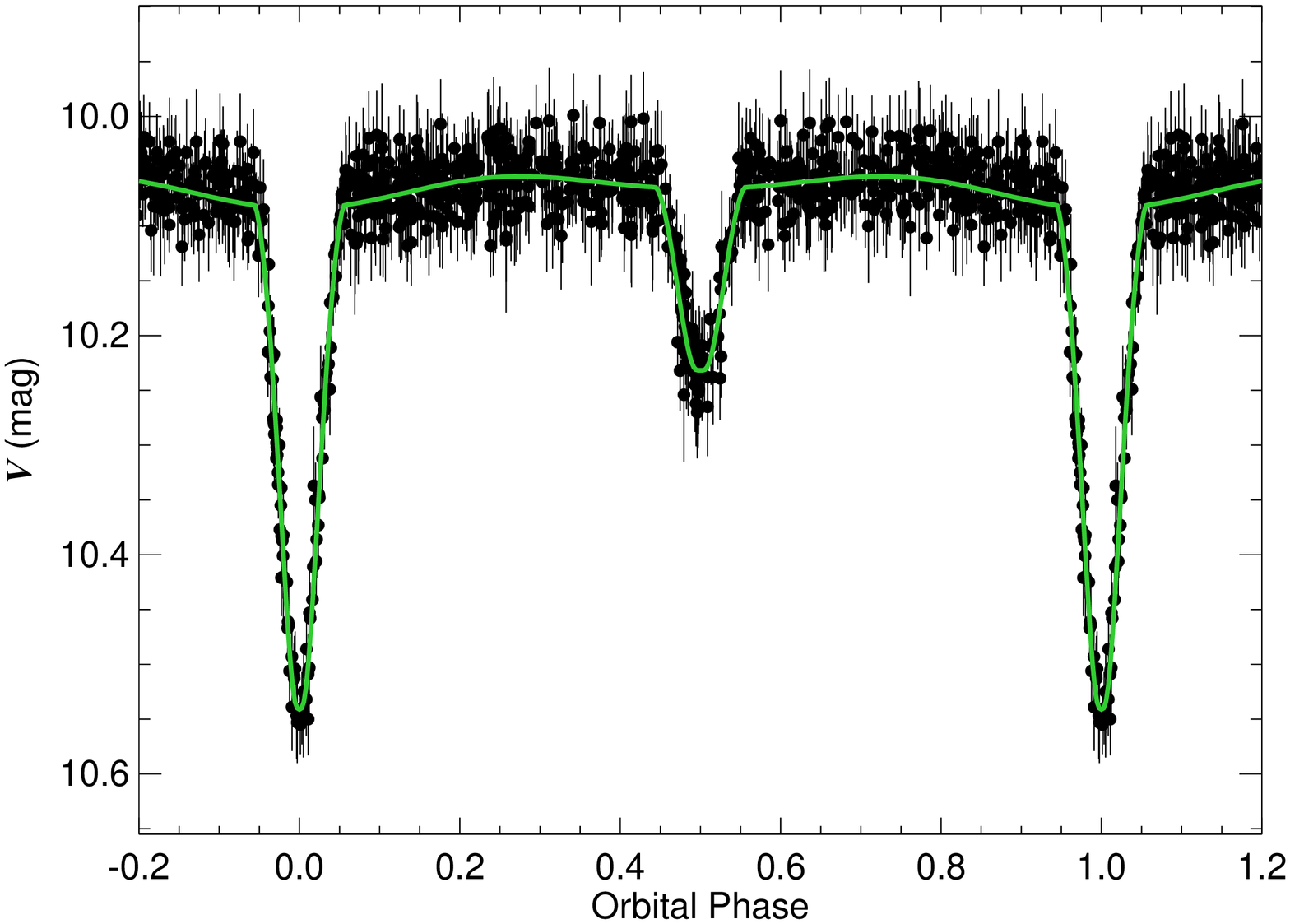}}
\end{center}
\caption{The ASAS $V-$band light curve for ASAS 070946$-$2005.5
  (NSV 3433). Filled circles with lines
  represent data with associated uncertainties. The best fit orbital solution
  listed in Table 2 is shown as a solid line passing through the data.}
\end{figure}
\clearpage

\begin{figure}
\begin{center}
{\includegraphics[angle=90,height=12cm]{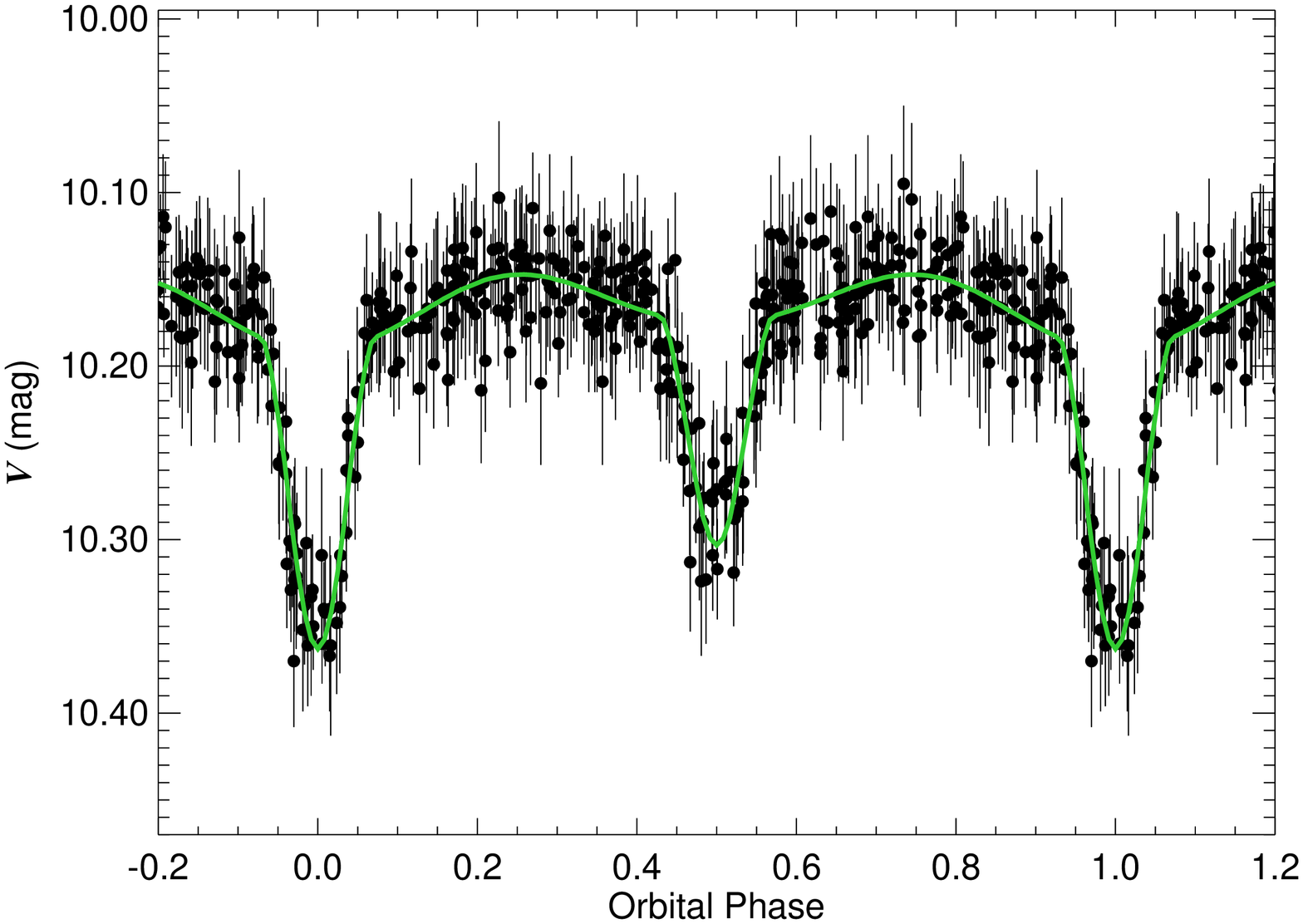}}
\end{center}
\caption{The ASAS $V-$band light curve for ASAS 071010$-$0035.1
  (HD 54780). Filled circles with lines
  represent data with associated uncertainties. The best fit orbital solution
  listed in Table 2 is shown as a solid line passing through the data.}
\end{figure}
\clearpage

\begin{figure}
\begin{center}
{\includegraphics[angle=90,height=12cm]{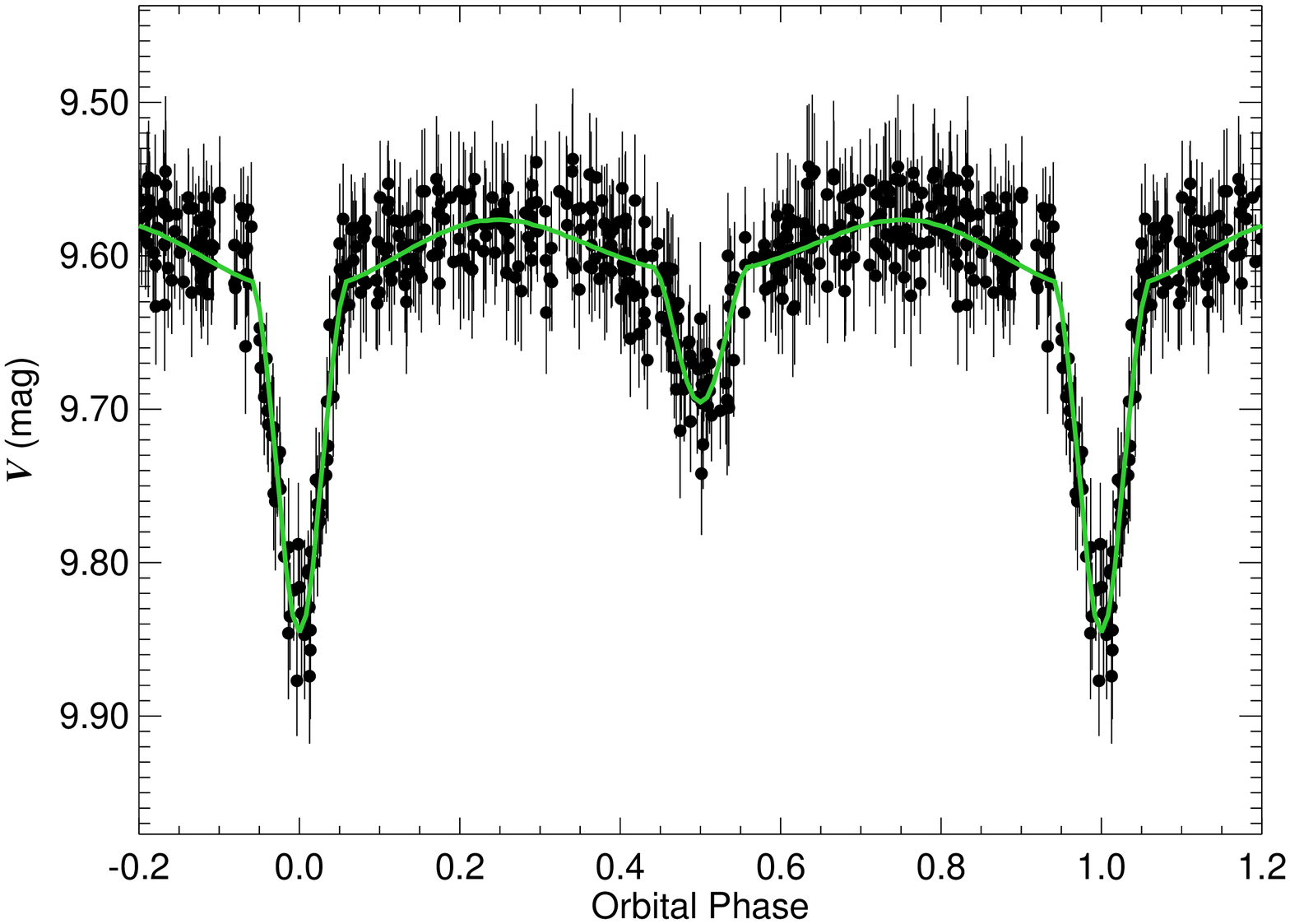}}
\end{center}
\caption{The ASAS $V-$band light curve for ASAS 071203$-$0139.1
  (HD 55236). Filled circles with lines
  represent data with associated uncertainties. The best fit orbital solution
  listed in Table 2 is shown as a solid line passing through the data.}
\end{figure}
\clearpage

\begin{figure}
\begin{center}
{\includegraphics[angle=90,height=12cm]{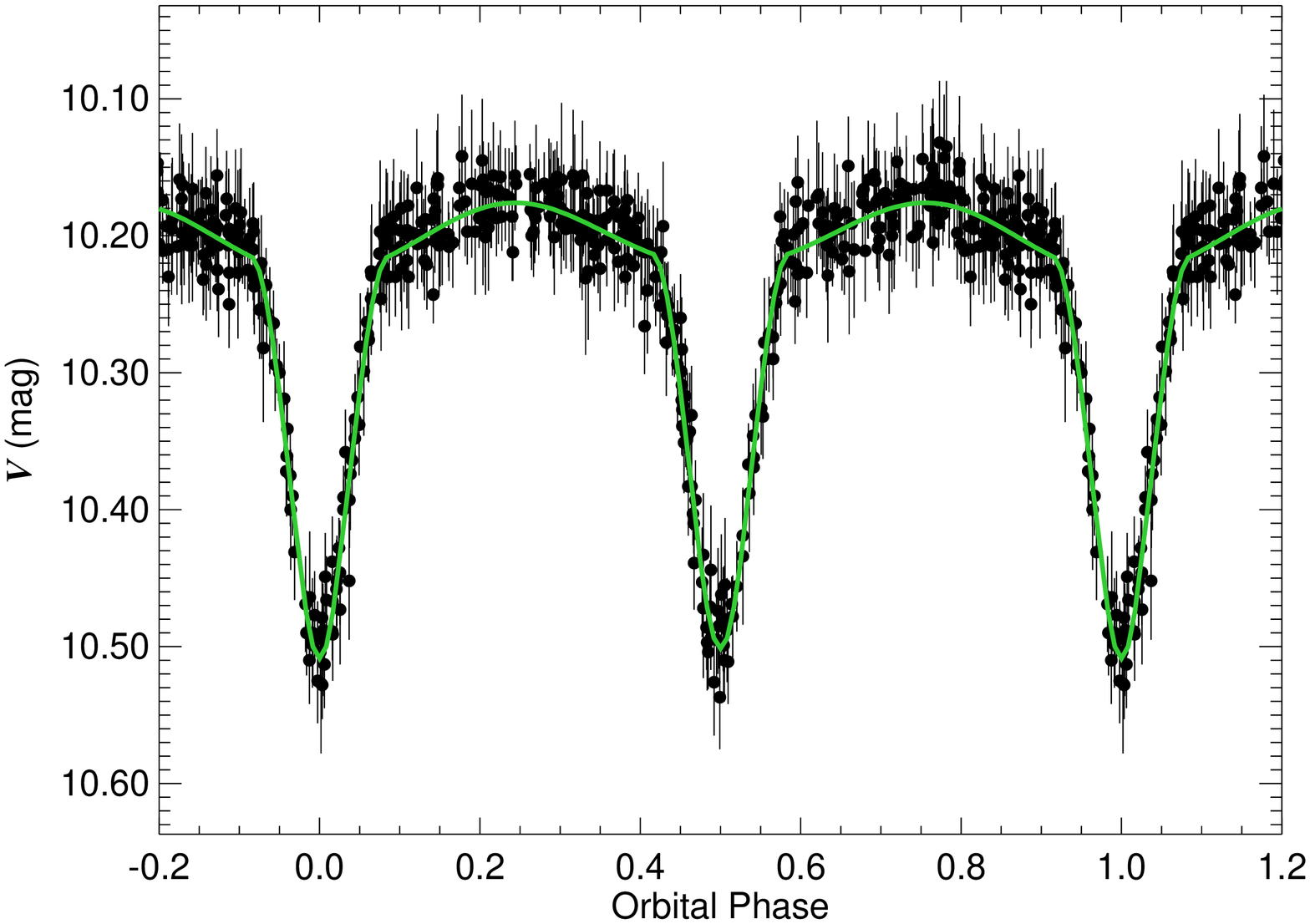}}
\end{center}
\caption{The ASAS $V-$band light curve for ASAS 071702$-$1034.9
  (HD 56544). Filled circles with lines
  represent data with associated uncertainties. The best fit orbital solution
  listed in Table 2 is shown as a solid line passing through the data.}
\end{figure}
\clearpage

\begin{figure}
\begin{center}
{\includegraphics[angle=90,height=12cm]{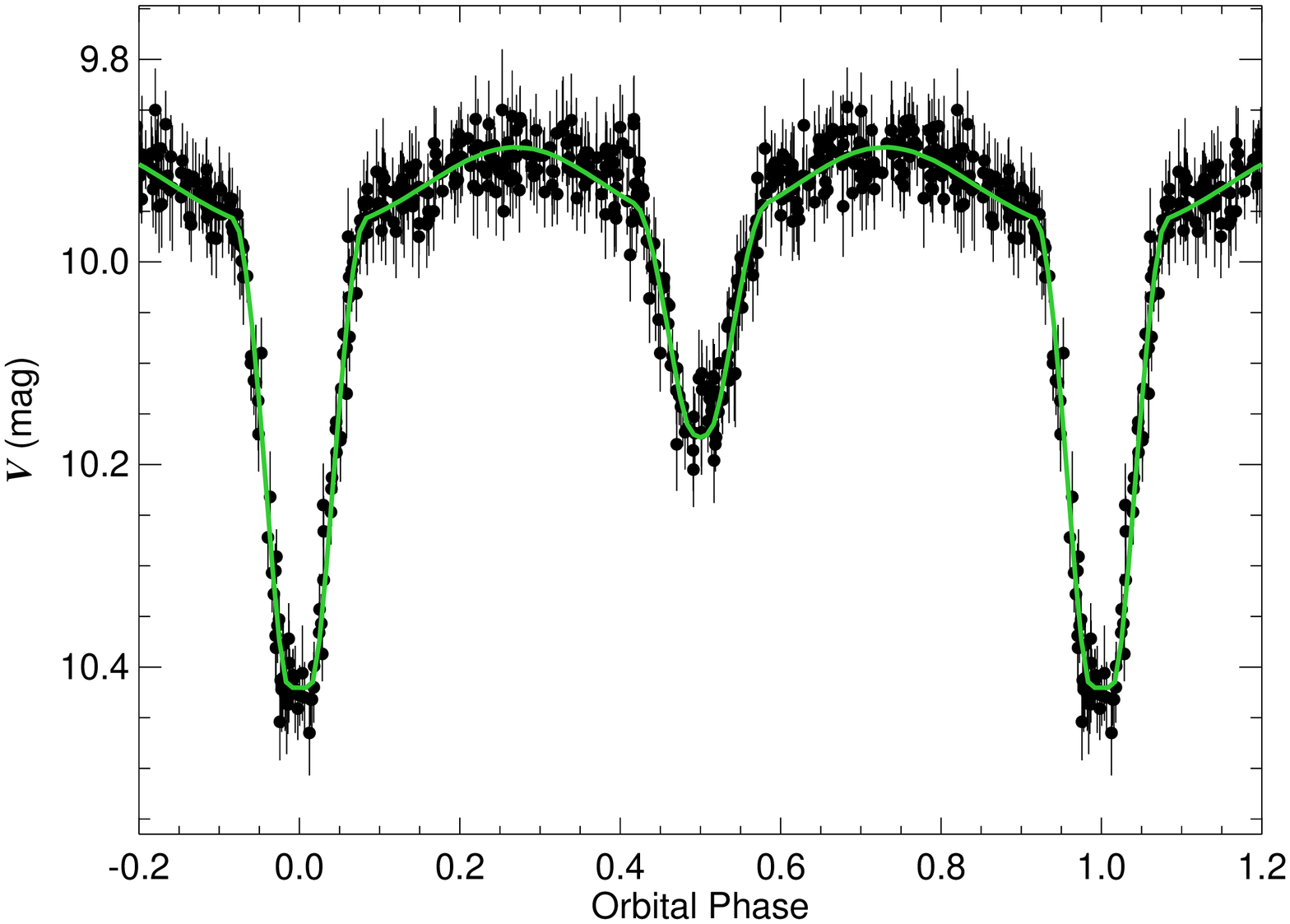}}
\end{center}
\caption{The ASAS $V-$band light curve for ASAS 072201$-$2552.6
  (CX CMa). Filled circles with lines
  represent data with associated uncertainties. The best fit orbital solution
  listed in Table 2 is shown as a solid line passing through the data.}
\end{figure}
\clearpage

\begin{figure}
\begin{center}
{\includegraphics[angle=90,height=12cm]{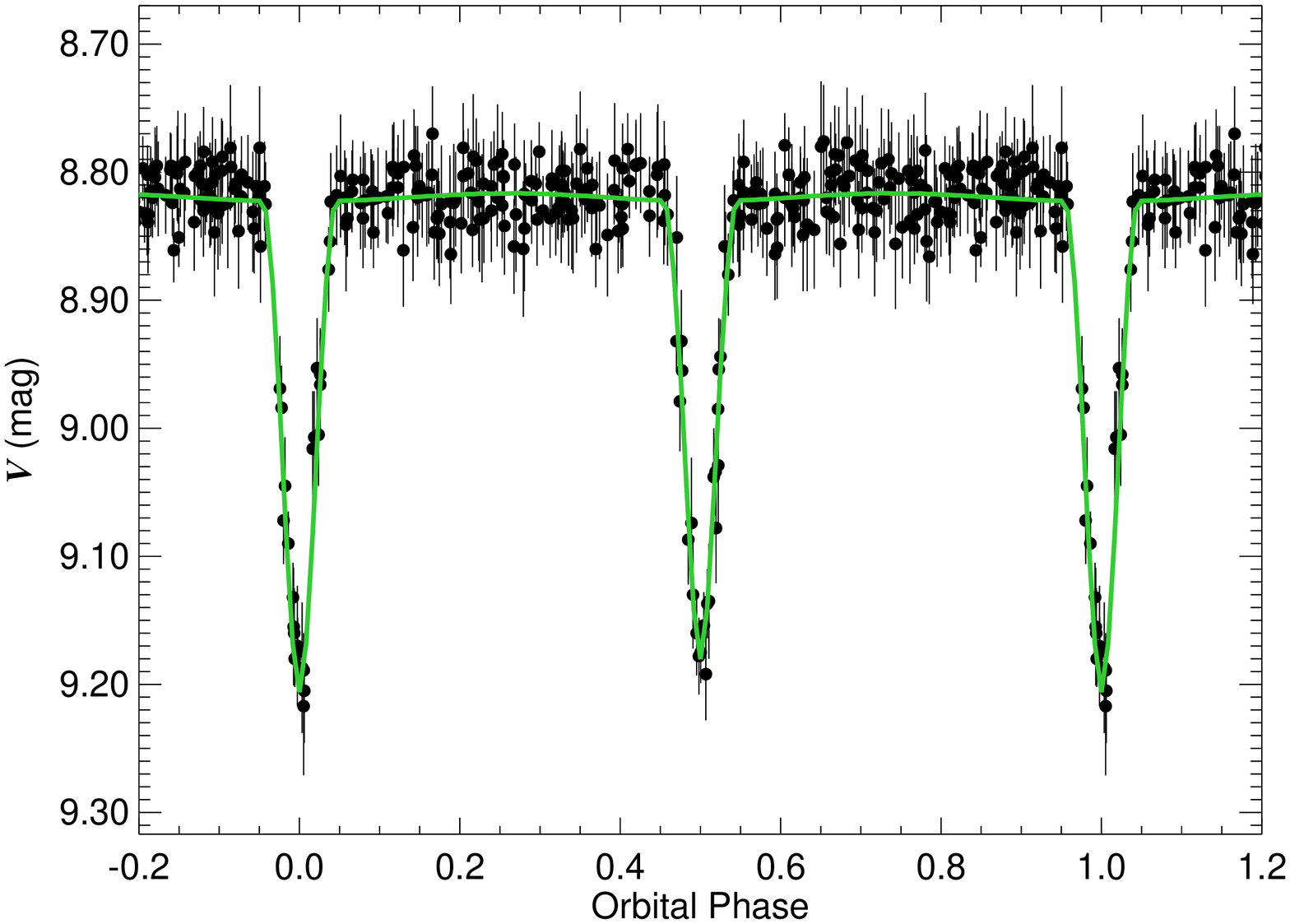}}
\end{center}
\caption{The ASAS $V-$band light curve for ASAS 073053+0513.7
  (HD 59607). Filled circles with lines
  represent data with associated uncertainties. The best fit orbital solution
  listed in Table 2 is shown as a solid line passing through the data.}
\end{figure}
\clearpage

\begin{figure}
\begin{center}
{\includegraphics[angle=90,height=12cm]{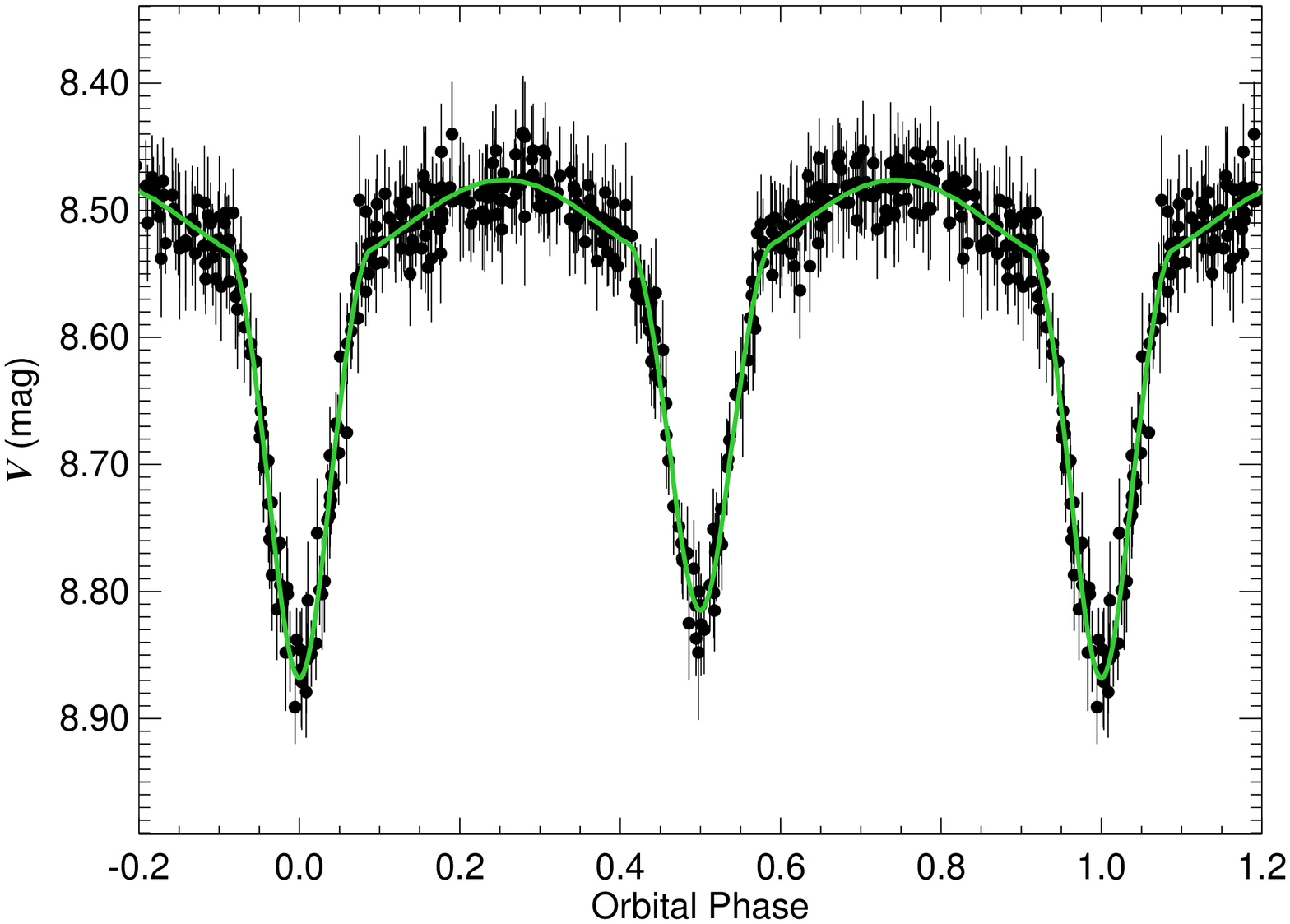}}
\end{center}
\caption{The ASAS $V-$band light curve for ASAS 073348$-$0940.9
  (HD 60389). Filled circles with lines
  represent data with associated uncertainties. The best fit orbital solution
  listed in Table 2 is shown as a solid line passing through the data.}
\end{figure}
\clearpage

\begin{figure}
\begin{center}
{\includegraphics[angle=90,height=12cm]{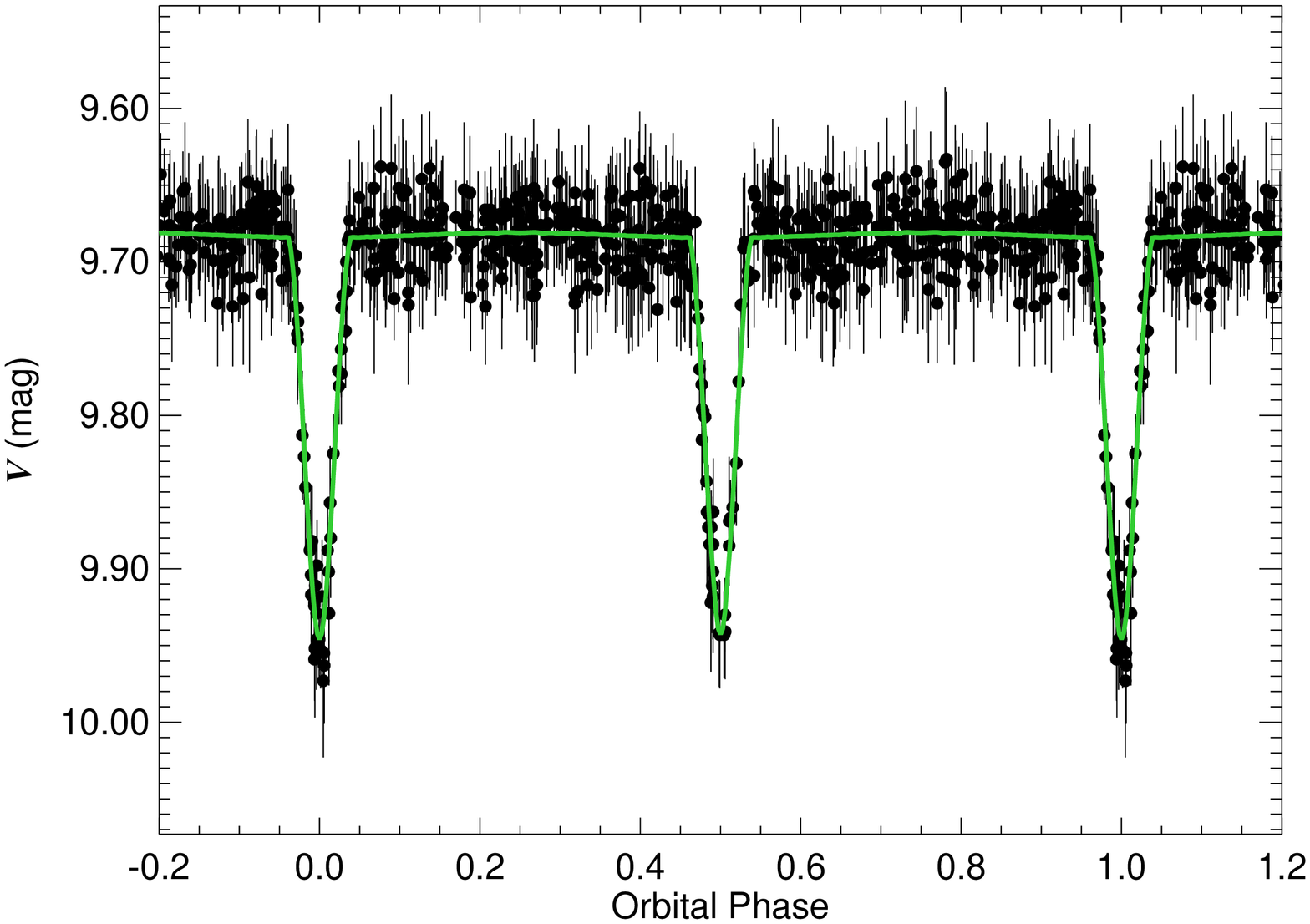}}
\end{center}
\caption{The ASAS $V-$band light curve for ASAS 074355$-$2517.9
  (HD 62607). Filled circles with lines
  represent data with associated uncertainties. The best fit orbital solution
  listed in Table 2 is shown as a solid line passing through the data.}
\end{figure}
\clearpage

\begin{figure}
\begin{center}
{\includegraphics[angle=90,height=12cm]{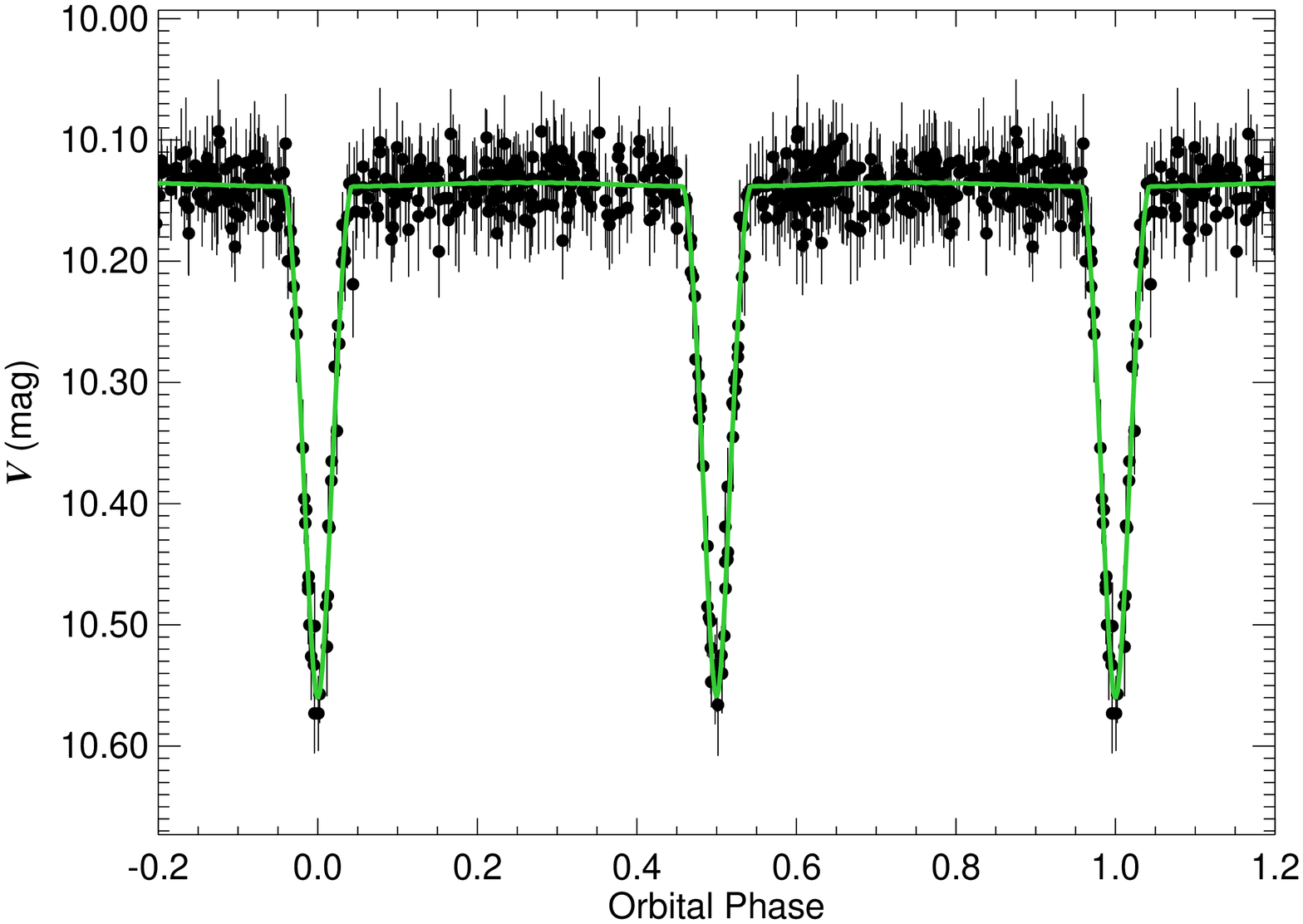}}
\end{center}
\caption{The ASAS $V-$band light curve for ASAS 074714$-$0519.8
  (HD 63141). Filled circles with lines
  represent data with associated uncertainties. The best fit orbital solution
  listed in Table 2 is shown as a solid line passing through the data.}
\end{figure}
\clearpage

\begin{figure}
\begin{center}
{\includegraphics[angle=90,height=12cm]{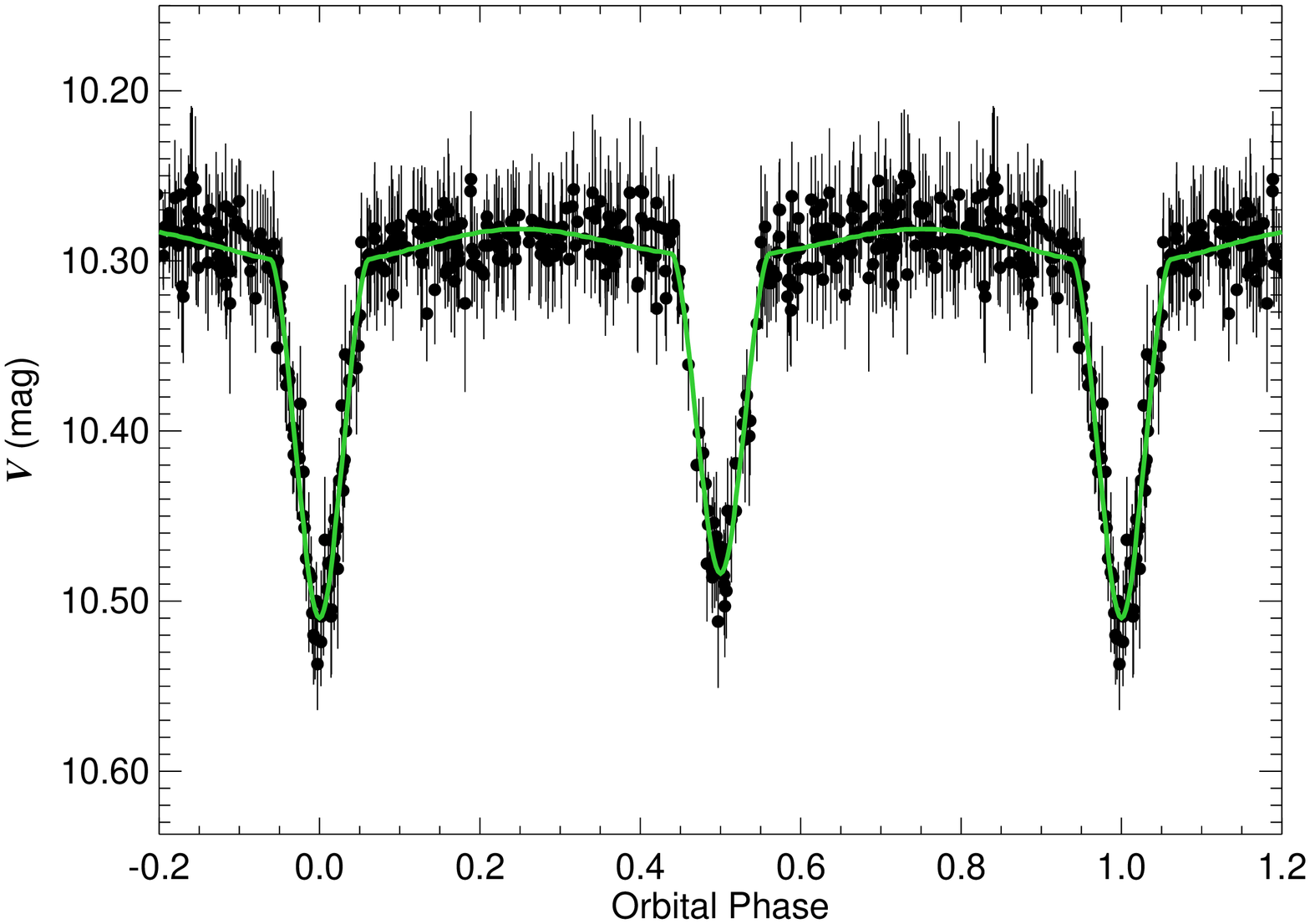}}
\end{center}
\caption{The ASAS $V-$band light curve for ASAS 074928$-$0721.6
  (BD $-06 $ 2317). Filled circles with lines
  represent data with associated uncertainties. The best fit orbital solution
  listed in Table 2 is shown as a solid line passing through the data.}
\end{figure}
\clearpage

\begin{figure}
\begin{center}
{\includegraphics[angle=90,height=12cm]{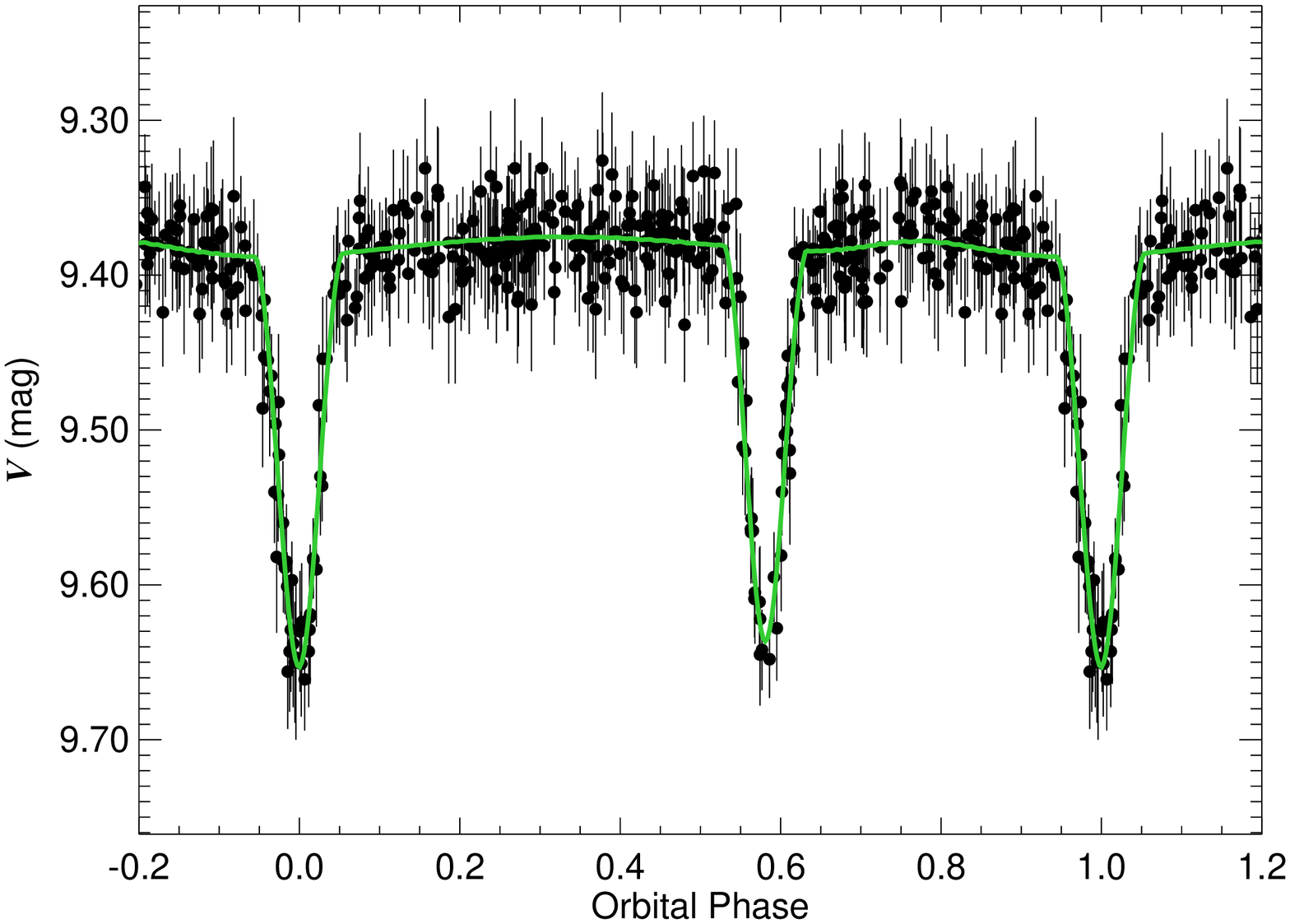}}
\end{center}
\caption{The ASAS $V-$band light curve for ASAS 075052+0048.0
  (HD 63818). Filled circles with lines
  represent data with associated uncertainties. The best fit orbital solution
  listed in Table 2 is shown as a solid line passing through the data.}
\end{figure}
\clearpage

\begin{figure}
\begin{center}
{\includegraphics[angle=90,height=12cm]{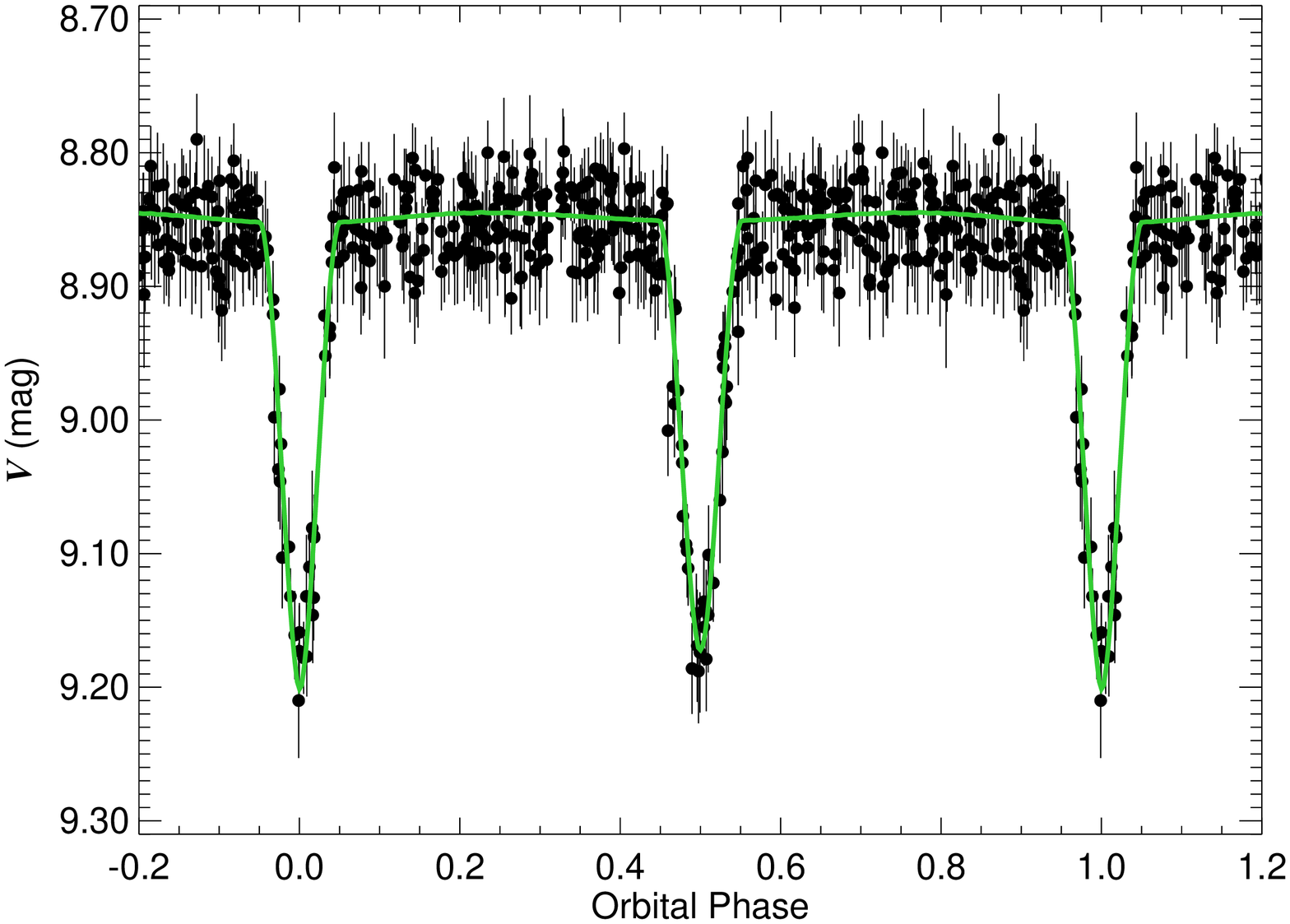}}
\end{center}
\caption{The ASAS $V-$band light curve for ASAS 080617$-$0426.8
  (V871 Mon). Filled circles with lines
  represent data with associated uncertainties. The best fit orbital solution
  listed in Table 2 is shown as a solid line passing through the data.}
\end{figure}
\clearpage

\begin{figure}
\begin{center}
{\includegraphics[angle=90,height=12cm]{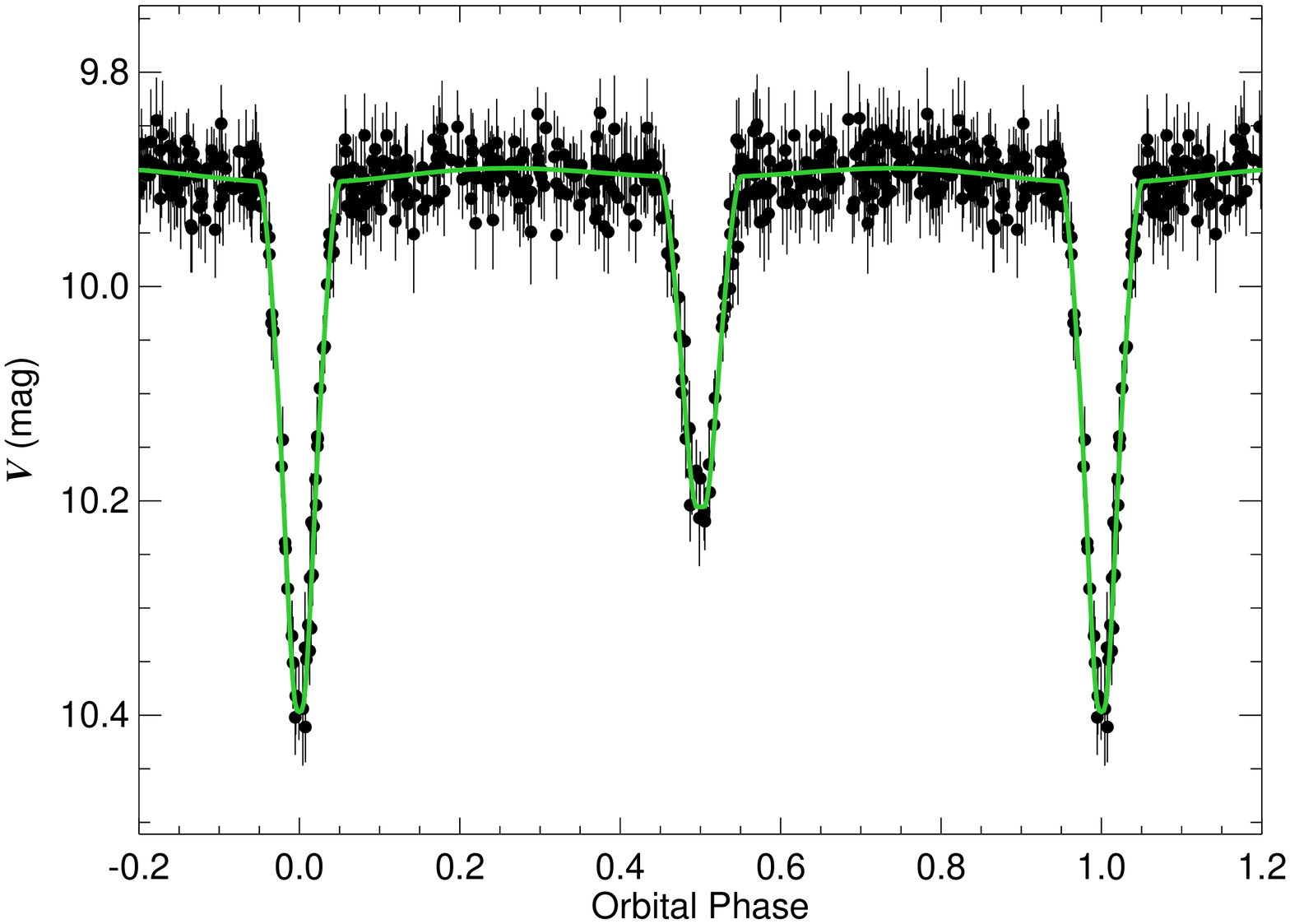}}
\end{center}
\caption{The ASAS $V-$band light curve for ASAS 081749$-$2659.7
  (HD 69797). Filled circles with lines
  represent data with associated uncertainties. The best fit orbital solution
  listed in Table 2 is shown as a solid line passing through the data.}
\end{figure}
\clearpage

\begin{figure}
\begin{center}
{\includegraphics[angle=90,height=12cm]{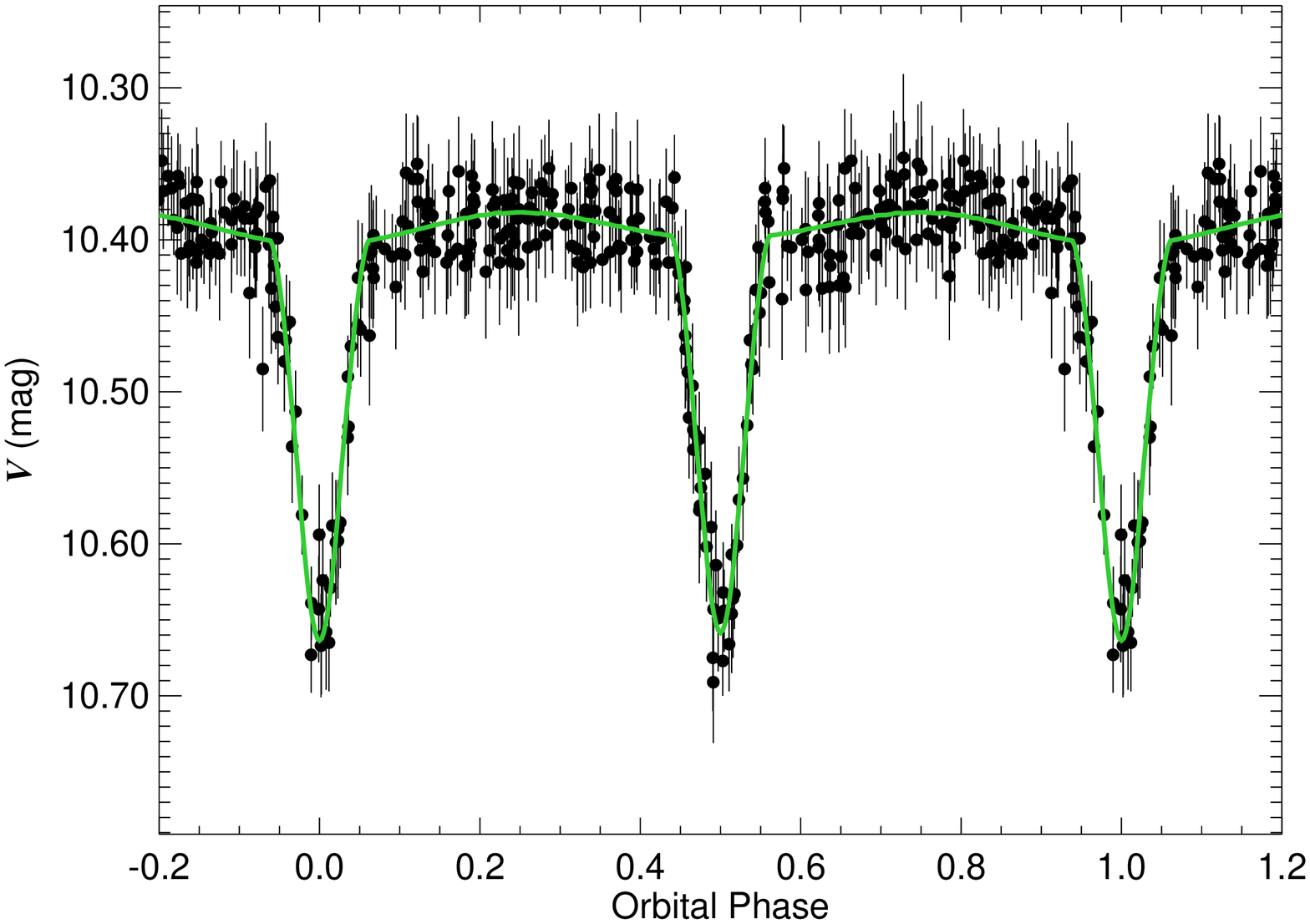}}
\end{center}
\caption{The ASAS $V-$band light curve for ASAS 083245+0247.3
  (BD +03 2001). Filled circles with lines
  represent data with associated uncertainties. The best fit orbital solution
  listed in Table 2 is shown as a solid line passing through the data.}
\end{figure}
\clearpage

\begin{figure}
\begin{center}
{\includegraphics[angle=90,height=12cm]{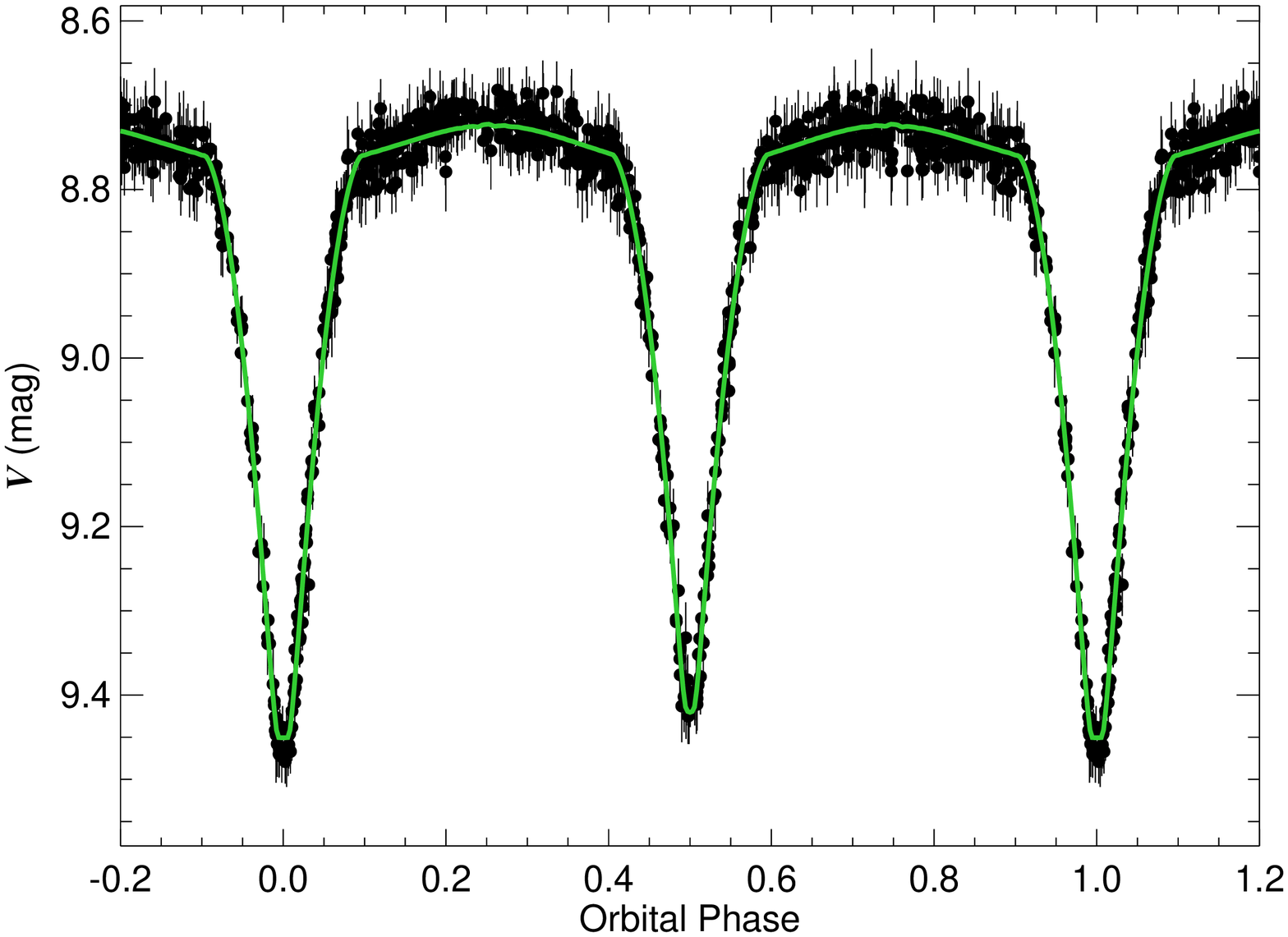}}
\end{center}
\caption{The ASAS $V-$band light curve for ASAS 084831$-$2609.8
  (TT Pyx). Filled circles with lines
  represent data with associated uncertainties. The best fit orbital solution
  listed in Table 2 is shown as a solid line passing through the data.}
\end{figure}
\clearpage

\begin{figure}
\begin{center}
{\includegraphics[angle=90,height=12cm]{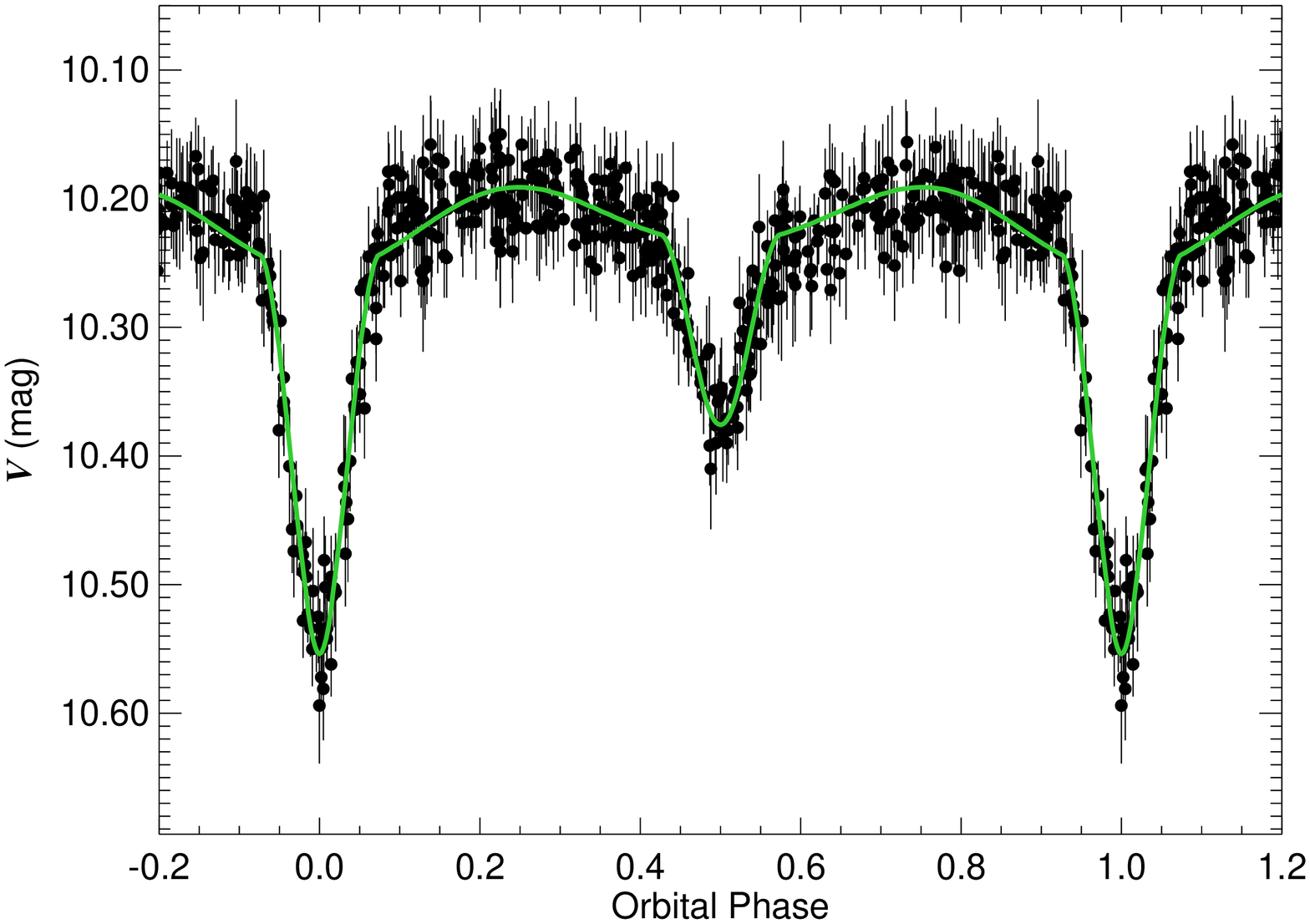}}
\end{center}
\caption{The ASAS $V-$band light curve for ASAS 101120$-$1956.3
  (HD 88409). Filled circles with lines
  represent data with associated uncertainties. The best fit orbital solution
  listed in Table 2 is shown as a solid line passing through the data.}
\end{figure}
\clearpage

\begin{figure}
\begin{center}
{\includegraphics[angle=90,height=12cm]{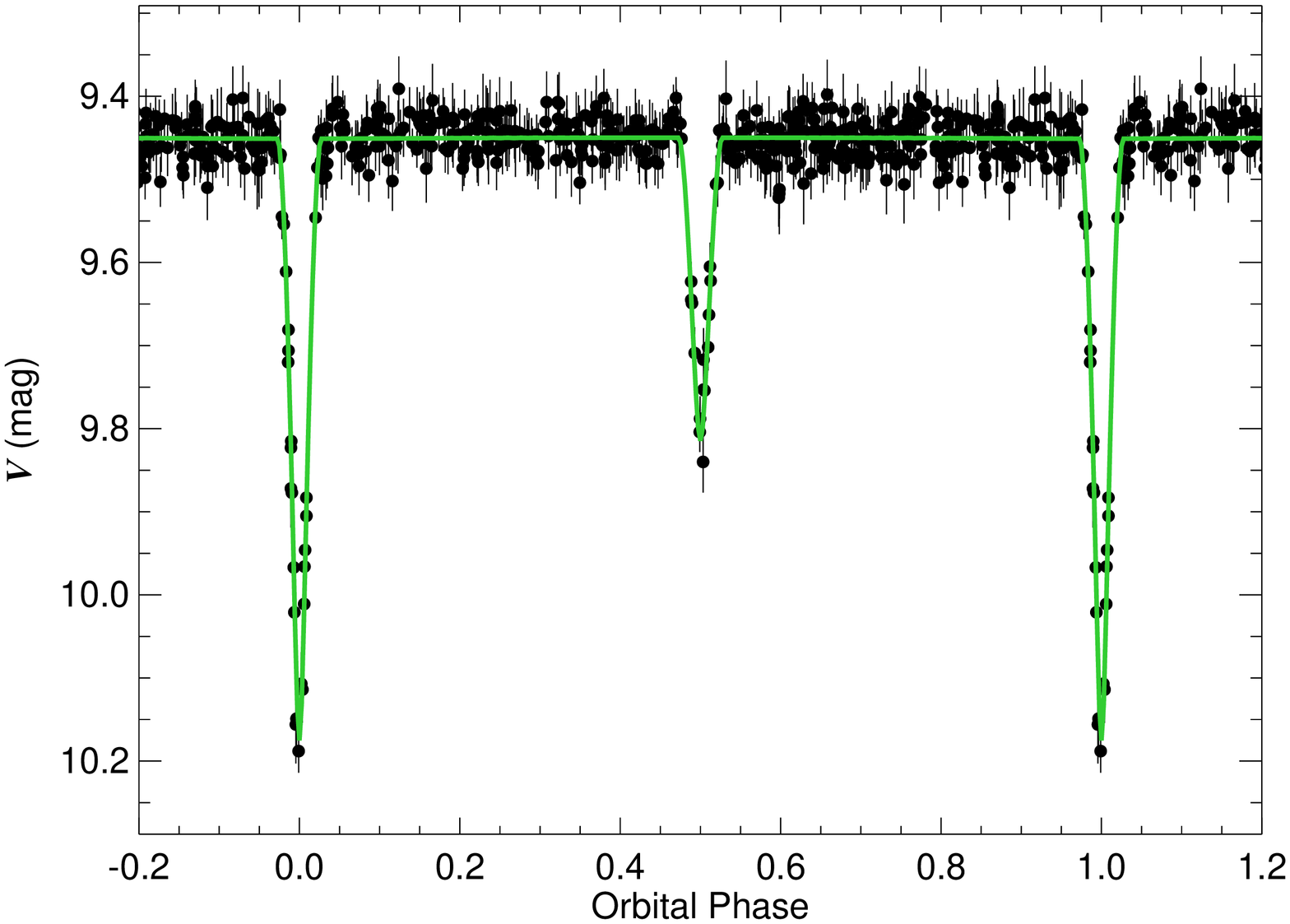}}
\end{center}
\caption{The ASAS $V-$band light curve for ASAS 135949$-$2745.5
  (HD 122026). Filled circles with lines
  represent data with associated uncertainties. The best fit orbital solution
  listed in Table 2 is shown as a solid line passing through the data.}
\end{figure}
\clearpage

\begin{figure}
\begin{center}
{\includegraphics[angle=90,height=12cm]{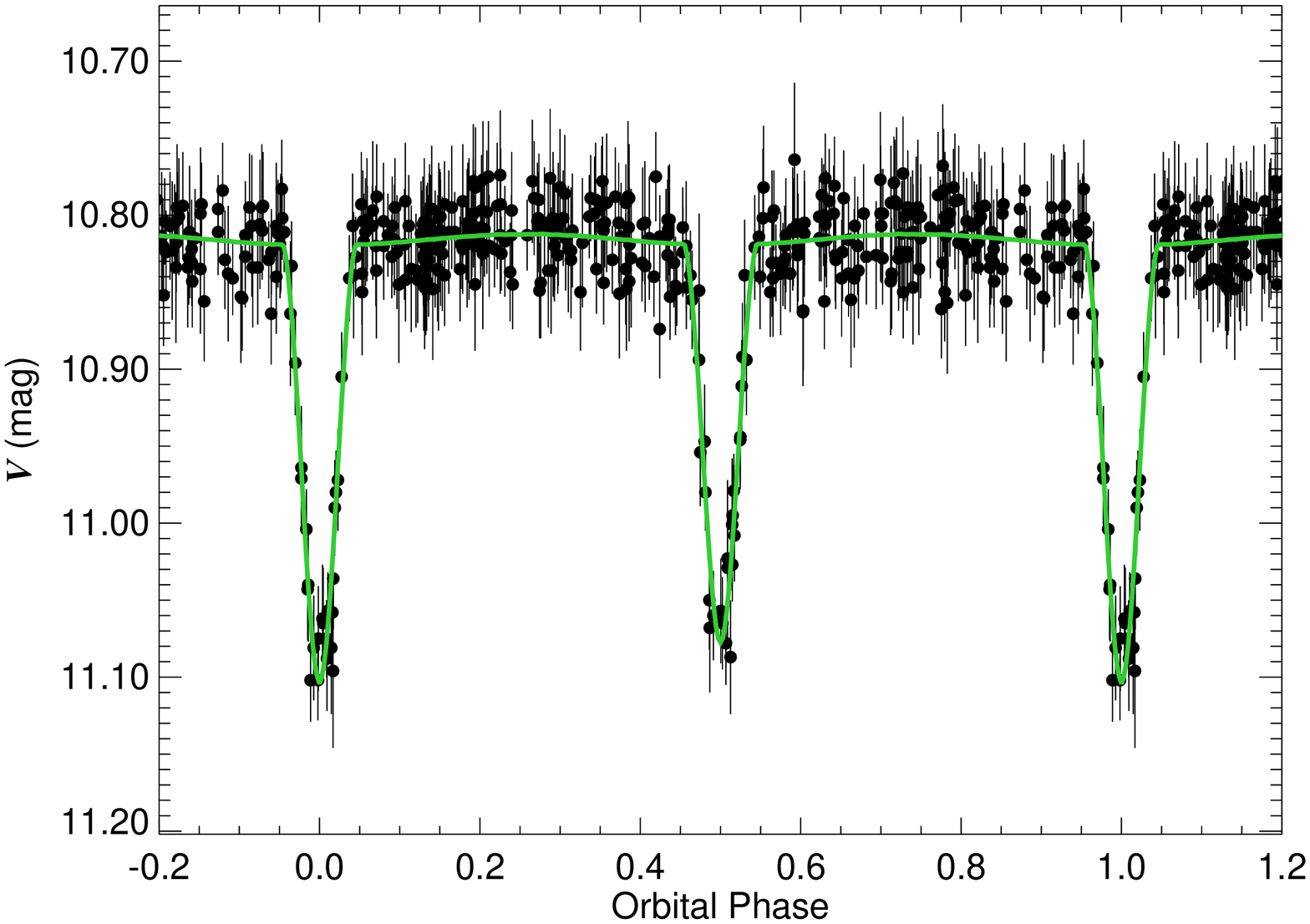}}
\end{center}
\caption{The ASAS $V-$band light curve for ASAS 160851$-$2351.0
  (TYC 6780$-$1523$-$1). Filled circles with lines
  represent data with associated uncertainties. The best fit orbital solution
  listed in Table 2 is shown as a solid line passing through the data.}
\end{figure}
\clearpage

\begin{figure}
\begin{center}
{\includegraphics[angle=90,height=12cm]{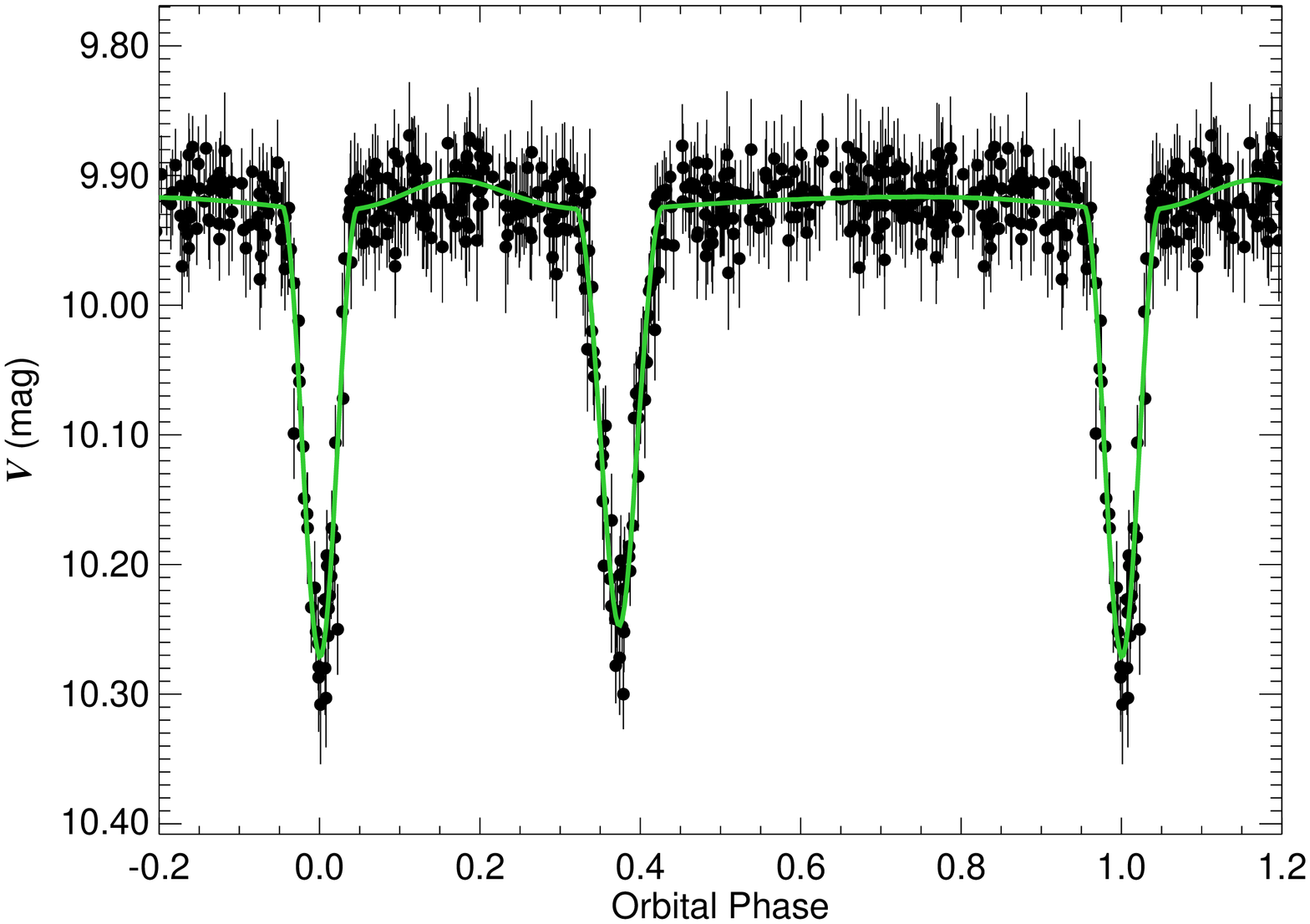}}
\end{center}
\caption{The ASAS $V-$band light curve for ASAS 165354$-$1301.9
  (HD 152451). Filled circles with lines
  represent data with associated uncertainties. The best fit orbital solution
  listed in Table 2 is shown as a solid line passing through the data.}
\end{figure}
\clearpage

\begin{figure}
\begin{center}
{\includegraphics[angle=90,height=12cm]{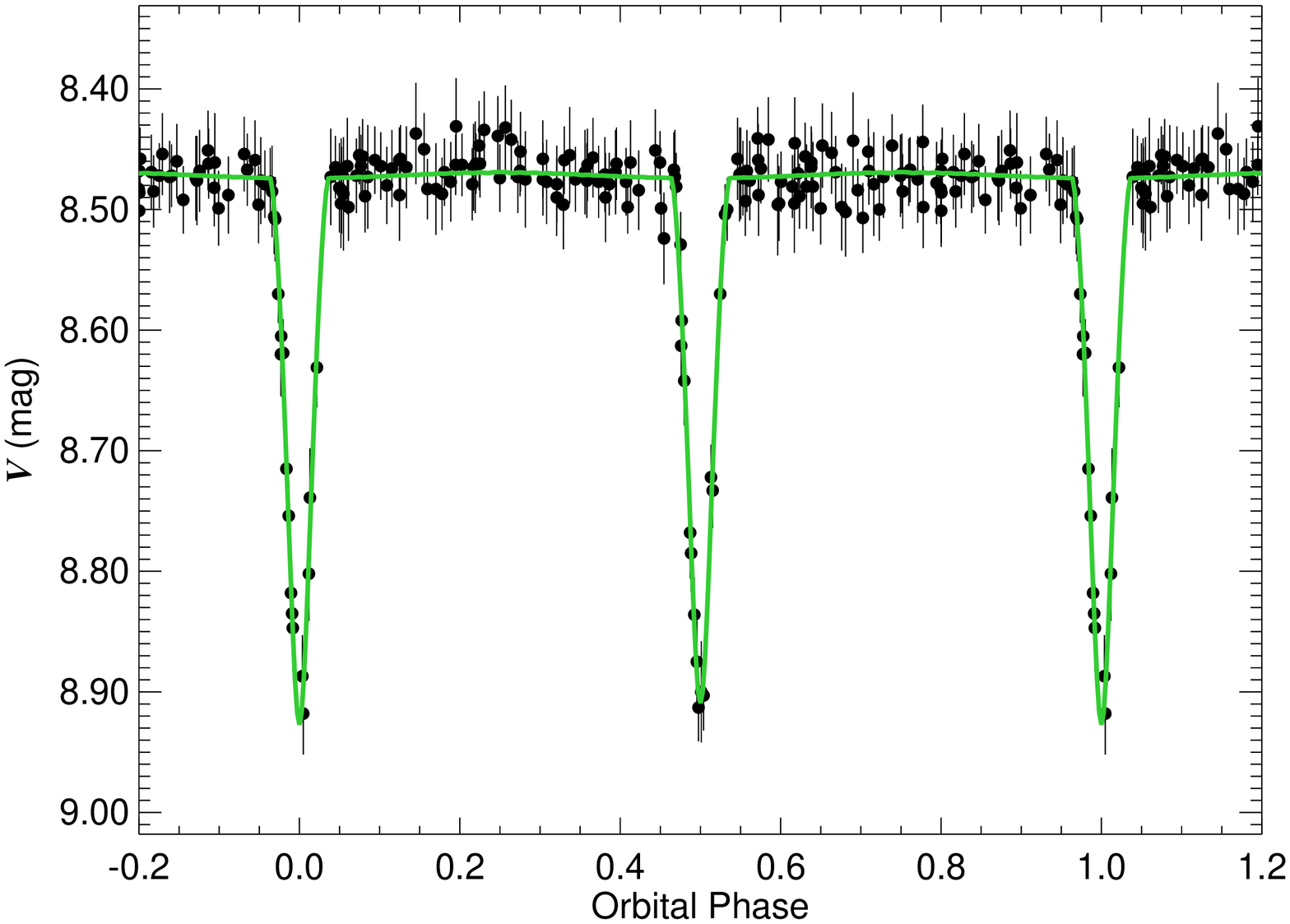}}
\end{center}
\caption{The ASAS $V-$band light curve for ASAS 170158+2348.4
  (HD 154010). Filled circles with lines
  represent data with associated uncertainties. The best fit orbital solution
  listed in Table 2 is shown as a solid line passing through the data.}
\end{figure}
\clearpage

\begin{figure}
\begin{center}
{\includegraphics[angle=90,height=12cm]{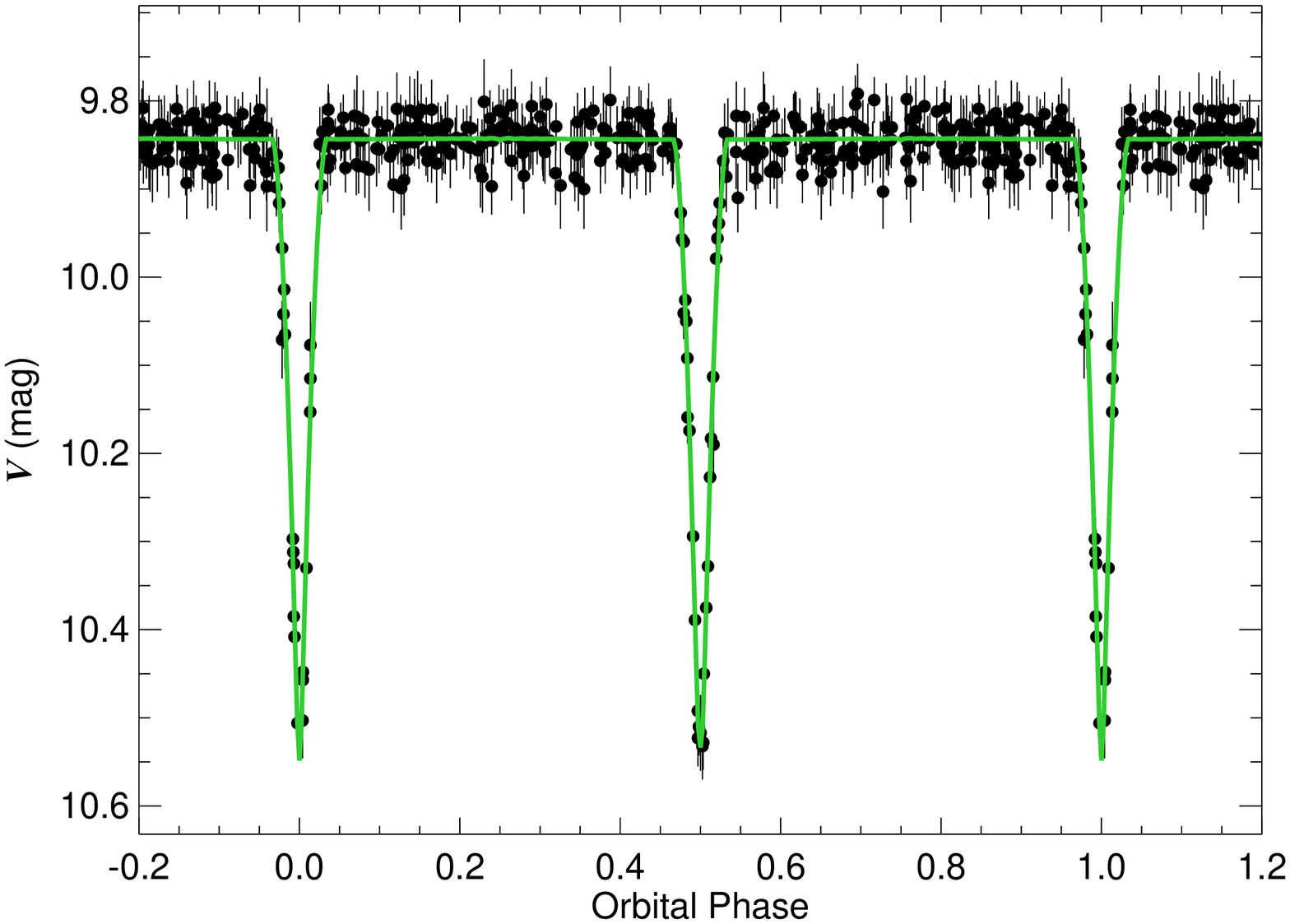}}
\end{center}
\caption{The ASAS $V-$band light curve for ASAS 173421$-$1836.3
  (HD 159246). Filled circles with lines
  represent data with associated uncertainties. The best fit orbital solution
  listed in Table 2 is shown as a solid line passing through the data.}
\end{figure}
\clearpage

\begin{figure}
\begin{center}
{\includegraphics[angle=90,height=12cm]{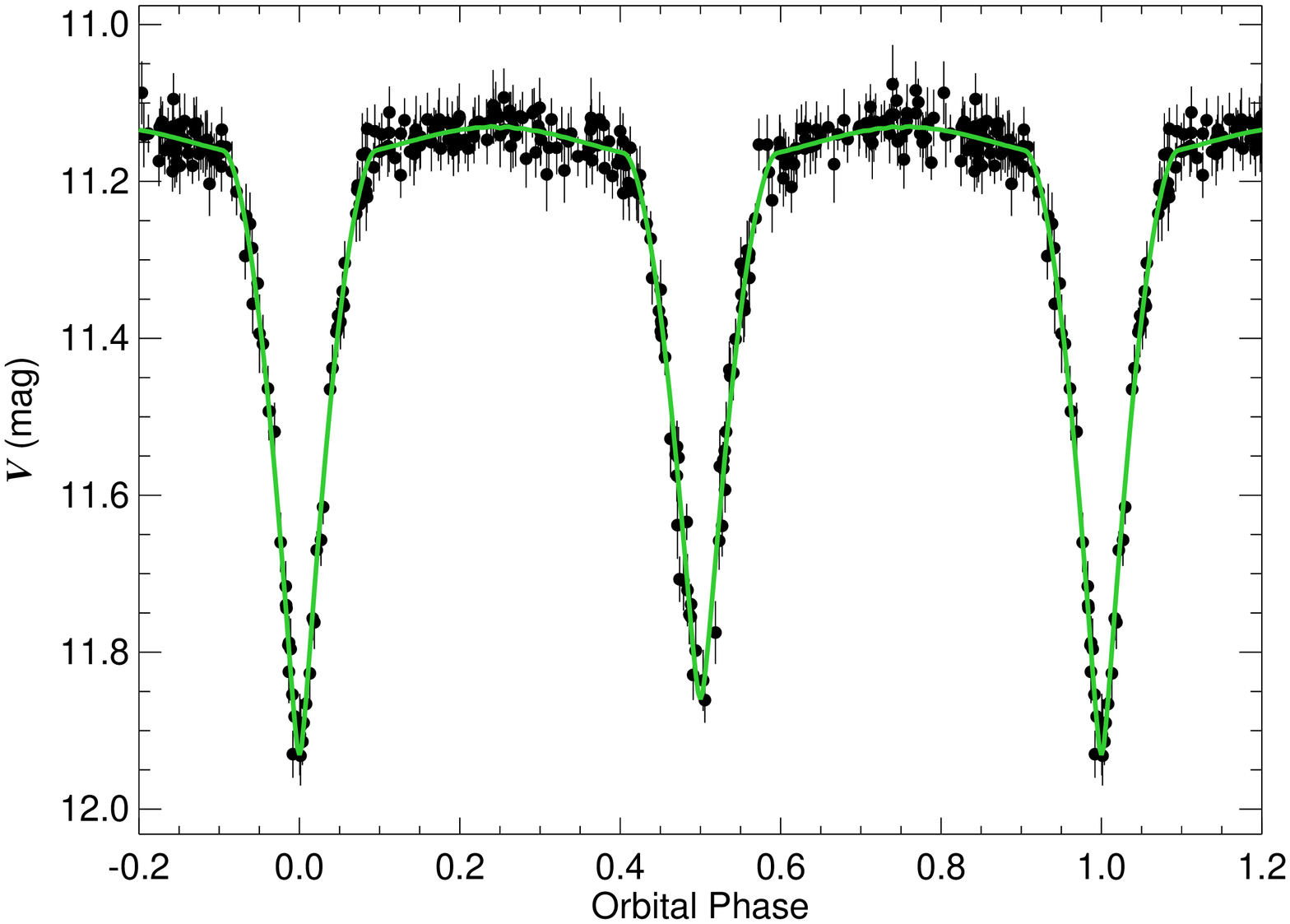}}
\end{center}
\caption{The ASAS $V-$band light curve for ASAS 174104+0747.1
  (V506 Oph). Filled circles with lines
  represent data with associated uncertainties. The best fit orbital solution
  listed in Table 2 is shown as a solid line passing through the data.}
\end{figure}
\clearpage

\begin{figure}
\begin{center}
{\includegraphics[angle=90,height=12cm]{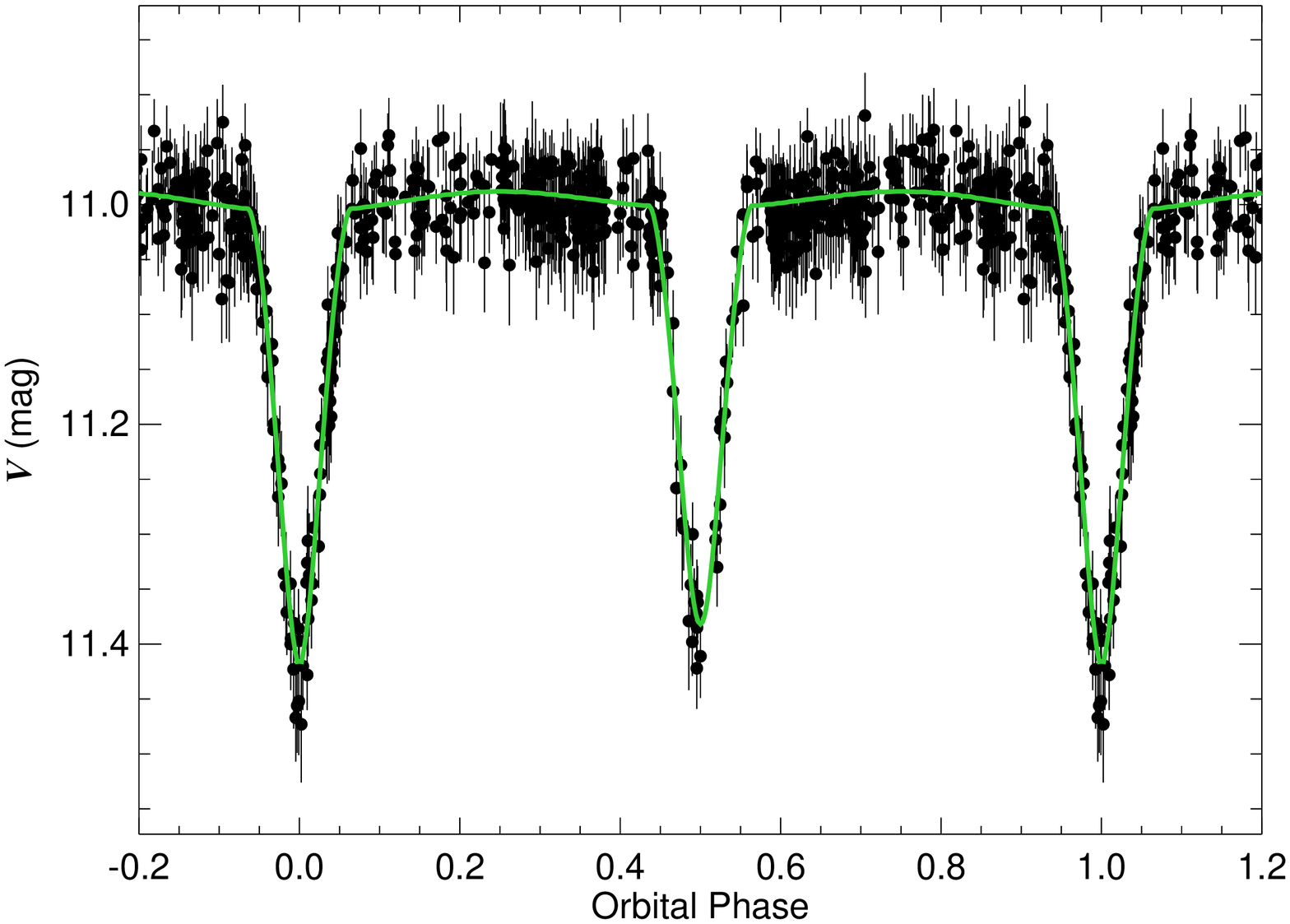}}
\end{center}
\caption{The ASAS $V-$band light curve for ASAS 175659$-$2012.2
  (HD 312444). Filled circles with lines
  represent data with associated uncertainties. The best fit orbital solution
  listed in Table 2 is shown as a solid line passing through the data.}
\end{figure}
\clearpage

\begin{figure}
\begin{center}
{\includegraphics[angle=90,height=12cm]{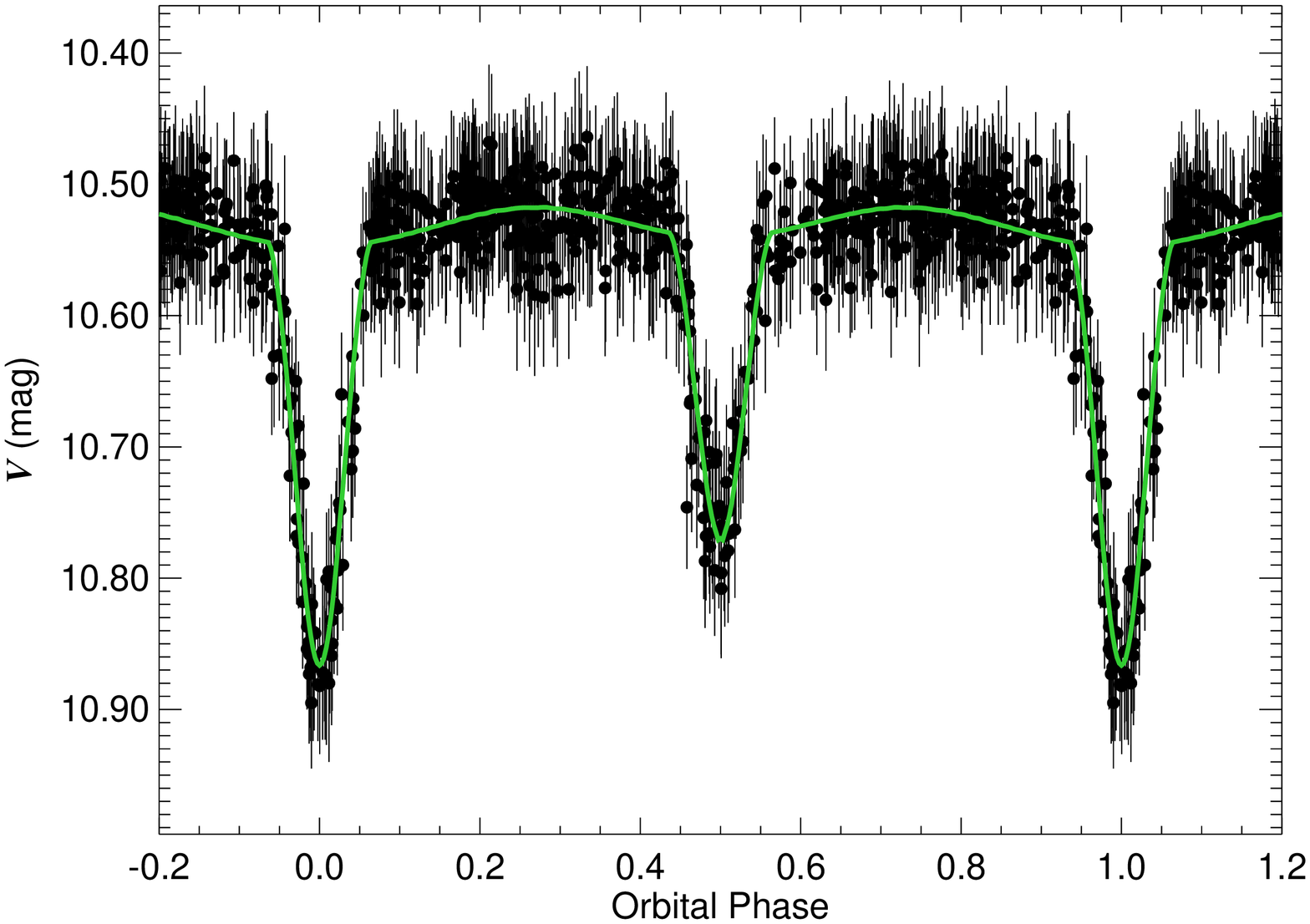}}
\end{center}
\caption{The ASAS $V-$band light curve for ASAS 175859$-$2323.1
  (HD 313508). Filled circles with lines
  represent data with associated uncertainties. The best fit orbital solution
  listed in Table 2 is shown as a solid line passing through the data.}
\end{figure}
\clearpage

\begin{figure}
\begin{center}
{\includegraphics[angle=90,height=12cm]{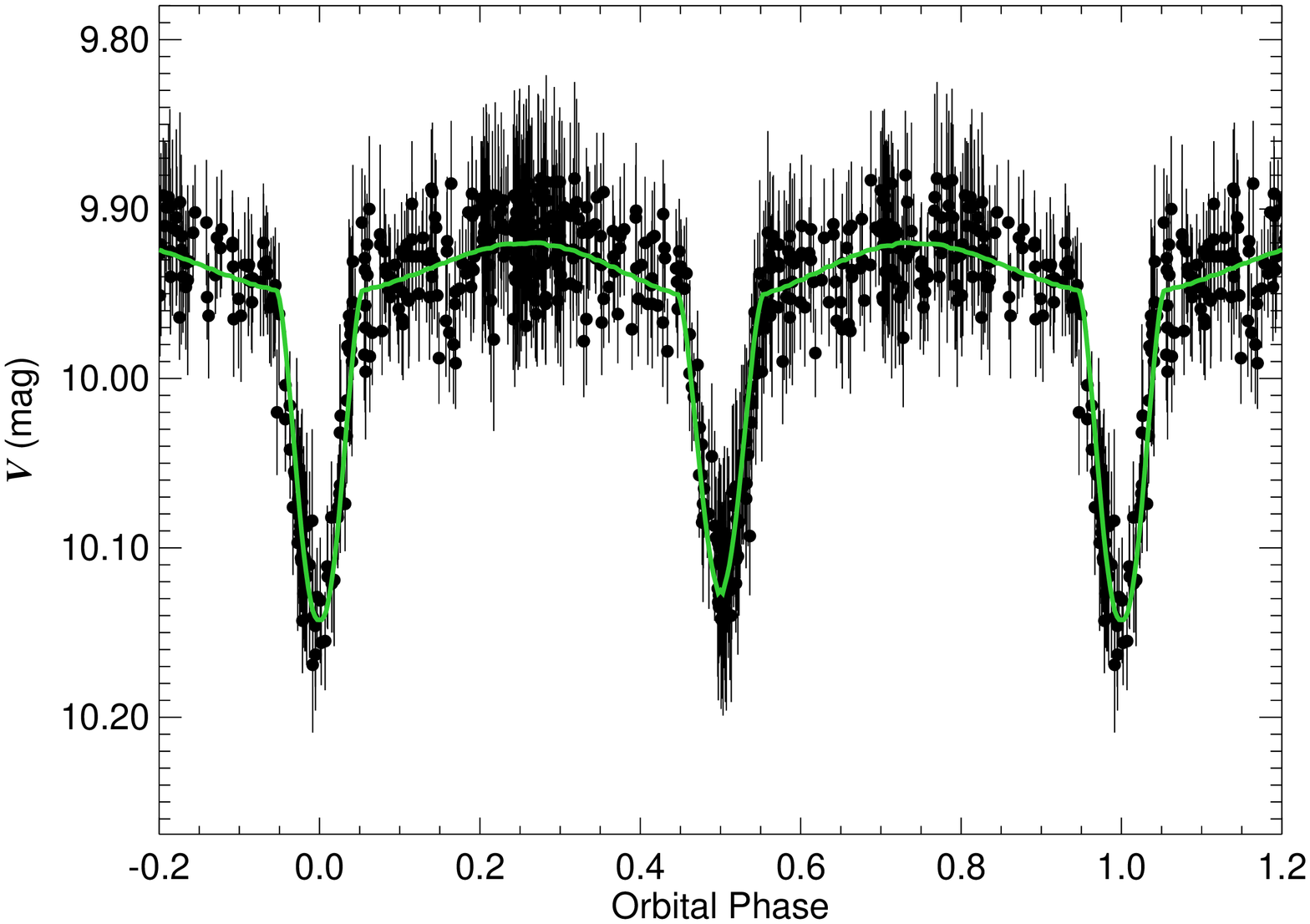}}
\end{center}
\caption{The ASAS $V-$band light curve for ASAS 180903$-$1824.5
  (HD 165890). Filled circles with lines
  represent data with associated uncertainties. The best fit orbital solution
  listed in Table 2 is shown as a solid line passing through the data.}
\end{figure}
\clearpage

\begin{figure}
\begin{center}
{\includegraphics[angle=90,height=12cm]{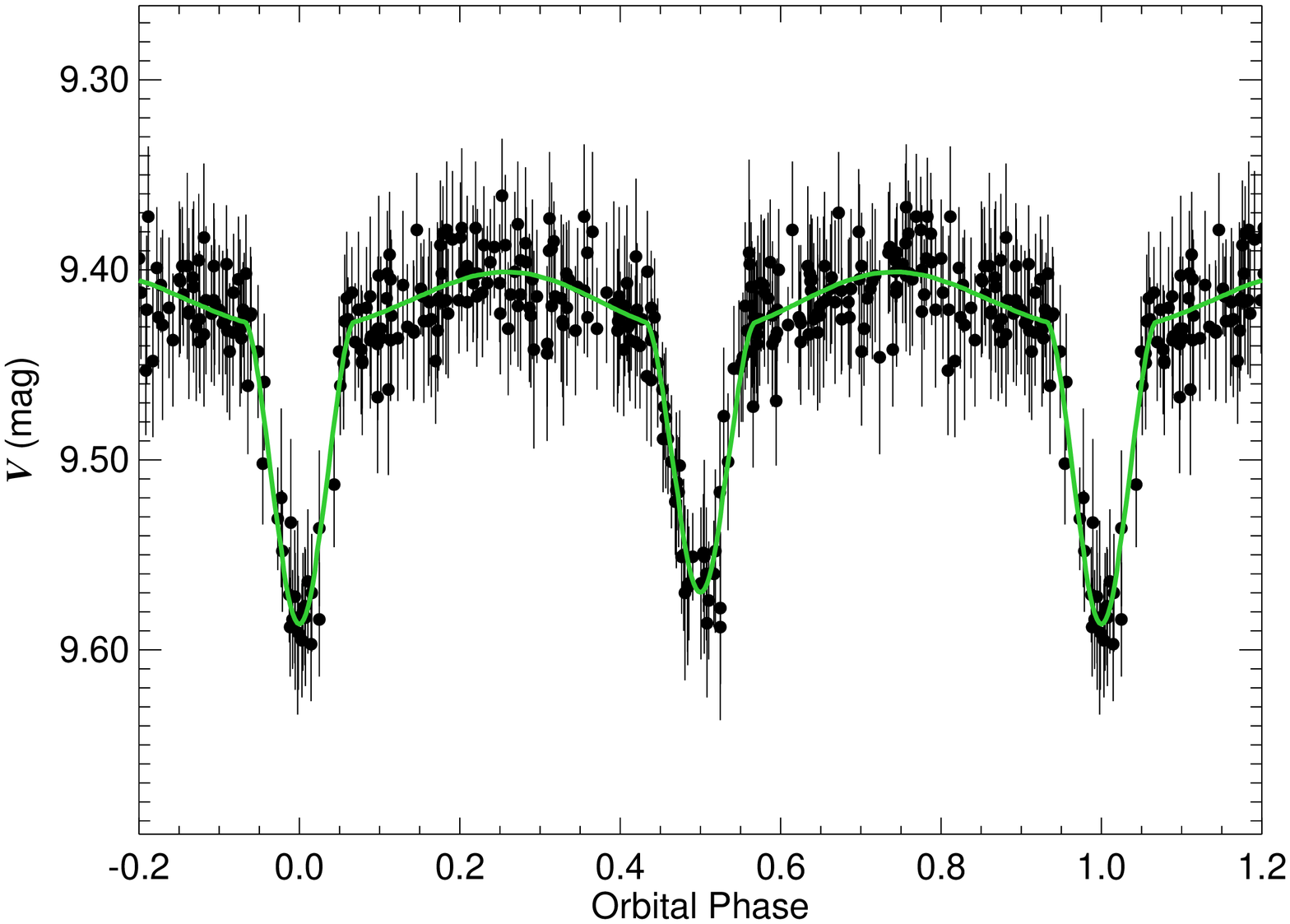}}
\end{center}
\caption{The ASAS $V-$band light curve for ASAS 181025+0047.7
  (HD 166383). Filled circles with lines
  represent data with associated uncertainties. The best fit orbital solution
  listed in Table 2 is shown as a solid line passing through the data.}
\end{figure}
\clearpage

\begin{figure}
\begin{center}
{\includegraphics[angle=90,height=12cm]{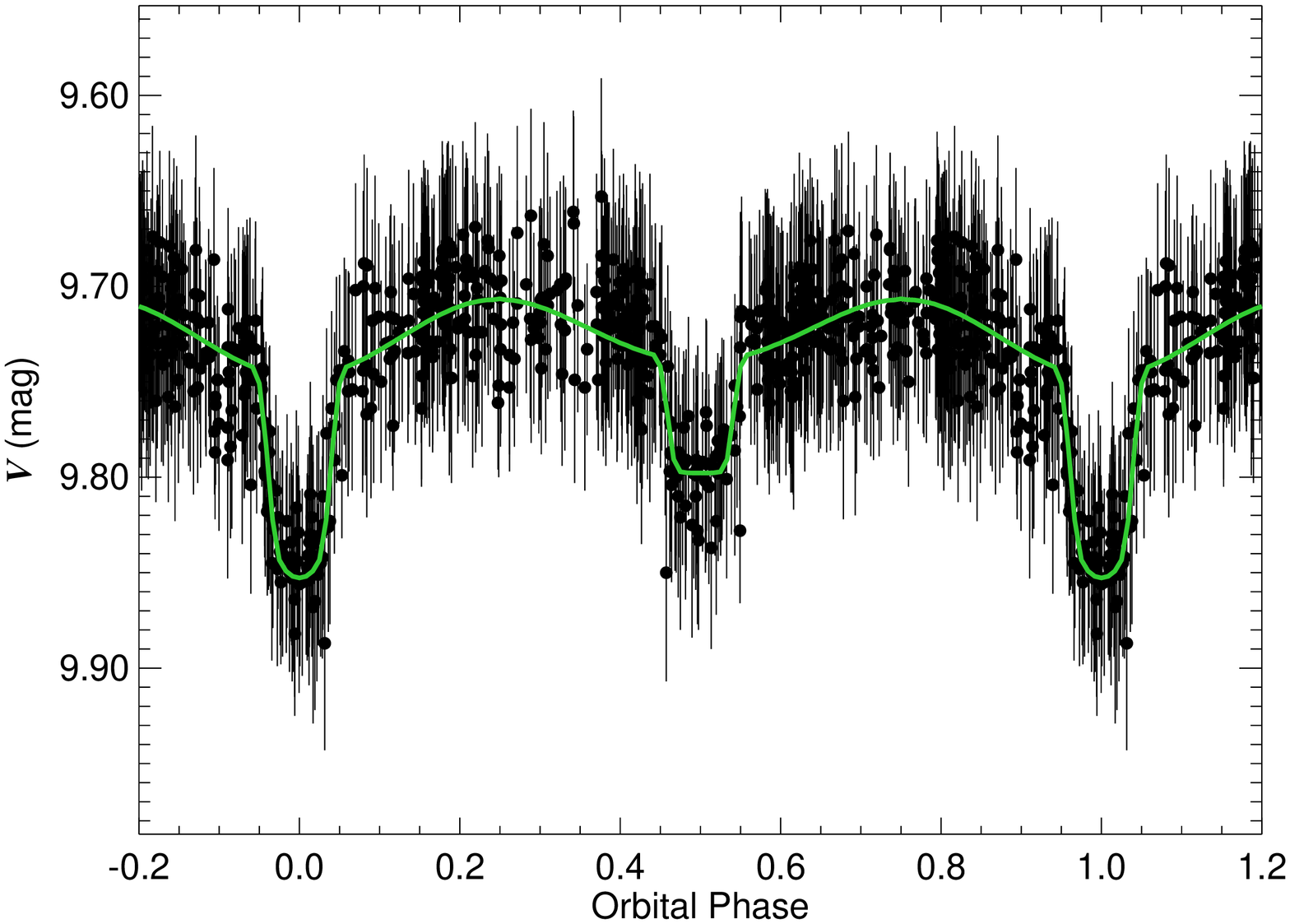}}
\end{center}
\caption{The ASAS $V-$band light curve for ASAS 181328$-$2214.3
  (HD 166851). Filled circles with lines
  represent data with associated uncertainties. The best fit orbital solution
  listed in Table 2 is shown as a solid line passing through the data.}
\end{figure}
\clearpage

\begin{figure}
\begin{center}
{\includegraphics[angle=90,height=12cm]{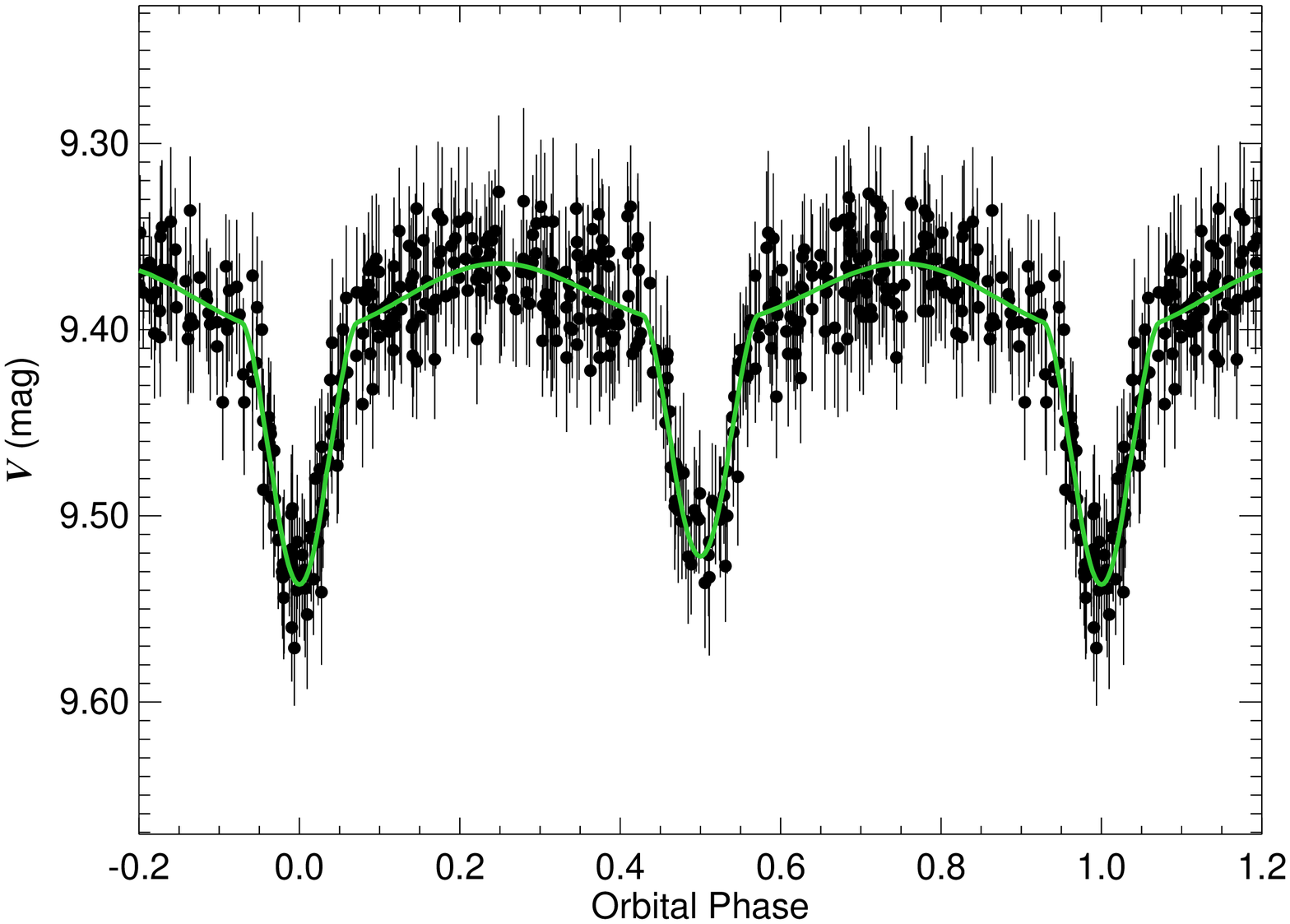}}
\end{center}
\caption{The ASAS $V-$band light curve for ASAS 181909$-$1410.0
  (HD 168207). Filled circles with lines
  represent data with associated uncertainties. The best fit orbital solution
  listed in Table 2 is shown as a solid line passing through the data.}
\end{figure}
\clearpage

\begin{figure}
\begin{center}
{\includegraphics[angle=90,height=12cm]{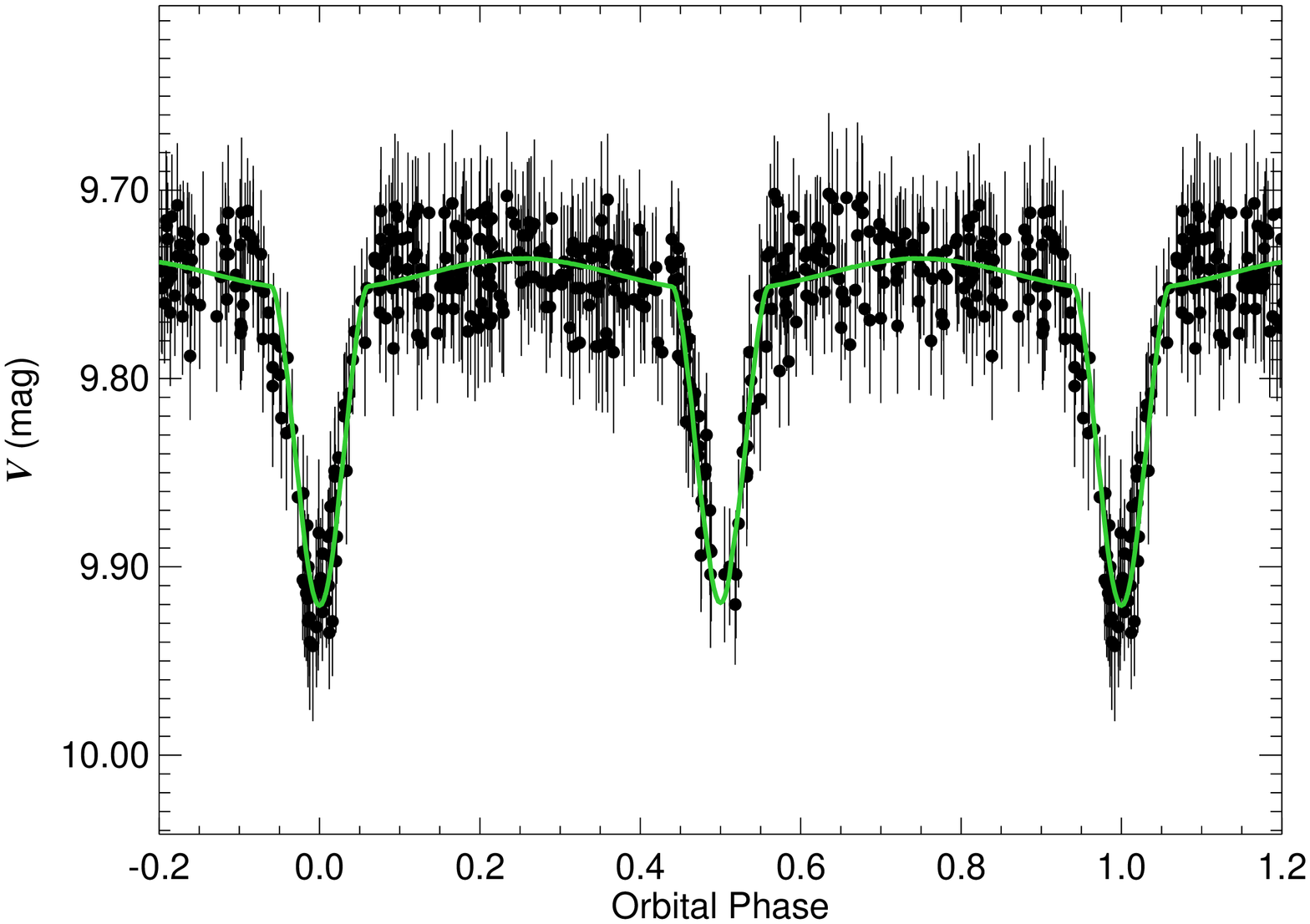}}
\end{center}
\caption{The ASAS $V-$band light curve for ASAS 183129$-$1918.8
  (BD $-$19 5039). Filled circles with lines
  represent data with associated uncertainties. The best fit orbital solution
  listed in Table 2 is shown as a solid line passing through the data.}
\end{figure}
\clearpage

\begin{figure}
\begin{center}
{\includegraphics[angle=90,height=12cm]{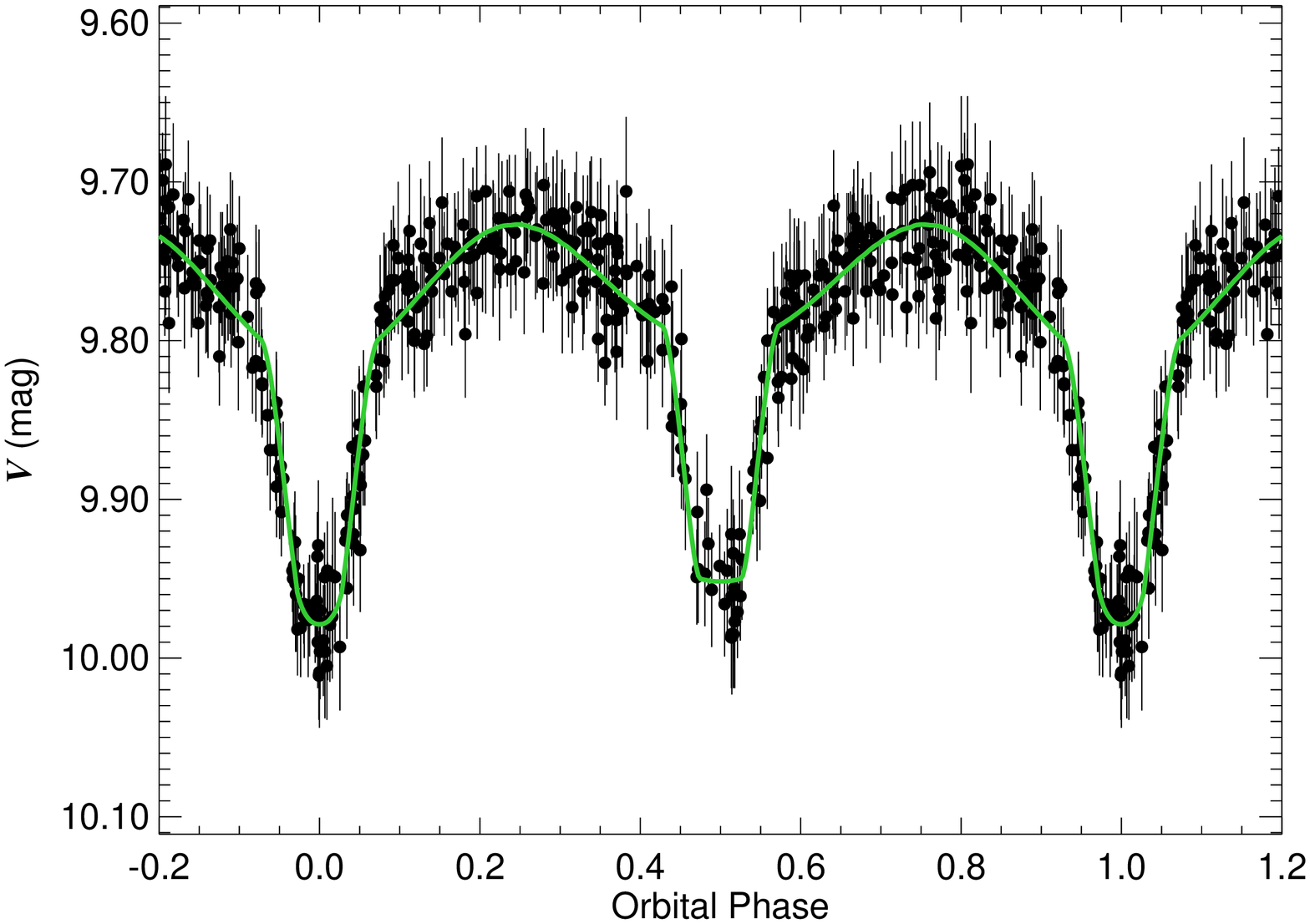}}
\end{center}
\caption{The ASAS $V-$band light curve for ASAS 183219$-$1117.4
  (BD $-$11 4667). Filled circles with lines
  represent data with associated uncertainties. The best fit orbital solution
  listed in Table 2 is shown as a solid line passing through the data.}
\end{figure}
\clearpage

\begin{figure}
\begin{center}
{\includegraphics[angle=90,height=12cm]{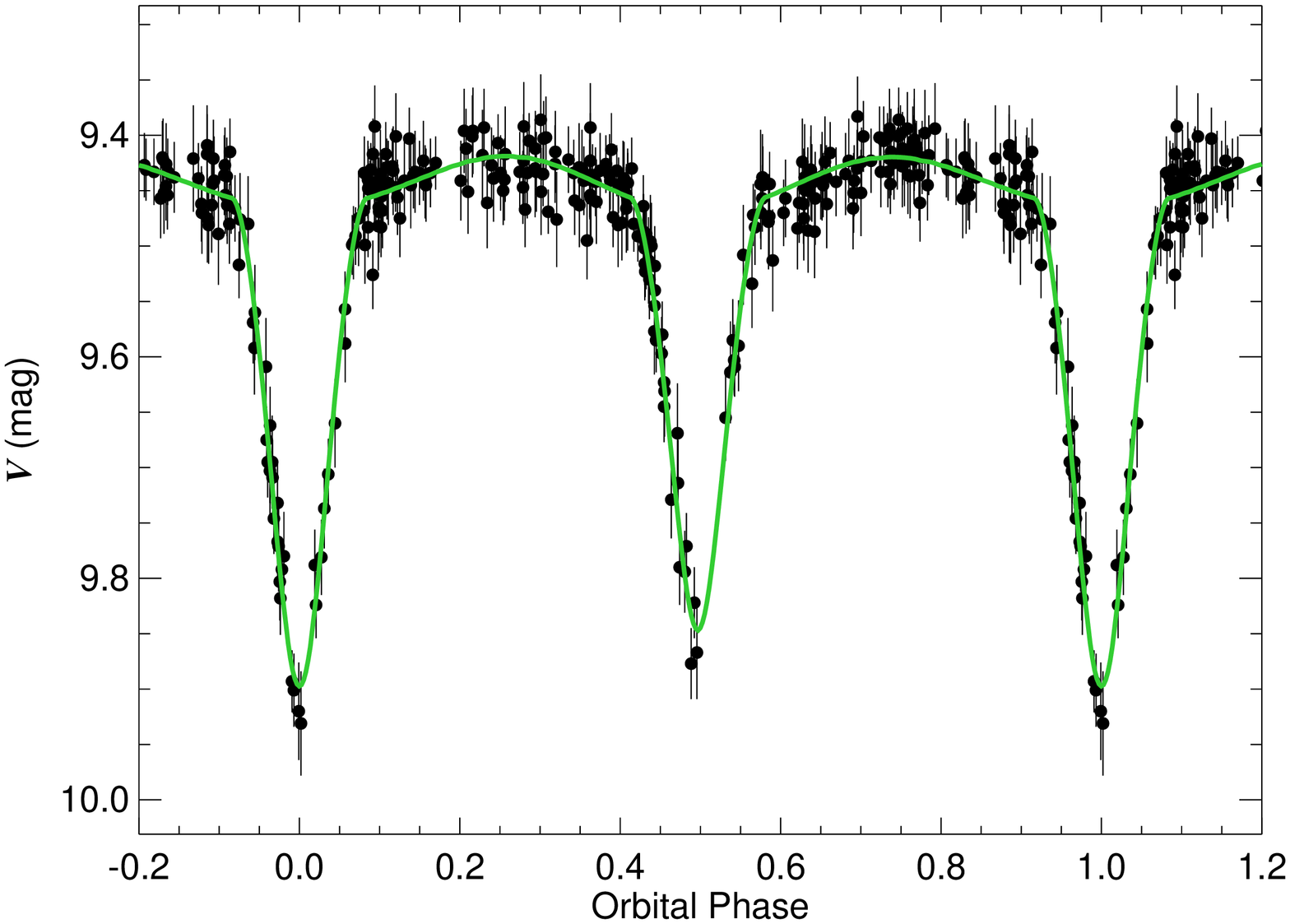}}
\end{center}
\caption{The ASAS $V-$band light curve for ASAS 184223+1158.9
  (BD +11 3569). Filled circles with lines
  represent data with associated uncertainties. The best fit orbital solution
  listed in Table 2 is shown as a solid line passing through the data.}
\end{figure}
\clearpage

\begin{figure}
\begin{center}
{\includegraphics[angle=90,height=12cm]{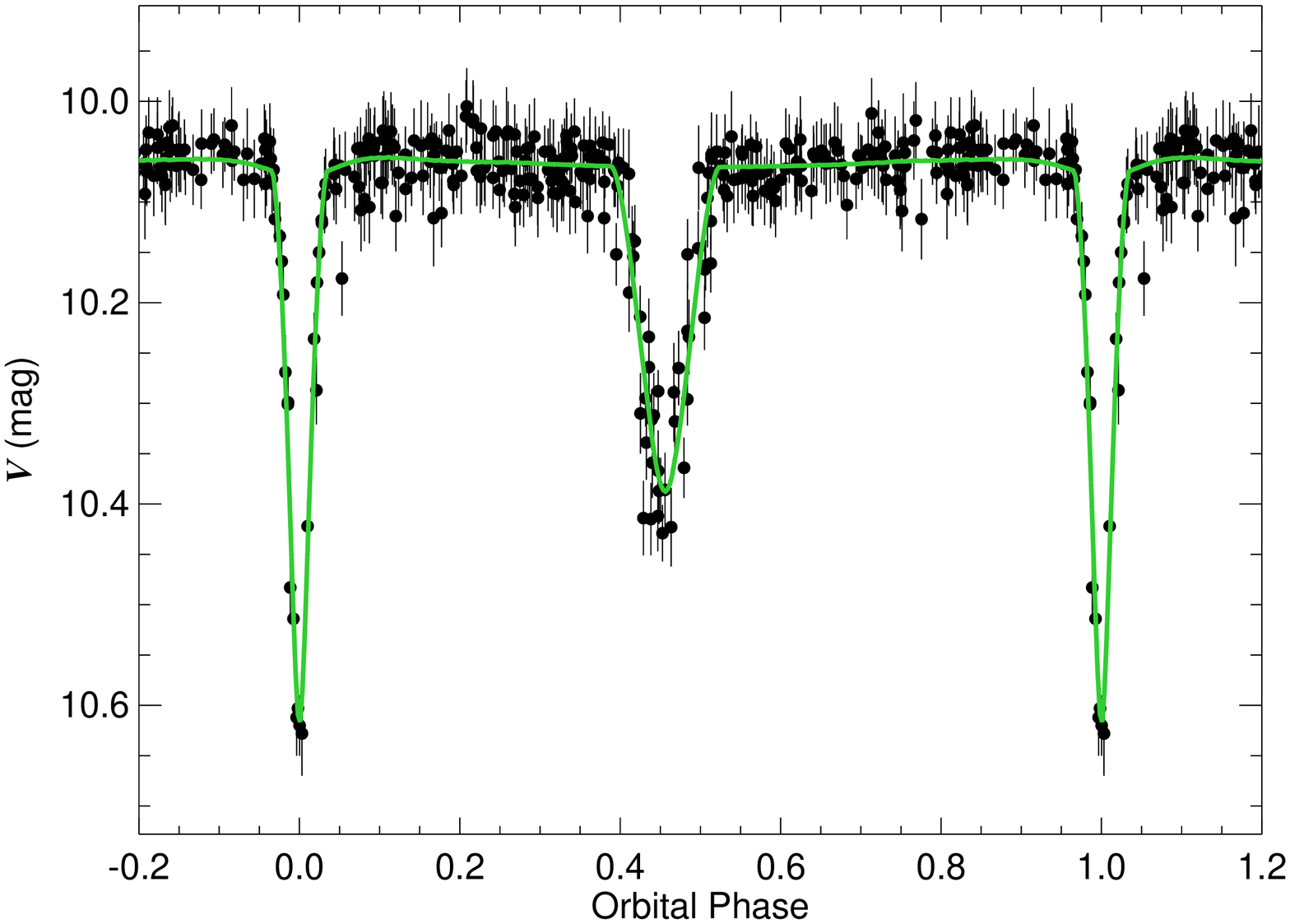}}
\end{center}
\caption{The ASAS $V-$band light curve for ASAS 184327+0841.5
  (TYC 1025$-$1524$-$1). Filled circles with lines
  represent data with associated uncertainties. The best fit orbital solution
  listed in Table 2 is shown as a solid line passing through the data.}
\end{figure}
\clearpage

\begin{figure}
\begin{center}
{\includegraphics[angle=90,height=12cm]{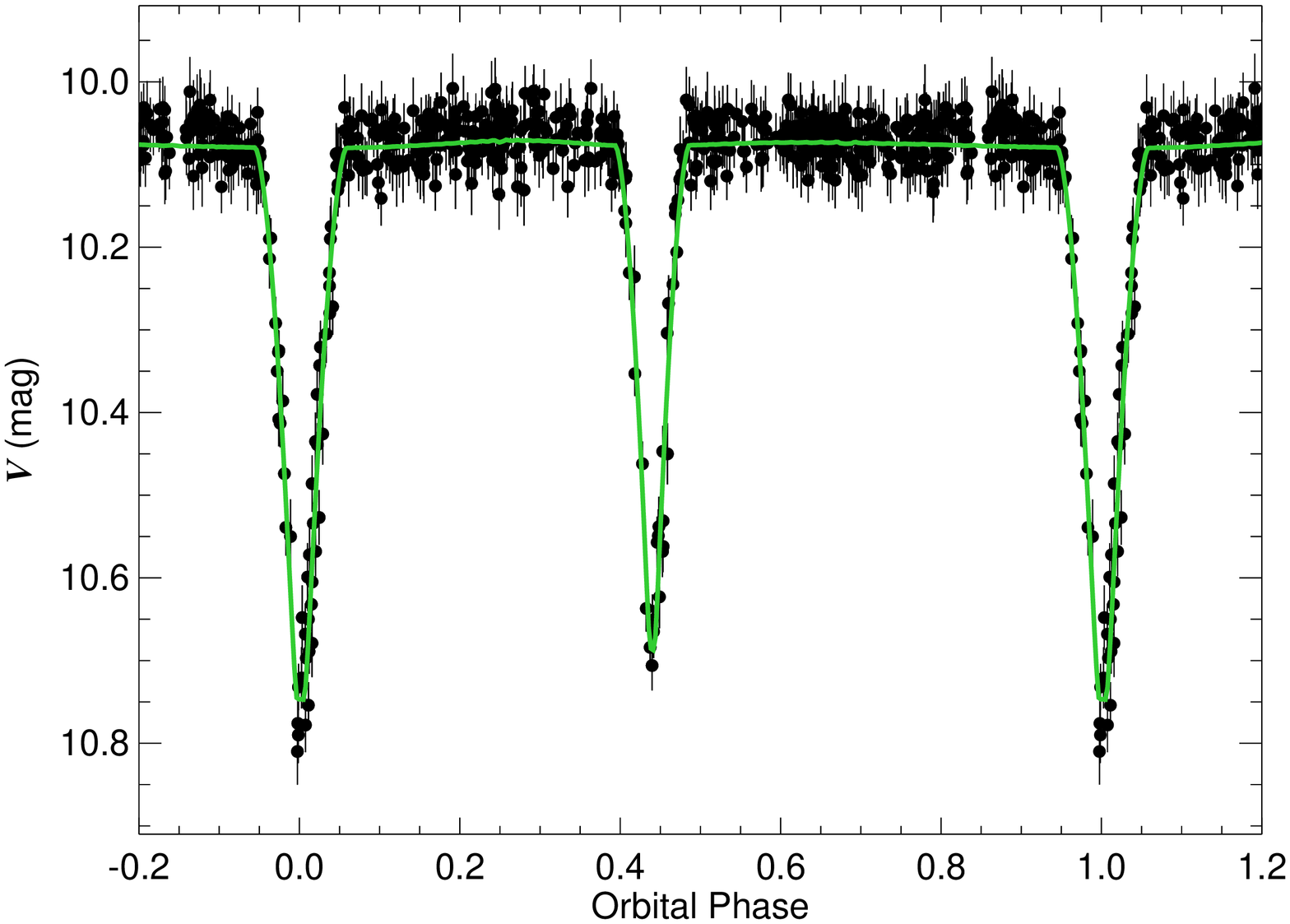}}
\end{center}
\caption{The ASAS $V-$band light curve for ASAS 184436$-$1923.4
  (YY Sgr). Filled circles with lines
  represent data with associated uncertainties. The best fit orbital solution
  listed in Table 2 is shown as a solid line passing through the data.}
\end{figure}
\clearpage

\begin{figure}
\begin{center}
{\includegraphics[angle=90,height=12cm]{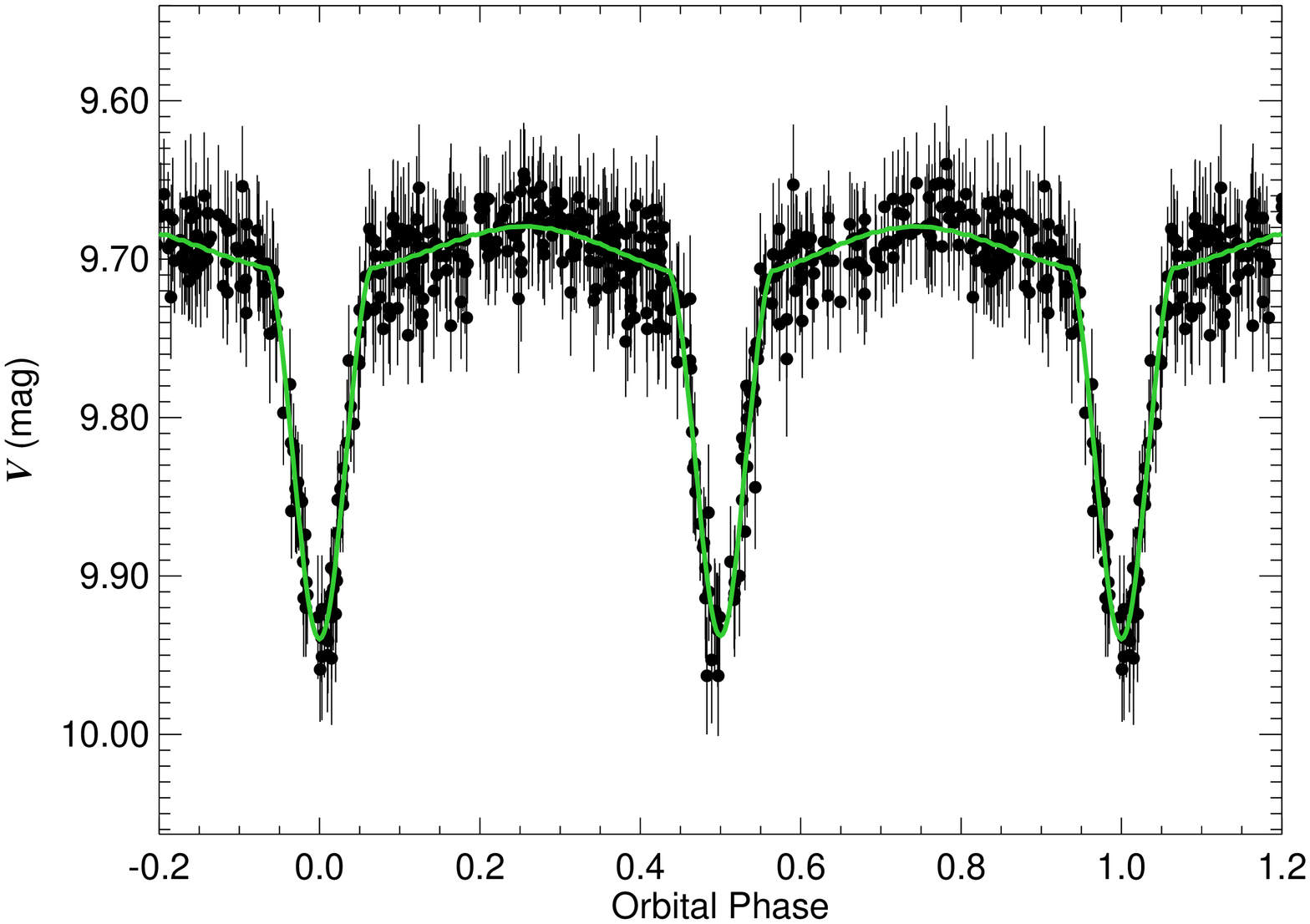}}
\end{center}
\caption{The ASAS $V-$band light curve for ASAS 185051$-$1354.6
  (HD 174397). Filled circles with lines
  represent data with associated uncertainties. The best fit orbital solution
  listed in Table 2 is shown as a solid line passing through the data.}
\end{figure}
\clearpage

\begin{figure}
\begin{center}
{\includegraphics[angle=90,height=12cm]{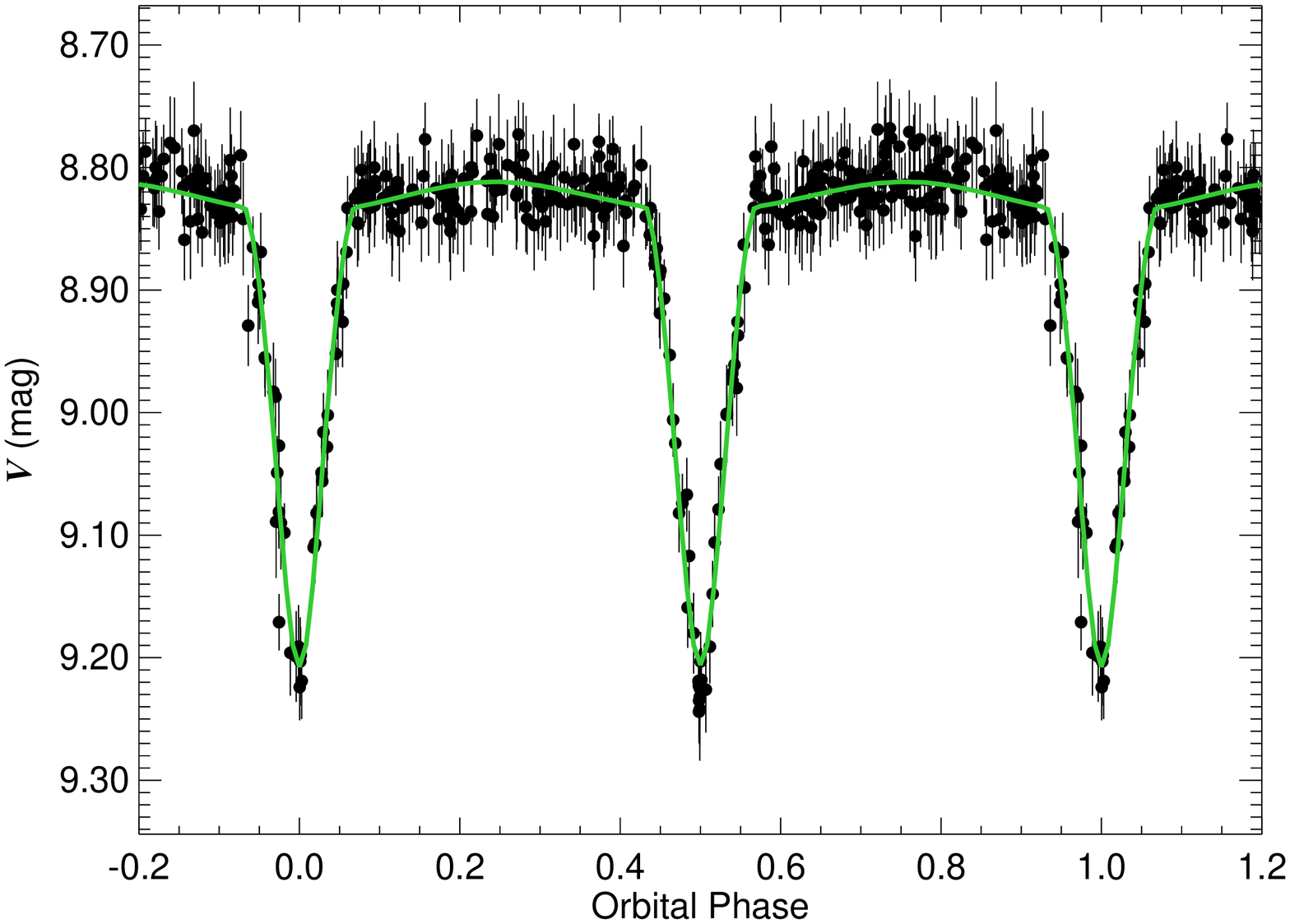}}
\end{center}
\caption{The ASAS $V-$band light curve for ASAS 194334$-$0904.0
  (V1461 Aql). Filled circles with lines
  represent data with associated uncertainties. The best fit orbital solution
  listed in Table 2 is shown as a solid line passing through the data.}
\end{figure}
\clearpage

\begin{figure}
\begin{center}
{\includegraphics[angle=90,height=12cm]{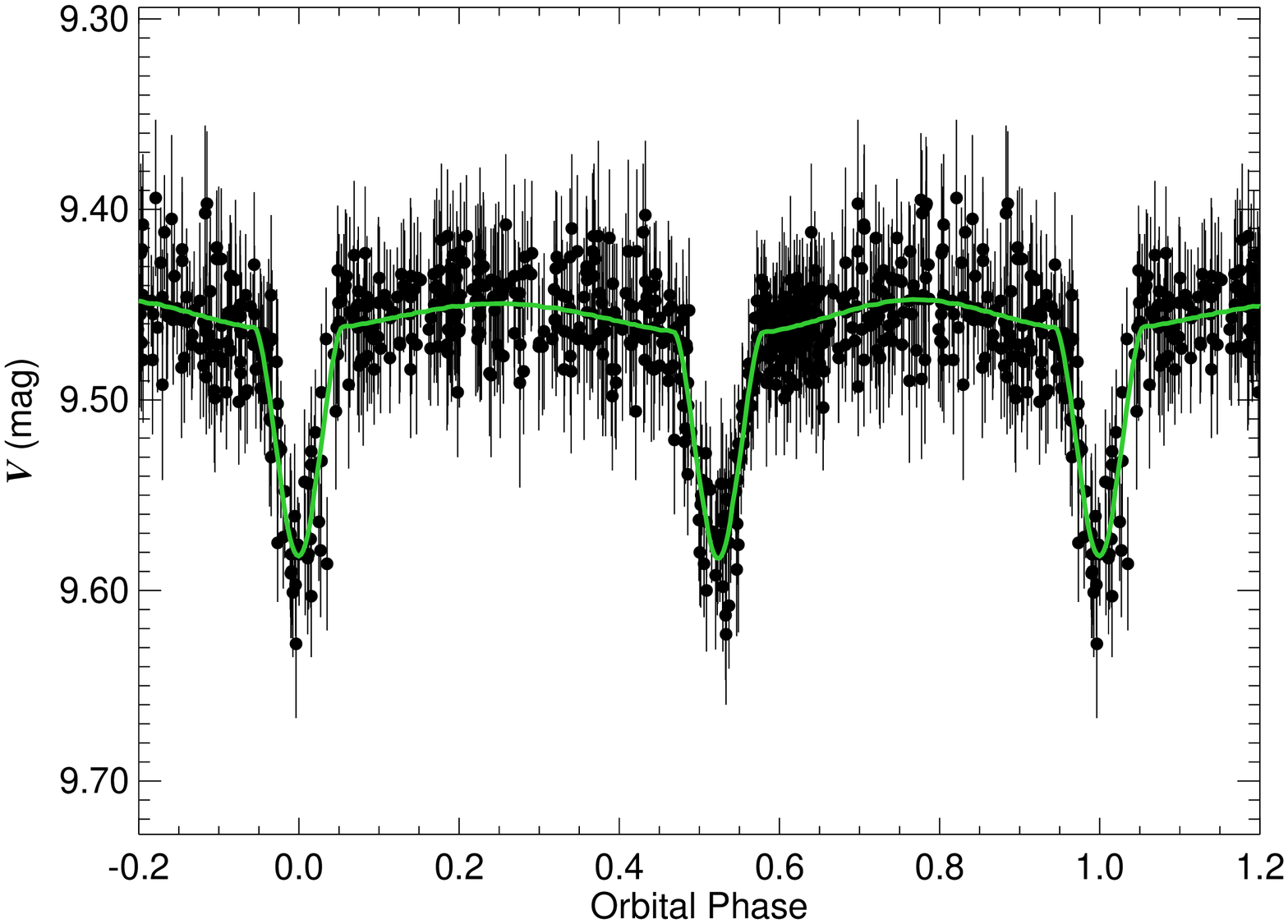}}
\end{center}
\caption{The ASAS $V-$band light curve for ASAS 195342+0205.4
  (HD 188153). Filled circles with lines
  represent data with associated uncertainties. The best fit orbital solution
  listed in Table 2 is shown as a solid line passing through the data.}
\end{figure}
\clearpage

\begin{figure}
\begin{center}
{\includegraphics[angle=90,height=12cm]{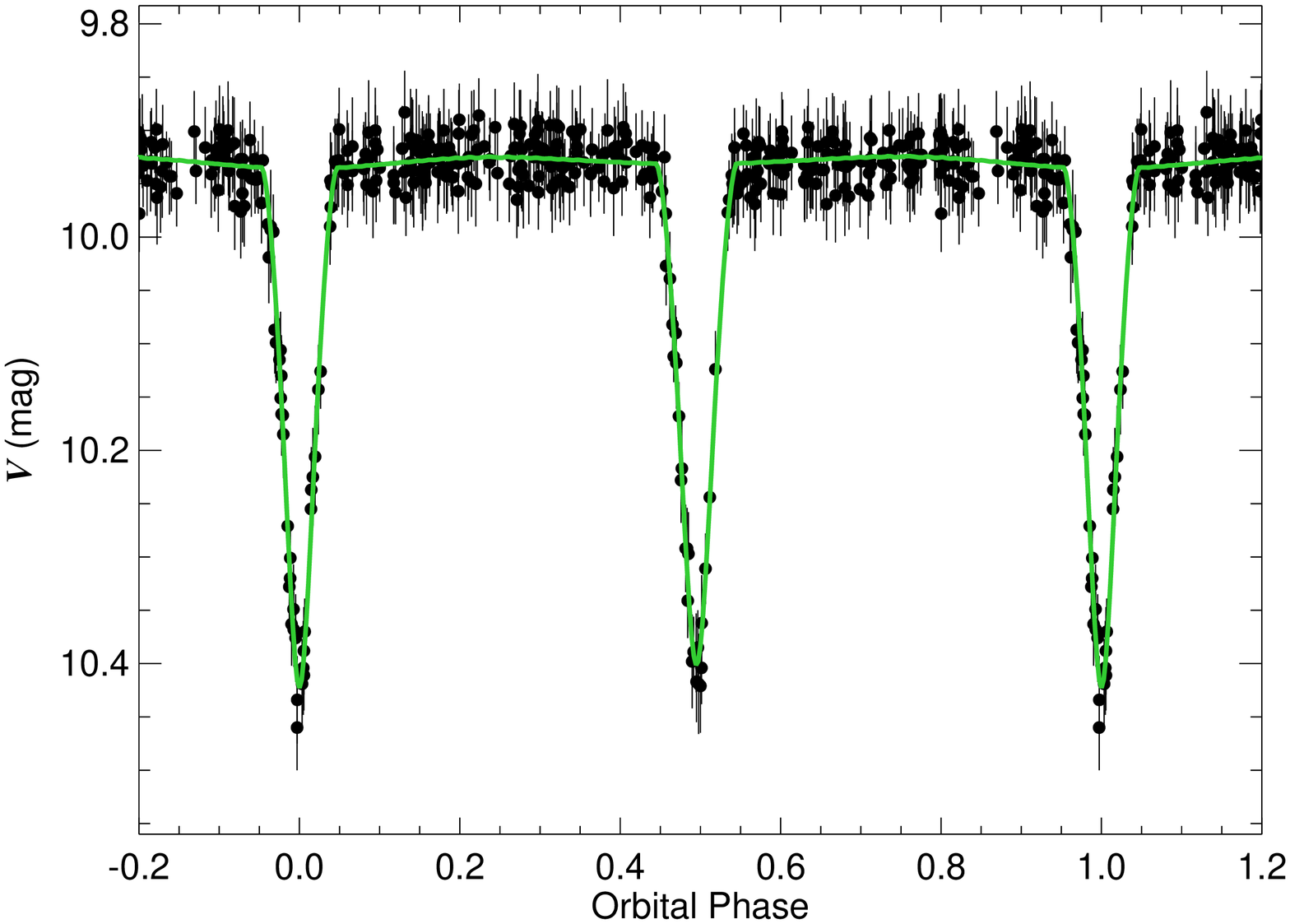}}
\end{center}
\caption{The ASAS $V-$band light curve for ASAS 195613+1630.9
  (HD 354110). Filled circles with lines
  represent data with associated uncertainties. The best fit orbital solution
  listed in Table 2 is shown as a solid line passing through the data.}
\end{figure}
\clearpage

\begin{figure}
\begin{center}
{\includegraphics[angle=90,height=12cm]{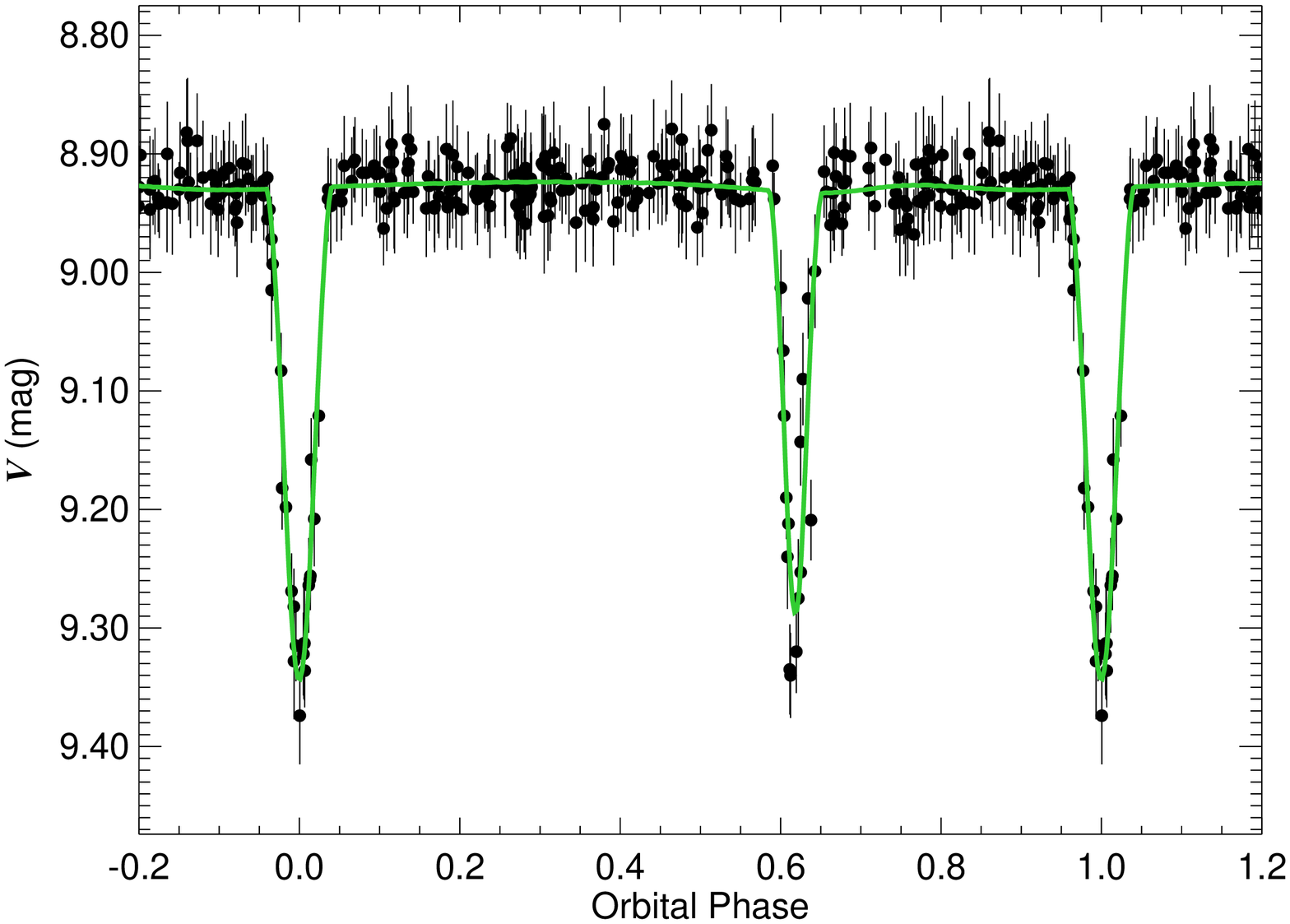}}
\end{center}
\caption{The ASAS $V-$band light curve for ASAS 205642+1153.0
  (HD 199428). Filled circles with lines
  represent data with associated uncertainties. The best fit orbital solution
  listed in Table 2 is shown as a solid line passing through the data.}
\end{figure}
\clearpage

\end{document}